\documentclass[12pt]{article}
\usepackage{amsfonts,amsthm,amsmath,amssymb,upgreek,bm}
\usepackage{hyperref}
\usepackage[paper=letterpaper,margin=.85in]{geometry}
\usepackage{graphicx}
\usepackage{units}
\usepackage{float}
\usepackage[export]{adjustbox}
\usepackage{xcolor}
\usepackage[shortlabels]{enumitem}
\usepackage[mathscr]{euscript}
\usepackage[normalem]{ulem}

\newcommand{\be}{\begin{equation}}
\newcommand{\ee}{\end{equation}}
\newcommand{\bea}{\begin{eqnarray}}
\newcommand{\eea}{\end{eqnarray}}
\renewcommand{\tilde}{\widetilde}
\renewcommand{\i}{\mathrm{i}}
\renewcommand{\d}{\mathrm{d}}

\newcommand{\initial}{|\hspace{-2pt}\cup\cup\hspace{1pt}\rangle}

\numberwithin{equation}{section}

\def\tr{\text{Tr}}

\begin{document}
\thispagestyle{empty}

\vspace*{2.5cm}
\begin{center}

{\bf {\LARGE Subleading Weingartens}}\\

\begin{center}

\vspace{1cm}

{\bf Douglas Stanford, Zhenbin Yang, and Shunyu Yao}\\
 \bigskip \rm

\bigskip 

Stanford Institute for Theoretical Physics,\\Stanford University, Stanford, CA 94305

\rm
  \end{center}

\vspace{2.5cm}
{\bf Abstract}
\end{center}
\begin{quotation}
\noindent

Haar integrals over the unitary group contain subleading terms that are needed for unitarity. We study analogous effects in the time evolution operators of JT gravity and Brownian SYK. In JT gravity with bulk matter we find an explanation for the first subleading terms, and in Brownian SYK we find configurations that can explain the full series. An important role is played by slightly off-shell modes that are exponentially amplified by chaos.

\end{quotation}

\setcounter{page}{0}
\setcounter{tocdepth}{2}
\setcounter{footnote}{0}
\newpage

\parskip 0.1in
 
\setcounter{page}{2}
\tableofcontents

\newpage

\section{Introduction}
In a chaotic quantum system, the time evolution operator $U(t) = e^{-\i H t}$ for late times is a complicated unitary matrix. In the crudest approximation, it is sometimes useful to model it as a {\it random} unitary matrix, drawn from the Haar measure on the unitary group. 

This model gets quite a lot wrong, such as the lack of a simple conserved $H$. However, it does manage to combine thorough scrambling with exact unitarity, and it has led to insights into black hole physics \cite{Hayden:2007cs,Sekino:2008he} and to computable models of scrambling and entropy dynamics \cite{Hosur:2015ylk,nahum2017quantum,von2018operator,nahum2018operator}. In these studies, the main workhorse formula is
\begin{align}\label{UUUU}
\int \hspace{-2pt}\mathrm{d}U\, U_{i_1}^{ \ j_1}U_{i_2}^{ \ j_2}(U^\dagger)_{k_1}^{ \ l_1}(U^\dagger)_{k_2}^{ \ l_2}& = \frac{1}{L^2 {-} 1}\left(\delta_{i_1}^{l_1}\delta_{i_2}^{l_2}\delta_{k_1}^{j_1}\delta_{k_2}^{j_2}+\delta_{i_1}^{l_2}\delta_{i_2}^{l_1}\delta_{k_1}^{j_2}\delta_{k_2}^{j_1} -\frac{1}{L}\delta_{i_1}^{l_2}\delta_{i_2}^{l_1}\delta_{k_1}^{j_1}\delta_{k_2}^{j_2}-\frac{1}{L}\delta_{i_1}^{l_1}\delta_{i_2}^{l_2}\delta_{k_1}^{j_2}\delta_{k_2}^{j_1}\right).
\end{align}
The coefficients here are known as ``Weingarten functions'' \cite{collins2003moments,collins2006integration}, depending on $L$, the dimension of the Hilbert space. The first two terms dominate for large $L$, and they have played the most important role in applications of (\ref{UUUU}), but they correspond to Wick contractions of $U$ and $\bar U$, which would also be there in a simpler ensemble that does not respect exact unitarity. By contrast, the last two terms, and the overall prefactor, are nontrivial consequences of unitarity. The main goal of this paper is to understand whether these subleading terms also have analogs in black hole physics. Our motivation for doing this is to better understand to what extent the simple semiclassical gravity description of a black hole incorporates exact unitarity.

As an example, in the random unitary model, one can compute the density matrix of the radiation of a partially evaporated black hole that starts out in state $|a\rangle$:
\be
\rho_R^{(a)} = \tr_{B}\Big(U |a\rangle\langle a| U^\dagger\Big).
\ee
Here $R$ represents the Hawking radiation after partial evaporation, and $B$ represents the remaining black hole. Using (\ref{UUUU}), the overlap of two such density matrices is
\be\label{rhoarhob}
\int \mathrm{d} U \ \tr(\rho^{(a)}_R\rho^{(b)}_R) = \frac{1}{1 - L^{-2}}\left[\frac{1}{L_R} + \frac{|\langle a|b\rangle|^2}{L_B}  - \frac{1}{L_RL_B^2}- \frac{|\langle a|b\rangle|^2}{L_BL_R^2}\right],
\ee
where $L_R$ and $L_B$ are Hilbert space dimensions and $L \equiv L_BL_R$. The first two terms on the RHS come from the leading ``Wick contraction'' terms in (\ref{UUUU}), and they have known gravitational analogs. In particular, they correspond to the contribution of two black holes and from a replica wormhole \cite{Almheiri:2019qdq,Penington:2019kki}; these trade off against each other in the (Renyi) Page curve \cite{Page:1993df,Penington:2019npb,Almheiri:2019psf}. The remaining terms are subleading, but important for recovering unitarity. For example, the third term in brackets ensures that when the black hole has evaporated fully ($L_B = 1$), the result is zero if the initial states were orthogonal. Does it have a gravity analog?

As another example, one can compute the out-of-time-order correlator (OTOC) with time evolution modeled by a random $U$ \cite{Roberts:2016hpo,Yoshida:2017non}: 
\begin{align}\label{OTOC}
\int\mathrm{d} U \ \big\langle (U^\dagger &W U) \, V \,(U^\dagger W U) \, V\big\rangle 
\\ &=\frac{1}{1-L^{-2}}\left[\langle WW\rangle\langle V\rangle\langle V\rangle + \langle W\rangle \langle W \rangle \langle VV\rangle - \langle W\rangle \langle W\rangle \langle V \rangle \langle V \rangle - \frac{1}{L^2}\langle WW\rangle\langle VV\rangle\right]\notag.
\end{align}
Here $\langle \cdot \rangle = \frac{1}{L}\text{Tr}(\cdot)$ means the expectation value in the maximally mixed density matrix. The four terms here arise, again, from the four terms in (\ref{UUUU}), and we view the last two terms as nontrivial consequences of unitarity. This connection to unitarity is a bit less obvious than for (\ref{rhoarhob}). However, in the models that we will study below, it is simpler to think about the OTOC, because it can be studied in equilibrium, without the complication of evaporation. So we will focus on finding gravitational analogs of the terms in (\ref{OTOC}) rather than (\ref{rhoarhob}).

The short summary is that in JT gravity coupled to matter fields, we find analogs of the four terms in brackets, including the important minus signs. In Brownian SYK we can go further and find configurations that explain the full formula, including the prefactor.

In \hyperref[sec:qualitativediscussion]{\bf section two}, we explain the origin of the three leading terms in (\ref{OTOC}) in a generic large $N$ chaotic quantum mechanics. This explanation is based on a simple two-dimensional integral that we propose as an effective description of the theory on an OTOC contour. At late times, the saddle point approximation in this two-dimensional integral breaks down, and the different terms in formula (\ref{OTOC}) come from the contribution of different regions of the resulting integral.

In \hyperref[sec:JT]{\bf section three}, we study the OTOC in JT gravity \cite{Teitelboim:1983ux,Jackiw:1984je,almheiri2015models} with matter fields. This system has a conserved energy, so at late times the time evolution operator does not converge to a random unitary. A better model for this case would be to replace \cite{Blommaert:2020seb}
\be\label{eqn: random energy}
U \rightarrow u^\dagger e^{-\i h t}u.
\ee
where $h$ is a diagonal matrix of energy eigenvalues, and $u$ is a unitary change of basis between the energy eigenbasis and the local basis where simple operators are simple. For a unitary $U$ of this type, the answer for (\ref{OTOC}) is much more complicated, but for the disk topology (order $L^{0}$) the prediction is the same as in (\ref{OTOC}). This can be explained by the effective theory described above, which can be derived explicitly for JT gravity.

In the case where one-point functions vanish, the prediction for the handle-disk topology (order $L^{-2}$) is also the same as (\ref{OTOC}). However, matching this to a gravity calculation is subtle for the following reason. In JT gravity, even if one point functions $\langle V\rangle$ are set to zero, unless one enforces this with a bulk gauge symmetry, there will be a nonzero answer for $\langle V\rangle\langle V\rangle$ at order $1/L^2$ \cite{Saad:2019pqd}. In this situation, the first, second and fourth terms of (\ref{OTOC}) all contribute at order $1/L^2$, and the net effect is that the handle-disk gives a {\it positive} contribution in JT gravity, obscuring the origin of the interesting minus sign. However, these different contributions can be distinguished in the bulk calculation as arising from different regions of the moduli space integral. Because of this, they can also be numerically separated from each other by including perturbative effects of bulk matter loops, which change $\langle V\rangle\langle V\rangle$ relative to $\langle VV\rangle$.

We conclude from this that at least some of the nontrivial unitarity-preserving subleading effects from the random unitary model are also present in JT gravity with matter fields.

 In \hyperref[sec:Brownian]{\bf section four}, we study Brownian SYK \cite{Saad:2018bqo}. This is a version of the SYK model \cite{Sachdev:1992fk,KitaevTalks,Kitaev:2017awl} where the couplings vary randomly with time, and it has a similar large $N$ collective field description. Due to the time-dependence, there is no conserved Hamiltonian, and at late times the time evolution operator converges to the Haar distribution on random unitaries (up to discrete global symmetries). So in this system, we expect to be able to recover (\ref{OTOC}) precisely.

The large $N$ dynamics of Brownian SYK on the OTOC contour reduce to a set of three ODEs
\be\label{ODEintro}
\dot x = y^3 z - y z^3, \hspace{20pt} \dot y = x^3 z - x z^3, \hspace{20pt} \dot z = x^3 y - x y^3
\ee
where $x,y,z$ represent the strength of three different patterns of correlation between the four time contours. The four terms in (\ref{OTOC}) correspond to different possibilities for the pattern of correlation at the beginning and end of the OTOC contour. The $1/L^2$ expansion that one gets by expanding the prefactor arises from a sum over different transitions between these patterns of correlation. These ingredients are sufficient to explain the exact formula (\ref{OTOC}).

In several appendices, we give further details on aspects of these two models.

\section{A first look at the minus sign}\label{sec:qualitativediscussion}

Let's begin by trying to reproduce the formula (\ref{OTOC}) for the OTOC computed with a random unitary, but evaluated in the leading large $L$ limit:
\begin{align}\label{OTOC2}
\int\mathrm{d} U \ \big\langle (U^\dagger &W U) \, V \,(U^\dagger W U) \, V\big\rangle\Big|_{L^0}=\langle WW\rangle\langle V\rangle\langle V\rangle + \langle W\rangle \langle W \rangle \langle VV\rangle - \langle W\rangle \langle W\rangle \langle V \rangle \langle V \rangle.
\end{align}
The first two terms on the RHS correspond to simple Wick contractions of the $U$ matrix, but the final term comes from a subleading term in the Weingarten formula (\ref{UUUU}) and is a nontrivial consequence of the unitarity of $U$. So it is interesting to understand how this final term arises.

We will try to give a general explanation for this, which could be valid in a generic large $N$ chaotic quantum-mechanical system. Consider such a theory on an OTOC contour as sketched at left below, where the long segments of the contour correspond to the forwards and backwards evolution, and where empty circles represent locations where operators will be inserted:
\be\label{fig:contourSketch}
\includegraphics[valign = c, width = .6\textwidth]{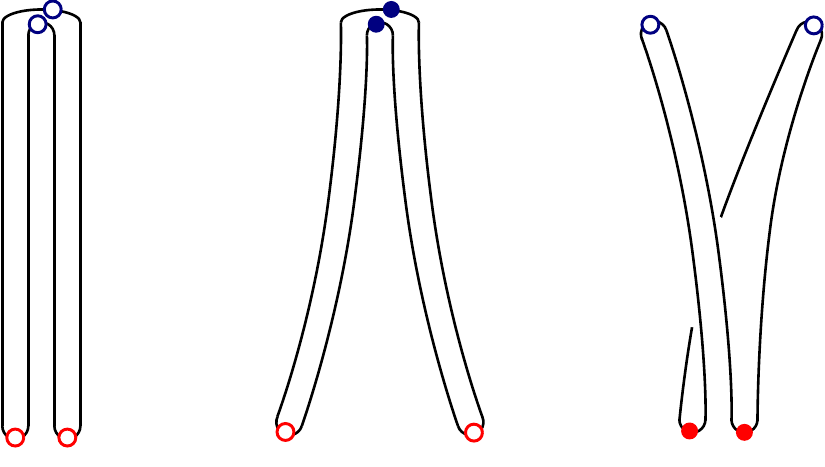}
\ee
In these figures, the horizontal direction is an abstract direction that represents correlation, where contours that are close together are understood as having large correlation. So, for example, in the leftmost diagram without any operator insertions turned on (empty circles), all four time contours remain close to each other and highly correlated.

In a chaotic theory, this highly correlated arrangement is unstable to small perturbations. If we insert a small perturbation by including operators at $t = 0$ (rightmost diagram, red circles filled), then at late times only one pairing of the contours will remain correlated. On the other hand, if we insert a small perturbation at $t = T$ (middle diagram), then at early times a different pairing will survive.

Considered as functions of time, these instabilities correspond to modes that grow exponentially either to the future (right diagram) or to the past (middle diagram). The rate of growth is given by the many-body Lyapunov exponent $\lambda_L$. We will define $x^+$ as the coefficient of the growing mode, normalized so that its largest value along the contour (at time $t = T$) is of order one. We similarly define $x^-$ as the coefficient of the mode that grows into the past, again normalized so that its maximum value is of order one (at time $t = 0$). So, to be clear, one mode is $O(1)$ in the future, and of order $e^{-\lambda_L T}$ in the past, and the other mode is $O(1)$ in the past and of order $e^{-\lambda_L T}$ in the future. At any given time, the product of the two modes will be small, of order $e^{-\lambda_L T}$, and we expect that the off-shell action for these modes will be
\be\label{actionConjecture}
I \propto N e^{-\lambda_L T} x^+x^-.
\ee
This action is a conjecture, but it can be checked in gravity, where it arises from the  action \cite{tHooft:1990fkf,Kabat:1992tb} that describes Dray 't Hooft shockwaves \cite{Dray:1984ha}, and in Brownian SYK, as we show in appendix \ref{app:kernel}.\footnote{Added in v2: this action has since been derived in SYK in independent work by Gu, Kitaev, and Zhang \cite{Gu:2021xaj}.} (In general, there could be more than one growing mode, and one would need to sum over them.)

These modes will couple to operators that are inserted at the locations of the red and blue circles in (\ref{fig:contourSketch}). The qualitative feature of this coupling is as follows
\begin{align}
W(T)W(T) &= \text{source for $x^-$ mode, affected by $x^+$ mode}\\
V(0)V(0) &= \text{source for $x^+$ mode, affected by $x^-$ mode}.
\end{align}
In particular, the late-time operators are strongly affected by the mode that is grows towards the future, and they act as sources for the mode that grows towards the past.  The early-time operators are strongly affected by the mode that grows towards the past, and they act as a source for the mode that grows towards the future.

We will use the notation $\langle WW\rangle_{x^+}$ for the correlation function of the $W$ operators in a configuration where the $x^+$ mode is nonzero, and similarly for $\langle VV\rangle_{x^-}$. Then we can write a formula for the OTOC in this effective theory as
\be\label{OTOCconj}
\langle W(T) V(0) W(T) V(0)\rangle = \int_{-\infty}^\infty \frac{\d x^+ \d x^-}{2\pi/a} \langle WW\rangle_{x^+}\langle VV\rangle_{x^-} e^{-\i a x^+ x^-}, \hspace{20pt} a = N e^{-\lambda_L T}.
\ee
We expect this theory to be accurate in the large $N$ limit, where fluctuations of other modes are suppressed. Indeed, before the scrambling time, even fluctuations of the $x^+$ and $x^-$ modes are suppressed by $N$. But after the scrambling time, so that $a \lesssim 1$, fluctuations in these modes become large, and they dramatically change the behavior of the correlator.

To write a detailed formula, one would need an expression for $\langle WW\rangle_{x^+}$ and $\langle VV\rangle_{x^-}$. However, at late times, all we really need to know to recover (\ref{OTOC2}) is that these correlators are smooth functions of $x^{\pm}$ that approach a limit for large values of their arguments. We will denote the values in this limit as $\langle WW\rangle_{\infty}$ and $\langle VV\rangle_{\infty}$. Then in the limit of small $a$, the integral is
\begin{align}\label{intanswer2}
\lim_{a \to 0} \int_{-\infty}^\infty \frac{\d x^+ \d x^-}{2\pi/a} &\langle WW\rangle_{x^+}\langle VV\rangle_{x^-} e^{-\i a x^+ x^-} \\ &= \langle WW\rangle_{\infty}\langle VV\rangle_0 + \langle WW\rangle_0\langle VV\rangle_{\infty} - \langle VV\rangle_{\infty} \langle WW\rangle_{\infty}\notag
\end{align}
Here, $\langle WW\rangle_0 = \langle WW\rangle$ is the unperturbed correlator, without any scrambling mode turned on. On the other hand $\langle WW\rangle_\infty$ is the correlator of the $W$ operators in a background with a very large amount of the growing $x^+$ mode turned on. This mode has the effect of de-correlating the two $W$ operators, so that $\langle WW\rangle_\infty \approx \langle W\rangle\langle W\rangle$, and similarly $\langle VV\rangle_{\infty} \approx \langle V\rangle\langle V\rangle$. This formula therefore reproduces (\ref{OTOC2}). The three contributions can be pictured as
\be\label{fig:contoursketch2}
\includegraphics[valign = c, width = .6\textwidth]{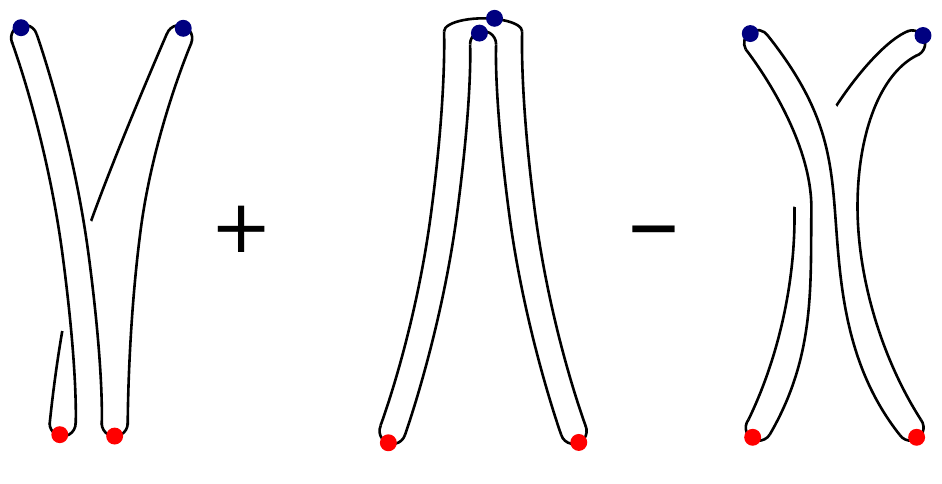}
\ee

Eq.~(\ref{intanswer2}) is just a mathematical identity as long as $\langle WW\rangle_{x^+}$ and $\langle VV\rangle_{x^-}$ are smoothly varying and approach a limit for large argument. One way to understand the formula, and the important minus sign, is to think about the simpler integral without any operator insertions:
\be
\int_{-\infty}^\infty \frac{\d x^+ \d x^-}{2\pi}  e^{-\i x^+x^- } = \int_{-\infty}^\infty \d x^+ \delta(x^+) = 1.
\ee
In the second expression, we have integrated out $x^-$, getting a delta function. This makes it seem like the integral is dominated by the region where $x^+$ is very small and $x^-$ is large. But of course we could have done the integral in the other order, and concluded the opposite. So why can't we add these two regions together and conclude that the answer to the integral is $1 + 1 = 2$? The resolution is that the answer to the integral is really $1 + 1 - 1 = 1$, where the final minus one comes from the region of the integral where both $x^+$ and $x^-$ are restricted to be nonzero. More precisely, we can consider the four regions, where ``small'' means restricted to $(-\epsilon,\epsilon)$ and ``big'' means restricted to $\mathbb{R}\setminus(-\epsilon,\epsilon)$:
\begin{align}
\{x^+ \text{ small} ,\  x^-\text{ big}\}\hspace{15pt} \{x^+ \text{ big} ,\  x^-\text{ small}\}\hspace{15pt} \{x^+ \text{ small} ,\  x^-\text{ small}\}\hspace{15pt} \{x^+ \text{ big} ,\  x^-\text{ big}\}.
 \end{align}
When $\epsilon$ itself is small, the contributions of these four regions are
\be\label{1101}
1 + 1 + 0 - 1.
\ee
The answer for the first three is straightforward. The integral over the final region is
\be
\int_{x^+,x^-\neq 0}{\d x^+ \d x^-\over 2\pi} e^{-\i x^+ x^-}=\int {\d x^+ \d x^-\over 2\pi} e^{-\i x^+x^-}{x^+\over x^+-\i \epsilon}{x^-\over x^--\i \epsilon}=-\epsilon\int_{0}^\infty \d x^- e^{-\epsilon x^-}=-1.
\ee
To make a connection with \ref{intanswer2}, we can rescale the $x^{\pm}$ in \ref{intanswer2} with $\sqrt{a}$ and then replace $\epsilon$ with $\sqrt{a}$ in the above discussion.

As we will see later, the same mechanism explains the important minus signs present at higher orders in the $L^{-2}$ expansion. From the perspective of the large $N$ integral, this mechanism is quite nontrivial. The true expansion parameter for the $x^+$ and $x^-$ modes is not $1/N$, but rather $e^{\lambda_LT}/N$. For large time, the large $N$ expansion breaks down for these modes, leading to large fluctuations. The minus sign comes from a specific region in the resulting integral.

\section{JT gravity}\label{sec:JT}

\subsection{Working at disk order}
In this section, we will review the computation of the OTOC in JT gravity, and see an explicit example of the effective description from the previous section. On the disk topology, the basic physics of the OTOC is that the insertion of boundary operators $VV$ and $WW$ create particles that intersect in a high energy gravitational scattering process in the bulk \cite{Shenker:2013pqa,kitaevfundamental,Shenker:2014cwa}:
\be
\includegraphics[valign = c, scale=0.8]{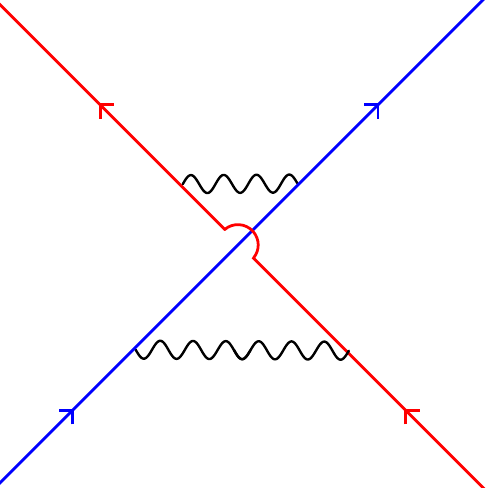}
\label{fig:bulkscattering}
\ee
The relative boost between the particles is controlled by the difference of boundary times, which for our case is just $T$. In the relevant range of energies, the gravitational scattering amplitude is (setting $\beta=2\pi$)
\be\label{Sgrav}
S_{\text{grav}}=e^{\i p_+ q_-/a}, \hspace{20pt} a = \frac{e^{-T}}{G_N}.
\ee
Here $p_+$ is a component of the momentum of the $W$ particles and $q_-$ is a component of the momentum of the $V$ particles, each measured in a frame where the respective particles are not boosted. The OTOC is given by folding this scattering amplitude against the wave functions of the $W$ and $V$ particles:\footnote{Here $\langle \cdot \rangle$ is normalized with the disk partition function.}
\be\label{p+q-}
\langle W(t)V(0)W(t)V(0)\rangle = \int_{-\infty}^0\d p_+\d q_- \langle W_L|p_+\rangle\langle p_+|W_R\rangle \langle V_L|q_-\rangle\langle q_- |V_R\rangle S_{\text{grav}}.
\ee
For the case of AdS${}_2$, the wave functions and further details are given for example in \cite{Maldacena:2016upp} (see also \cite{Haehl:2021dto} for application to six-point functions).

How is this formula related to the effective description above? The scattering matrix $S_{\text{grav}}$ is obtained by integrating out the ``shock wave'' modes that represent null shifts along the horizons where the red and blue particles propagate \cite{Dray:1984ha,tHooft:1990fkf,Kabat:1992tb}. We will refer to these displacements as $x^+$ and $x^-$, intentionally using the same notation that we used in the general discussion above. The path integral weighting (off-shell action) for these shock wave modes is $\exp(-\i a x^+x^-)$, where $a$ is defined in (\ref{Sgrav}). The formula for the OTOC in terms of an integral over these modes is a special case of (\ref{OTOCconj}):
\begin{align}\label{firstline}{}
\langle W(t) V(0)W(t)V(0)\rangle &= \int_{-\infty}^\infty \frac{\d x^+ \d x^-}{2\pi/a} e^{-\i ax^+x^-}\langle WW\rangle_{x^+}\langle VV\rangle_{x^-}\\
&= \int_{-\infty}^\infty \frac{\d x^+ \d x^-}{2\pi/a} e^{-\i ax^+x^-}\frac{1}{(2 + \frac{x^+}{\sqrt{2}} + \i \epsilon)^\Delta}\frac{1}{(2 + \frac{x^-}{\sqrt{2}} + \i \epsilon)^\Delta}.\label{secondLine}
\end{align}
In the first line (\ref{firstline}), we wrote a general expression, where $\langle WW\rangle_{x^+}=\langle W e^{-\i x^+ \hat p_+}W\rangle$ is the correlator in a background with a shock wave parametrized by $x^+$, and similarly for $\langle VV\rangle_{x^-}=\langle V e^{-\i x^- \hat q_-}V\rangle$. In the second line we wrote explicit formulas for these correlators in the case of AdS${}_2$, assuming that the operators are conformal primaries of dimension $\Delta$, and with a particular normalization.

For this specific case (\ref{secondLine}), the late time (small $a$) limit of this expression is actually zero. We can get a more nontrivial result by studying a type of exponentiated OTOC
\be\label{eqn:eVVeWW}
\langle \mathcal{C}e^{W_1(T)W_3(T)} e^{V_2(0) V_4(0)}\rangle  = \sum_{n,m = 0}^\infty \frac{1}{n!m!}\langle W_1(T)^n V_2(0)^m W_3(T)^n V_4(0)^m\rangle.
\ee
Here $\mathcal{C}$ means that the operators are ordered in the sense of the OTOC contour after we expand the exponential downstairs, as indicated at right. In order to simplify the answer, we would like to remove self contractions between operators at the same point along the contour, e.g.~within $W_1^n$. These can be suppressed by replacing $W_1W_3$ and $V_2V_4$ by a sum of $K$ operators: 
${1\over K}\sum_{i=1}^K W_1^i W_3^i$ and ${1\over K}\sum_{i=1}^K V_2^i V_4^i$. 
Large $K$ suppresses the unwanted terms. However, the $i$ labels makes the formulas a bit ugly to read, so we will omit them in the formulas. We will further assume that the one-point functions vanish $\langle W\rangle = \langle V\rangle = 0$. With these assumptions, one can write a formula for (\ref{eqn:eVVeWW}) by bringing the operators in (\ref{secondLine}) into the exponent:
\be
\langle \mathcal{C}e^{W_1(0)W_3(0)} e^{V_2(T) V_4(T)}\rangle=\int_{-\infty}^{\infty}{\d x^+\d x^-\over 2\pi/a} \exp\left(-\i a x^+x^- + {1\over (2+{x^+\over \sqrt{2}}+\i \epsilon)^{2\Delta}}+{1\over (2+{x^-\over \sqrt{2}}+\i \epsilon)^{2\Delta}}\right)
\ee
Now applying (\ref{intanswer2}), we get the following nontrivial late-time behavior
\begin{align}\label{eqn:diskRUresult}
\langle \mathcal{C}e^{W_1(T)W_3(T)} e^{V_2(0) V_4(0)}\rangle&\rightarrow e^{2^{-2\Delta}} + e^{2^{-2\Delta}} - 1 \\&= \langle e^{W_1(T)W_2(T)}\rangle+\langle e^{V_2(0)V_4(0)}\rangle-1.
\label{diskRHRHRHR}
\end{align}
This matches the prediction of (\ref{OTOC2}) for this correlator.

\subsection{Handle-disk without dynamical matter fields}\label{sec:handle-disk}
Next, we will look for a gravitational analog of the fourth term in (\ref{OTOC}), which we reproduce here
\begin{align}\label{OTOCReproduced}
\int\mathrm{d} U \ \big\langle (U^\dagger &W U) \, V \,(U^\dagger W U) \, V\big\rangle  
\\ &=\frac{1}{1-L^{-2}}\left[\langle WW\rangle\langle V\rangle\langle V\rangle + \langle W\rangle \langle W \rangle \langle VV\rangle - \langle W\rangle \langle W\rangle \langle V \rangle \langle V \rangle - \frac{1}{L^2}\langle WW\rangle\langle VV\rangle\right]\notag.
\end{align}
Like the third term, the fourth term on the RHS is interesting because it arises from the subleading terms in the Weingarten formula (\ref{UUUU}) and is therefore a nontrivial consequence of unitarity. It is also notable because it is the only term that survives if all one-point functions vanish. So in this section we will try to isolate this piece by assuming that $\langle W\rangle=\langle V\rangle=0$, although we will be forced to re-examine this assumption a little later.

\begin{figure}[h]
\begin{center}
\includegraphics[width = .65\textwidth]{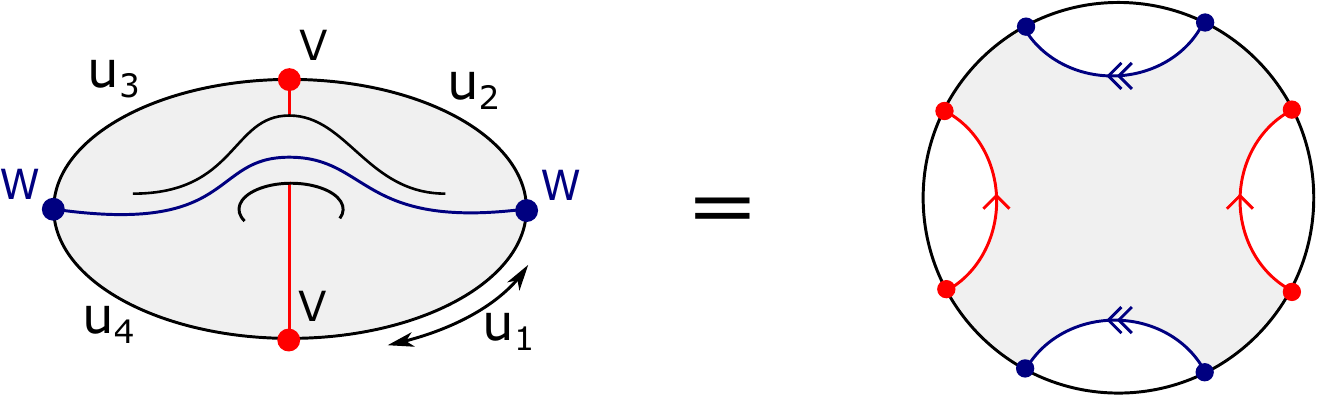}
\caption{Cutting the handle-disk geometry on the red and blue geodesics leads to the shaded portion of the disk geometry shown at right. The renormalized proper distance along the boundary between the operators is $u_1,u_2,u_3,u_4$, which will eventually be continued to Lorentzian values.}
\label{fig:handle-disk}
\end{center}
\end{figure}

The fourth term on the RHS of (\ref{OTOCReproduced}) has three prominent features. The first is that the sign is negative, and based on our previous discussion, it is tempting to associate it with an integral over slightly off-shell geometries. The second is the overall ${1\over L^2}$ factor, which suggests that it arises from a genus one correction to the disk in 2d gravity.\footnote{The parameter $L$ is interpreted in the bulk as being proportional to $e^{S_{0}}$, where different topologies are weighted by $e^{\chi S_{0}}$ where $\chi$ is the Euler characteristic. So a suppression by two factors of $L$ relative to the leading terms corresponds to two units of Euler characteristic, which is accomplished by inserting a handle.} This is the topology of a disk with a handle glued in, and we will refer to it as the ``handle-disk.'' The last feature is that the answer involves a product of two point functions, but somehow does not decay with time. On the disk topology, the $WW$ and $VV$ propagators would intersect and scatter with each other, leading to decay at late time where the kinematics is highly boosted. But on the handle-disk, the topology allows the propagators to avoid each other, and therefore avoid the scattering that would otherwise lead to decay (see figure \ref{fig:handle-disk}).\footnote{The propagators can also avoid each other on the unorientable crosscap spacetime. If this geometry is allowed (which it should be in a theory with time-reversal symmetry) then it would dominate over the handle-disk.} In the rest of this section, we will work out the OTOC on the handle-disk geometry.

On AdS${}_2$ and its quotients, free-field propagators with points on the boundaries are given exactly by a sum over geodesics of $e^{-\Delta \ell}$, where $\ell$ is the regularized length of the geodesic, and $\Delta$ is the conformal dimension, which is related to the mass of the free field. In principle, one needs to sum over geodesics, and a four point function on the handle-disk will involve a sum over pairs of geodesics, one connecting the $W$ operators together, and another connecting the $V$ operators. In cases where the the two geodesics intersect, one has to take the gravitational scattering into account, and because the large boost implies a large scattering, we expect such contributions to be suppressed at late times.\footnote{At genus two, Phil Saad has explained to us that there is a nondecaying contribution where the geodesics manage to intersect in an unboosted configuration. However we do not expect such a contribution at genus one.} So we are interested in
\be
\int_{\text{moduli space and bdy wiggles}}e^{-(\text{gravitational action})}\sum_{\gamma_W,\gamma_V \text{ nonintersecting}}e^{-\Delta (\ell_W + \ell_{V})}
\ee 
where $\gamma_W$ is a geodesic connecting the $W$ insertions and $\gamma_V$ connects the $V$ insertions.

To do the computation, we will let the Euclidean boundary times between the operators be $u_1,u_2,u_3,u_4$. In the end, we will be interested in continuing these to complex values to implement Lorentzian time evolution, but it is easiest to picture the geometry if we begin with real values. 

If we cut the geometry along non-intersecting $\gamma_W$ and $\gamma_V$ (which, on this topology, are therefore also non-self-intersecting), the result will be a piece of the hyperbolic plane, shown shaded at right in figure \ref{fig:handle-disk}. The path integral in the central shaded region is a standard JT path integral on the disk topology, bounded by an alternating sequence of geodesics and asymptotic boundary segments of renormalized lengths $u_i$. The general answer for this path integral can be worked out using formulas from \cite{Yang:2018gdb} (see e.g.~\cite{Penington:2019kki}) and it is
\be\label{eqn:middleregion}
\begin{split}
Z(u_1,\ell_1,u_2,\ell_2...,u_k,\ell_k)&=2^k \int \d s \rho(s)e^{-\sum_{i=1}^k u_i {s^2\over2 N}}K_{2\i s}(4 N e^{-{\ell_1\over 2}})...K_{2\i s}(4 N e^{-{\ell_k\over 2}});\\
\rho(s)&={s\over 2\pi^2}\sinh (2\pi s).
\end{split}
\ee
Here ${s^2\over 2N}$ has the interpretation of the boundary energy on a particular segment of the asymptotic boundary. This energy turns out to be equal for all of the boundary segments, which can be understood as a result of the ``Gauss-law'' constraint in gravity and the absence of any matter excitations in the central region of the geometry.

Locally, the answer for the path integral on the handle-disk is obtained by ``gluing'' this computation together along the red and blue geodesics by integrating over the lengths $\ell_W$ and $\ell_V$. This integral is weighted by an appropriate measure, which includes a factor $e^{-\Delta(\ell_W+\ell_V)}$ to account for the matter propagators. We also include a factor $e^{-S_0}$ from the topological part of the JT gravity action:
\be\label{unrestrictedIntegral}
\langle WVWV\rangle_{\text{handle-disk}} = 4e^{-S_0}\int_{-\infty}^\infty \d\ell_V\d \ell_W e^{-\Delta(\ell_V+\ell_W)} Z(u_1,\ell_W,u_2,\ell_V,u_3,\ell_W,u_4,\ell_W).
\ee
In general, one needs to worry about tricky global issues in this type of computation, because an unrestricted integral over $\ell_W$ and $\ell_V$ counts the same geometry more than once. This is due to the fact that there are multiple ways to cut open the exact same handle-disk and get a topologically trivial geometry like the one on the right in figure \ref{fig:handle-disk}. So in fact, the unrestricted integral over $\ell_W$ and $\ell_V$ counts not just each geometry, but also each way of cutting the geometry open. In the present case we get lucky, as shown in a closely related cases by Saad \cite{Saad:2019pqd} and by Blommaert \cite{Blommaert:2020seb}. The ``redundant'' sum over ways of cutting the geometry open corresponds precisely to a sum over non-intersecting geodesics connecting the red and blue dots, which is something that we actually {\it are} supposed to be summing over, in order to compute the propagators. So the redundancy is perfectly canceled by the sum over nonintersecting geodesics, and the full answer is just an unrestricted integral over $\ell_W$ and $\ell_V$.

The answer for the integral (\ref{unrestrictedIntegral}) is
\be\label{eqn:WVWVHD}
\langle WVWV\rangle_{\text{handle-disk}}=\int \frac{\d s}{e^{S_0}{\rho(s)}}e^{-\sum_i u_i {s^2\over2 N}} \langle s|VV|s\rangle\langle s|WW|s\rangle
\ee
where we have defined the function $\langle s|\mathcal{O}\mathcal{O}|s\rangle$ as
\be
\langle s|\mathcal{O}\mathcal{O}|s\rangle=8{\rho(s)}\int_{-\infty}^\infty \d\ell K^2_{2\i s}(4 N e^{-\ell/2}) e^{-\Delta\ell}=N^{-2\Delta}{\Gamma(\Delta)^2\over 2^{2\Delta+1}\Gamma(2\Delta)}|\Gamma(\Delta-2\i s)|^2{\rho(s)}.
\ee
The meaning of this function is that in a microcanonical version of the thermofield double state at an energy corresponding to $s$, the two point function of operators on the two sides is given by $\langle s|\mathcal{O}\mathcal{O}|s\rangle$. A final detail is that to compute the normalized answer, we should divide by the partition function without operators inserted. To leading order this is just the disk partition function
\be
\langle 1\rangle_{\text{disk}} = e^{S_0} \int \d s \rho(s)e^{-\beta {s^2\over2 N}}.
\ee
where for this case, $\beta = \sum_{i} u_i$.

Let's now take stock of this answer and try to compare to the formula (\ref{OTOCReproduced}). First of all, we are interested in a Lorentzian configuration where e.g.
\be
u_1 = u_3 = \frac{\beta}{4} + \i T, \hspace{20pt} u_2 = u_4 = \frac{\beta}{4} - \i T.
\ee
However, the answer (\ref{eqn:WVWVHD}) depends only on the sum of the $u$ parameters, and therefore is independent of $T$. So in particular, it does not decay for large $T$.\footnote{The exact correlator on the handle-disk will not be independent of $T$, because for early times, configurations with intersecting geodesics will be relevant. However, it will approach this $T$-independent value for large $T$.} If we freeze the value of $s$ in both the numerator and the normalizing denominator by transforming from $\beta$ to a microcanonical ensemble, then we find
\be
\frac{\langle WVWV\rangle_{\text{handle-disk}}}{\langle 1\rangle_{\text{disk}}} = \frac{1}{({\rho(s)}e^{S_0})^2} \langle s|VV|s\rangle  \langle s|WW|s\rangle.
\ee
The effective dimension of the Hilbert space at energy determined by $s$ is $L \sim {\rho(s)} e^{S_0}$, and the two point function of operators at this energy is $\langle s|\mathcal{O}\mathcal{O}|s\rangle$, so the RHS can be interpreted as
\be\label{matches}
\frac{1}{L^2}\langle VV\rangle\langle WW\rangle.
\ee
This is the same term that appears in the fourth term of the formula (\ref{OTOCReproduced}), so the full answer we get in JT gravity is some form of average of this result over energies.

\begin{figure}[h]
\begin{center}
\includegraphics[width = .6\textwidth]{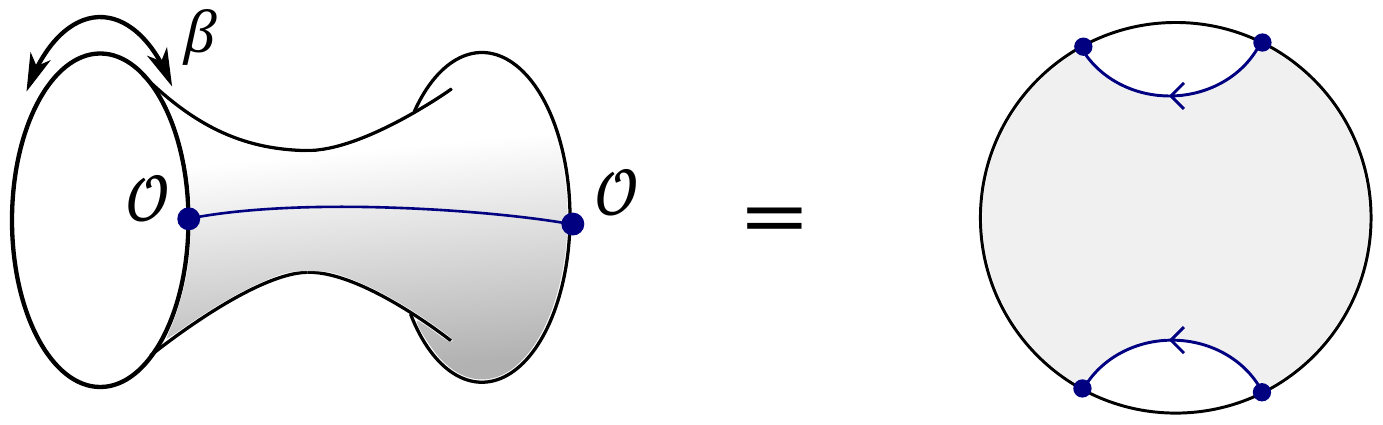}
\caption{The cylinder geometry that gives a nonzero value for $\langle \mathcal{O}\rangle^2$.}
\label{fig:cylinder}
\end{center}
\end{figure}

However, the minus sign is missing! We will now explain this apparent discrepancy. In focusing on the fourth term in (\ref{OTOCReproduced}), we made an assumption that one point function of the operators vanish: $\langle W\rangle=\langle V\rangle=0$. But in gravity, this is hard to achieve as pointed out by Saad \cite{Saad:2019pqd}. The reason is that even if we assume that the disk one point function is zero, when we consider one point function squared, there can be wormhole contribution leading to a non-trivial answer.
The corresponding geometry is a cylinder, shown in figure \ref{fig:cylinder},
which can be computed with a similar cutting and gluing procedure to relate it to a portion of the hyperbolic plane. Again, the sum over geodesics perfectly cancels the redundancy associated to the mapping class group, so the gluing is an unrestricted integral over $\ell$, giving \cite{Saad:2019pqd}
\be
\frac{\langle \mathcal{O}\rangle^2_{\text{cylinder}}}{(\langle 1\rangle_{\text{disk}})^2}=\frac{\int \d s\  e^{-{\beta s^2\over N}}\langle s|\mathcal{O}\mathcal{O}|s\rangle}{\left(e^{S_0}\int\d s \rho(s) e^{-\frac{\beta s^2}{2N}}\right)^2}.
\ee
Freezing to a given value of $s$, we can interpret this as the statement that in pure JT gravity with probe matter fields
\be\label{OOvsOO}
\langle \mathcal{O}\rangle^2 = \frac{1}{L^2}\langle \mathcal{O}^2\rangle.
\ee
Inserting this into (\ref{OTOCReproduced}), we find that the first, second and fourth terms contribute at the same order, combining to give
\be\label{combine}
\langle WW\rangle\langle V\rangle^2 + \langle W\rangle^2\langle VV\rangle - \frac{1}{L^2}\langle WW\rangle\langle VV\rangle = \frac{1}{L^2}\langle WW\rangle\langle VV\rangle.
\ee
This is consistent with the result (\ref{matches}) above. The final coefficient $+1$ is actually the result of a combination of two subtle effects: (i) the explicit $L^{-2}$ term from the subleading Weingarten coefficient (ii) even if we try to set one-point functions to zero, without imposing a symmetry one finds that they end up being nonzero, with a square of order $L^{-2}$.

This equation $1 + 1 - 1 =1 $ sounds reminiscent of (\ref{1101}), and in fact we can understand it in the same way, where the different contributions arise from different regions of an integral over nearly-zero modes. To do that, we need to retreat from the exact answer and examine more carefully the integral over the moduli space. This is done in appendix \ref{App:handle-disksoftmode}, where we find that at late time, the handle-disk path integral is again dominated by two exponentially soft modes, weighted by the Dray-t'Hooft shockwave action. These two modes represent the geodesic lengths of the two closed cycles that intersect the $W$ or $V$ geodesics.\footnote{See discussion below equation (\ref{eqn:effectivetemp}).} Physically, the reason they appear is that at late time the bulk region of the handle-disk includes the relevant part of the geometry of the disk on an OTOC contour.

\subsection{Handle-disk with dynamical bulk matter}
One way to resolve the RHS of (\ref{combine}) into the three terms on the LHS is to modify the JT theory by introducing dynamical bulk matter fields $\tilde{\phi}_i$. The integral over the moduli space then needs to be reweighted by the partition function of these fields. This doesn't affect the disk two point function $\langle \mathcal{O}\mathcal{O}\rangle$, but it does modify the cylinder computation of $\langle \mathcal{O}\rangle^2$. This will break the relationship (\ref{OOvsOO}) and allow us to distinguish the three contributions in (\ref{combine}).

Unfortunately, there is a problem with this theory \cite{Saad:2019lba}, which is that in a region of moduli space where some periodic direction is very short (on the cylinder or any other geometry), we get a divergence from the propagation of highly excited states around the periodic direction, making the theory ill-defined. We will assume that this ``tachyon'' divergence can be regulated somehow, for example by a ``capping off'' transition as in the SYK model \cite{Maldacena:2018lmt}. To simplify the analysis, we will work in a special limit where the details of this are unimportant. This is the limit where the $W$ and $V$ operators are very heavy $\Delta \gg N$, so that in the computation of $\langle W\rangle^2$, the length of the closed geodesic is stabilized at a large value.

Let's see how this simplifies the cylinder computation of $\langle \mathcal{O}\rangle^2$. The geodesic connecting the two boundaries becomes small, and the ``cut-open'' cylinder approaches the full hyperbolic disk with a boundary of length $2\beta$, as shown at left here:
\be\label{fig:diskLargeDelta}
\includegraphics[valign = c, width = .6\textwidth]{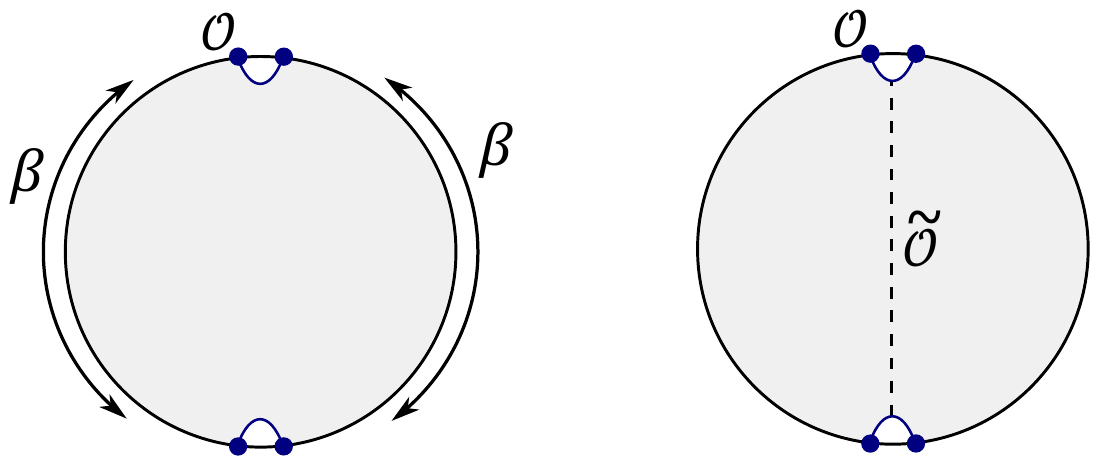}
\ee
This is reflected in the behavior of the answer
\be
\langle \mathcal{O}\rangle_{\text{cylinder }\beta}^2 \approx C e^{-S_0}\langle 1\rangle_{\text{disk } 2\beta}, \hspace{20pt} \Delta \gg N.
\ee
Here we used that for large $\Delta$, the microcanonical two point function $\langle s|\mathcal{O}\mathcal{O}|s\rangle$ defined in (\ref{eqn:WVWVHD}) simplifies to a multiple of the density of states $C\rho(s)$, where the constant is $C=N^{-2\Delta}{\Gamma(\Delta)^4\over 2^{2\Delta+1}\Gamma(2\Delta)}$. 

Now we can add the loops of the $\tilde{\phi}$ field, propagating along the dashed line at right in (\ref{fig:diskLargeDelta}). When $\Delta$ is large, the dashed line is long, and we can approximate the partition function of the massive field on the cylinder as
\be
1 + e^{-\mu\,b} + \dots 
\ee
where $b$ is the length of the closed geodesic. The ``$1$'' represents the contribution of the vacuum, and the exponential term represents a single closed loop propagating around the cylinder. If we work in the regime where this term is small, but we include a large number $c$ of matter fields $\tilde{\phi_i}$, then we can arrange that the weighting is 
\be
(1 + e^{-\mu\,b} + \dots)^c \approx \exp\left(c e^{-\mu b}\right).
\ee
In the large $\Delta$ limit where we are working, the ``cut-open'' disk approximates the ordinary disk, and this insertion can be approximated as a boundary insertion
\be
\exp\left(\gamma\tilde{\mathcal{O}}_i(0)\tilde{\mathcal{O}}_i(\beta)\right).
\ee
The coefficient $\gamma$ depends on the fact that $b$ is the length to a finite point determined by the geodesics of the $WW$ insertions, whereas the two point function of operators is defined with respect to a regularized length between boundary points. The difference between the two is a multiplicative renormaliation of the operators, by a coefficient that has a well-stabilized value in the large $\Delta$ limit where we are working. We will suppress some of these details and simplify the notation by just referring to the insertion from now on as
\be
\exp\left(\tilde{\mathcal{O}}(0)\tilde{\mathcal{O}}(\beta)\right).
\ee
The upshot of all of this is that the corrected value of the one-point function squared in the presence of matter loops is
\be\label{onePtCorrected}
\langle \mathcal{O}\rangle_{\text{cylinder }\beta}^2\approx Ce^{-S_0}\langle e^{\tilde{\mathcal{O}}(0)\tilde{\mathcal{O}}(\beta)}\rangle_{\text{disk } 2\beta}.
\ee

Now we repeat this discussion on the handle-disk. When the $W$ and $V$ operators have large $\Delta$, this also approximates a disk, as shown at left here:
\be\label{fig:handleDiskLargeDelta}
\includegraphics[valign = c,width = .6\textwidth]{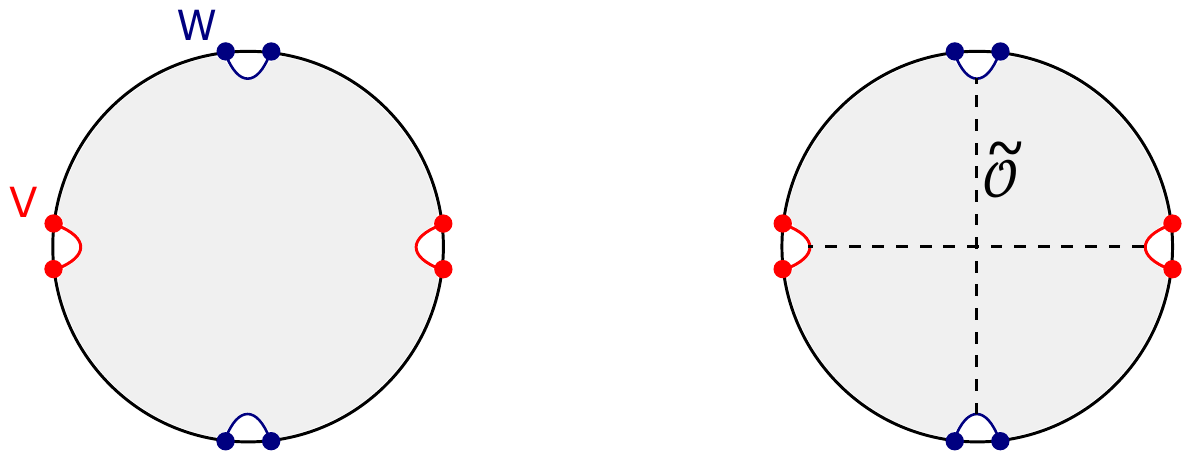}
\ee
This fact is also reflected in the limiting behavior of the explicit formula (\ref{eqn:WVWVHD})
\be
\langle WVWV\rangle_{\text{handle-disk}}\approx C^2 e^{-2S_0}\langle 1\rangle_{\text{disk}}, \hspace{20pt} \Delta \gg N.
\ee
If now we add bulk matter, we can excite loops around either of the cycles shown with dashed lines in (\ref{fig:handleDiskLargeDelta}).\footnote{Loops could also  wind around both cycles, but these are highly suppressed at late time.} Replacing these by exponentiated operator insertions as above, we find
 \begin{align}\label{firsthandledisk}
\langle WVWV\rangle_{\text{handle-disk}}&\approx C^2 e^{-2S_0}\langle e^{\mathcal{\tilde{O}}(0)\mathcal{\tilde{O}}(u_1+u_4)}e^{\mathcal{\tilde{O}}(u_1)\mathcal{\tilde{O}}(u_1+u_4+u_3)}\rangle_{\text{disk}}\\
&\approx C^2 e^{-2S_0}\left(\langle e^{\mathcal{\tilde{O}}(0)\mathcal{\tilde{O}}(u_1+u_4)}\rangle_{\text{disk}}+\langle e^{\mathcal{\tilde{O}}(u_1)\mathcal{\tilde{O}}(u_1+u_4+u_3)}\rangle_{\text{disk}}-\langle 1\rangle_{\text{disk}}\right).\label{finalhandledisk}
\end{align}
In going to the second line, we used the result from the disk computation (\ref{diskRHRHRHR}). 

The positive answer for the handle disk has now been split into three terms, two positive and one negative. Qualitatively, these match the three terms in the random unitary formula
\be\label{consistentWith}
\langle WW\rangle\langle V\rangle^2 + \langle W\rangle^2\langle VV\rangle - \frac{1}{L^2}\langle WW\rangle\langle VV\rangle.
\ee
A quick way to see this is that we have already shown that without the effect of the $\tilde{\mathcal{O}}$ operators, the fixed energy version of this formula matches (\ref{matches}). The first two terms in (\ref{finalhandledisk}) are modified by the $\tilde{\mathcal{O}}$ operators in the same way as the one point function (\ref{onePtCorrected}), and the final term is unmodified, which is consistent with (\ref{consistentWith}) because the disk two point function is not modified by the matter loops.

The random unitary model does not predict the particular $u$-dependence in (\ref{finalhandledisk}), but in appendix \ref{App:REB}, we discuss a more complicated random model where one can consider the time dependence, and we find that crude features of this time dependence match.

The first two terms in (\ref{finalhandledisk}) depend on the details of the bulk matter. In this sense, these contributions are not very ``universal.'' Of course, a similar situation applies to the square of the one point function, for exactly the same reason. However, the final term does not depend on details of the bulk matter. The reason for this is that the final term corresponds to a region of the moduli space where both the ``horizontal'' and ``vertical'' cycles are long, and there are simply no short closed geodesics anywhere on the geometry, and the matter fields are not excited.

\section{Brownian SYK}\label{sec:Brownian}
In the previous section, we focused on obtaining the third and fourth terms in the OTOC formula (\ref{OTOC}). Because JT gravity is dual to a system with a time-independent Hamiltonian, if we try to go to higher orders, we expect to find terms in JT that have no analog in the random unitary formula, see appendix \ref{App:REB}. However, if we study a system with a time-dependent Hamiltonian, then at late times it really is possible for $U(T)$ to approximate a random unitary, and we could hope to reproduce the entire formula from large $N$ collective fields, including the interesting prefactor $1/(1-L^{-2})$. In this section we will study Brownian SYK \footnote{For recent studies about the Brownian SYK model, see: \cite{Sunderhauf:2019djv,Jian:2020krd}}, and we will find a geometric series of configurations that can indeed explain the entire formula.

Brownian SYK is an ensemble of quantum systems with time-dependent Hamiltonians
\be
H(t) = \i^{\frac{q}{2}}\sum_{1\le a_1<\dots<a_q \le N} J_{a_1\dots a_q}(t) \psi_{a_1}\dots \psi_{a_q}.
\ee
Here the $\psi_a$ operators are Majorana fermions, satisfying $\{\psi_a,\psi_b\} = \delta_{ab}$. The ensemble is defined by giving the statistics of the couplings, which are taken to be drawn from a Gaussian distribution with mean zero and with
\be\label{couplingDist}
\langle J_{a_1\dots a_q}(t)J_{a_1'\dots a_q'}(t')\rangle = \delta_{a_1a_1'}\dots \delta_{a_qa_q'}\frac{(q{-}1)!}{N^{q-1}} J^2(t,t').
\ee
The RHS contains a function $J^2(t,t')$ that determines the correlation of the couplings at different times. In the standard SYK theory, the couplings are time-independent, so $J^2(t,t')$ is just a constant, called $J^2$. At the opposite extreme, one has Brownian SYK, where the couplings at different times are completely uncorrelated with each other:
\be
J^2(t,t') = \delta(t-t')J.
\ee
We will set up the initial computation with a general $J^2(t,t')$ and specialize later.

\subsection{Fermions on an OTOC contour}
We are interested in computing an OTOC in this ensemble of theories. One way to think of the computation of the OTOC is as a path integral over a single set of fermion fields on a time contour that folds forward and backwards in physical time, with operator insertions at the turnaround points. This contour is shown at left below, parametrized by a variable $s$ that increases in the direction of the arrows. The dots indicate the locations where operators will eventually be inserted:
\be
\includegraphics[width = .4\textwidth, valign =c]{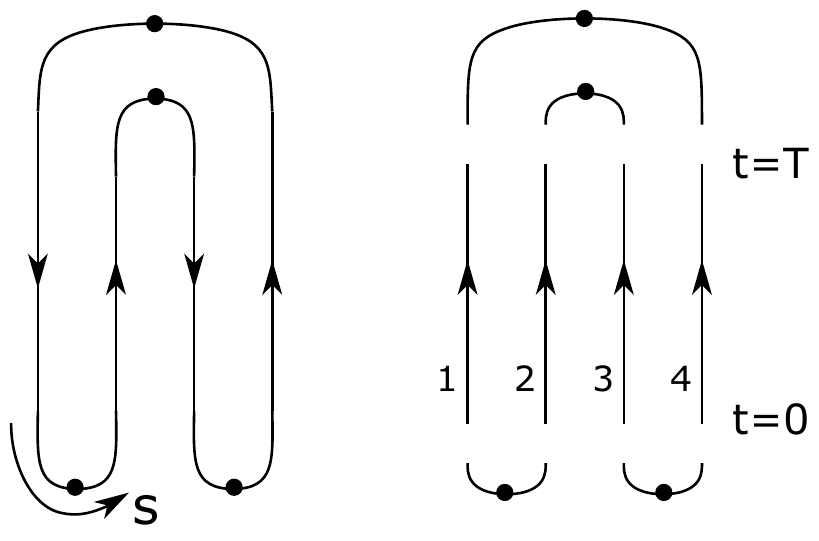}
\ee
At right we have a different way of thinking of the same problem, as a matrix element of a time evolution operator for four copies of the system (central region), evolving in physical time $t$ between an initial state at $t = 0$ and a final state at $t = T$. We now think of all four of these contours evolving forwards in the same physical time, but roughly speaking, the Hamiltonian for contours 1 and 3 will have a minus sign relative to that for contours 2 and 4.

To write down a path integral formula, the picture at the left is a convenient starting point. We have a single set of fermion fields $\psi_a(s)$ propagating around the full contour, with antiperiodic boundary conditions. Apart from operator insertions, the path-integral weighting is $e^{-I}$ with
\be\label{actionPsi}
I = \int \d s\left[\frac{1}{2}\psi_a(s)\partial_s\psi_a(s) + \frac{\d t}{\d s} \i^{1+\frac{q}{2}}J_{a_1\dots a_q}(t(s))\psi_{a_1\dots a_q}(s)\right].
\ee
In this expression and below, we define $\psi_{a_1\dots a_q}$ as the product $\psi_{a_1}\psi_{a_2}\dots \psi_{a_q}$, and we sum repeated sets of indices $a_1\dots a_q$ over values with $1\le a_1<a_2<\dots <a_q\le N$.

One can also use a path integral defined in terms of fermions that are functions of the physical time $t$, but the price for doing this is that one needs four sets of fermions, one for each of the forwards or backwards contours. We  introduce a $j$ index to label the contour, so the notation is
\be
\psi_a^{(j)} = \text{fermion of flavor $a \in \{1,\dots N\}$ on contour $j \in \{1,\dots 4\}$}.
\ee
In order to write down the path integral in terms of these variables, it is convenient to start with (\ref{actionPsi}) and work out the relationship between the $\psi^{(j)}(t)$ fermions and the $\psi(s)$ fermion that propagates around the entire contour. One consistent convention for this is as follows. Relative to one of the ``turnaround'' points in the initial state, the relationship between the fermion fields in the two pictures is
\be\label{sign1}
\psi^{(1)} = +\i \psi, \hspace{20pt} \psi^{(2)} = \psi.
\ee
There is a similar relationship for the other turnaround, involving $\psi^{(3)}$ and $\psi^{(4)}$. A mnemonic for the factor of $\i$ is that $s$ and $t$ increase in opposite directions on contour 1, and the $\pi$ rotation of the time direction introduces a factor of $\i$. Relative to one of the turnaround points in the final state, we have
\be\label{sign2}
\psi^{(1)} = -\i \psi, \hspace{20pt} \psi^{(4)} = \psi.
\ee

Using these relations between the $\psi^{(j)}$ and $\psi$ fermion fields, one can rewrite (\ref{actionPsi}) as
\be
I = \int_0^T\d t \left[\frac{1}{2}\psi_a^{(j)}(t)\partial_t \psi_a^{(j)}(t) + \i^\frac{q}{2}J_{a_1\dots a_q}(t)s_j\psi^{(j)}_{a_1\dots a_q}(t)\right].
\ee
where
\be\label{sjphases}
s_j = \begin{cases}+\i & \text{$j \in \{2,4\}$ is a ``forwards contour''} \\ -\i\cdot\i^q & \text{$j\in \{1,3\}$ is a ``backwards contour'' }\end{cases}.
\ee
One can also deduce the boundary conditions for the $\psi^{(j)}$ fermions by noting that the fermion $\psi$ is continuous right at the turnaround point itself. Then (\ref{sign1}) and (\ref{sign2}) imply
\begin{align}\label{bc1}
\psi^{(1)} &= +\i \psi^{(2)}, \hspace{20pt} \psi^{(3)} = +\i\psi^{(4)}\hspace{20pt} \text{at time $t = 0$}\\
\psi^{(1)} &= - \i\psi^{(4)}, \hspace{20pt} \psi^{(2)} = -\i \psi^{(3)}\hspace{20pt} \text{at time $t = T$}.\label{bc2}
\end{align}

\subsection{Disorder averaged theory and collective fields}
We are interested in the average of the OTOC over the ensemble of theories defined by the distribution for couplings (\ref{couplingDist}). To evaluate the average, one can simply integrate over the couplings. This is a Gaussian integral, and after doing it we find\footnote{Here we are suppressing a subtlety related to the $f_q$ function in appendix A.3 of \cite{Saad:2018bqo}.}
\begin{align}
I &=\frac{1}{2}\int_0^T\d t \ \psi_a^{(j)}\partial_t \psi_a^{(j)} - \i^q\frac{(q{-1})!}{2N^{q-1}}\int\hspace{-8pt}\int_0^T\d t\d t' J^2(t,t')  s_j s_{j'}\psi^{(j)}_{a_1\dots a_q}(t)\psi^{(j')}_{a_1\dots a_q}(t') \\
& = \frac{1}{2}\int_0^T\d t \ \psi_a^{(j)}\partial_t \psi_a^{(j)} - \frac{N}{2q}\int\hspace{-8pt}\int_0^T\d t\d t' J^2(t,t')  s_j s_{j'}\left(\frac{1}{N}\psi^{(j)}_a(t)\psi^{(j')}_a(t')\right)^q.\label{disorderAvgAction}
\end{align}
One can now rewrite this action using the standard SYK trick of introducing a collective field
\be
G_{ij}(t,t') = \frac{1}{N}\psi^{(i)}_a(t)\psi^{(j)}_a(t')
\ee
and the corresponding Lagrange multiplier $\Sigma_{ij}(t,t')$ that enforces this constraint, and then integrating out the fermions. After writing the fermion interaction in terms of $G$, the remaining fermion action is quadratic, with kernel $\partial_t - \Sigma$. The fermions can be integrated out to give a Pfaffian, and the final action is
\be\label{actionGSigma}
-\frac{I}{N} = \log \text{Pf}\left(\partial_t - \Sigma\right) - \frac{1}{2}\int\hspace{-8pt}\int_0^T \d t \d t'\left[\Sigma_{jj'}(t,t')G_{jj'}(t,t') - \frac{J^2(t,t')}{q}s_js_{j'}G_{jj'}(t,t')^q\right].
\ee
The boundary conditions enter this expression only through the Pfaffian.

\subsection{Large $N$ equations of motion}
The equations of motion for the action (\ref{actionGSigma}) can be written as
\begin{align}
\delta_{ij}\delta(t-t') &= \partial_{t}G_{ij}(t,t') - \int \d t''\Sigma_{ik}(t,t'')G_{kj}(t'',t')\\
\Sigma_{ij}(t,t') &= J_{ij}^2(t,t')s_is_j G_{ij}(t,t')^{q-1}.\label{SigmaSecond}
\end{align}
These are similar to the standard Schwinger-Dyson equations of the SYK model, except that we now have a matrix of correlators for the different contours, and we also have the $s_j$ phases (\ref{sjphases}) to account for the Lorentzian directions of the different contours. 

From these equations, one can derive another equation that will be very convenient for studying the Brownian limit.  To derive this, start by writing the first equation in matrix notation as
\be\label{diff1}
1 = \partial^{(1)}G - \Sigma G
\ee
where we are thinking of time and the contour index together as a matrix index. Here, $\partial^{(1)}$ means a derivative with respect to the first argument of $\Sigma(t,t')$. Taking the transpose, and using the fact that $G$ and $\Sigma$ are antisymmetric, we find
\begin{align}
1 &= (\partial^{(1)}G)^T - (\Sigma G)^T = \partial^{(2)}G^T - G^T \Sigma^T\\
&= -\partial^{(2)}G - G \Sigma.\label{diff2}
\end{align}
Finally, taking (\ref{diff1}) minus (\ref{diff2}) gives
\be\label{diagInTime}
 (\partial^{(1)} + \partial^{(2)})G = [\Sigma,G].
\ee

This equation is valid in standard SYK or in Brownian SYK, but it is especially powerful in Brownian SYK, where it closes on the components of $G,\Sigma$ that are diagonal in time. The reason for this is that in Brownian SYK, we have
\be
J^2(t,t') = J \delta(t-t'),
\ee
which implies that $\Sigma$ is diagonal in time. Therefore (\ref{diagInTime}) closes on the diagonal components of $G$. To write this closed equation, we introduce the notation $\sigma$ and $g$ for the ``time-diagonal'' components of $\Sigma$ and $G$:
\be
\Sigma_{ij}(t,t') = \delta(t-t')\sigma_{ij}(t), \hspace{20pt} G_{ij}(t,t) =  g_{ij}(t).
\ee
Then (\ref{diagInTime}) and (\ref{SigmaSecond}) imply 
\begin{align}\label{simplifiedEOM}
\partial_t g(t) &= [\sigma(t),g(t)]\\
\sigma_{ij}(t) &= \begin{cases} Js_is_j g_{ij}(t)^{q-1}\hspace{20pt} & i\neq j \\ 0 & i= j\end{cases}.\label{SigmaEOM}
\end{align}
In other words, for Brownian SYK, the equations of motion reduce to ordinary ODEs, rather than the integro-differential equations of the standard SYK theory.

\subsection{Reducing to a smaller set of variables}\label{sec:reducing}
What about the unequal-time components of $G$? In principle, their saddle point values can be obtained by solving (\ref{diff1}) once the equal-time components are found. But in fact, the unequal-time components are not needed in the theory at all: they can be integrated out trivially.

To see how this works, notice that in the Brownian limit, the $J(t,t')G(t,t')^q$ term in (\ref{actionGSigma}) is only nonzero for equal times, due to the delta function in $J(t,t')$. This means that the only place in the action where $G(t,t')$ with {\it unequal} times appears is when it multiplies $\Sigma(t,t')$. We can then integrate out the unequal-time $G(t,t')$ variable, getting a delta function constraint that sets the unequal-time components of $\Sigma(t,t')$ to zero. One is left with an exact description of the theory in terms of only the $g(t)$ and $\sigma(t)$ variables introduced above.

There is a further more minor simplification that will also be useful. The original SYK theory has a $(-1)^F$ symmetry, but after disorder averaging, the action (\ref{disorderAvgAction}) separately conserves fermion parity within each flavor $a$. In other words, we have the symmetry operators $(-1)^{F_a} = -\prod_{j = 1}^4\sqrt{2}\psi_a^{(j)}$ for each value of $a$. The boundary conditions (including with the operator insertions we will choose) correspond to initial and final states that have $+1$ eigenvalues under each of these separate fermion parity operators, so we can work entirely within the subspace of $+1$ eigenvalues.

The $+1$ eigenvalue condition implies that for each value of $a$,
\be
\psi_a^{(1)}\psi_a^{(2)}\psi_a^{(3)}\psi_a^{(4)} = -\frac{1}{4}.
\ee
In terms of $g_{ij}$, this means that in addition to antisymmetry, the $g_{ij}$ matrix should satisfy
\be\label{gconditions}
g_{12} = g_{34}, \hspace{20pt} g_{14} = g_{23}, \hspace{20pt} g_{13} = -g_{24}.
\ee
In the path integral, this constraint arises because the Pfaffian term is exactly independent of some linear combinations of components of $\sigma_{ij}$. Integrating over these components imposes a set of delta function that restrict $g$ to be antisymmetric and to satisfy the the conditions (\ref{gconditions}). 

The upshot is that we can restrict to a $g$ matrix parametrized by three functions $x(t), y(t), z(t)$
\be\label{gForm}
g_{ij} = \delta_{ij}\frac{\text{sgn}(0)}{2} + \frac{1}{2}\left(\begin{array}{cccc} 0 & -\i x & y & -\i z \\
\i x & 0 & -\i z & -y \\
-y & \i z  & 0 & -\i x\\
\i z  & y & \i x & 0  \end{array}\right).
\ee
In writing this, we introduced a $1/2$ and some factors of $\i$ to simplify later equations. The equation of motion (\ref{simplifiedEOM}) in terms of these variables becomes three ODEs, which read
\begin{align}
\dot{x} &=  \frac{J}{2^{q-2}}(y^{q-1}z - y z^{q-1}) \notag\\
\dot{y} &= \frac{J}{2^{q-2}}(x^{q-1}z - x z^{q-1}) \label{odes}\\
\dot{z} &= \frac{J}{2^{q-2}}(x^{q-1} y - x y^{q-1}).\notag
\end{align}
The interpretation of the $x$, $y$ and $z$ variables is that they measure the instantaneous correlation between the different contours. For example, the $x$ variable gives the correlation between contours 1 and 2, or equivalently (becuase of the $(-1)^{F_a} = 1$ condition) between contours 3 and 4. If $x = 0$, the correlation vanishes, and if $x = \pm 1$, the two contours are maximally entangled. At large $N$, the dynamics of the Brownian SYK model reduces to the coupled dynamics of the three different types of this correlation.

\subsection{Conserved quantities and qualitative behavior of solutions}\label{sec:conservation}
The evolution equations (\ref{odes}) have two conserved quantities associated to them 
\begin{align}\label{conservedQuantity1}
\frac{\d}{\d t}(x^2 -y^2+z^2) = 0\\
\frac{\d}{\d t}(x^q-y^q+z^q) = 0.\label{conservedQuantity2}
\end{align}
As we will see, for the solutions of interest, the first conserved quantity is always exactly equal to one, so the solutions will live in the space defined by
\begin{align}
x^2-y^2+z^2&=1\\
x^q-y^q+z^q &= r\label{defr}
\end{align}
for some constant $r$. This is a one-dimensional set within the three-dimensional space of $x,y,z$. It will be useful to understand its properties. For now, we will focus on real values of $x,y,z$, and the cases where $r \le 1$.

For $r = 1$, the set consists of the components
\be\label{r1}
\{\pm 1, a,\pm a\} \cup \{a,\pm a,\pm 1\}, \hspace{20pt} a \text{ arbitrary, signs independent}
\ee
The points of intersection between the various components of this set (for example $\{1,1,1\}$ or $\{1,0,0\}$) are fixed points of the equations of motion, and nontrivial solutions evolve from one fixed point in the infinite past to another one in the infinite future. These $r = 1$ solutions represent an OTOC, in which one of the pairs of operators is treated in a probe approximation.

For $r < 1$, the conservation rules forbid the solution to intersect the locus with $x = 0$ or $z = 0$, but they allow $y = 0$. The compact part of the solution space consists of four topological circles, one for each possible sign of the $x$ and $z$ coordinates. As a function of $t$, the coordinates $\{x(t),y(t),z(t)\}$ in a particular solution will move around one of these circles periodically. The period depends on the value of $r$. As $r$ approaches one, the solutions pass close to the fixed points mentioned above, and the period becomes large. These $r<1$ solutions correspond to fully backreacted OTOC computations, and the solutions that wind around the circle generate the $1/L^2$ expansion.

\subsection{Boundary conditions}
On the OTOC contour without any operator insertions, the initial condition consists of two ``contour turnarounds,'' which we write as a state $\initial$ in the fermion Hilbert space, where
\be\label{initialConse}
(\psi^{(1)} -\i \psi^{(2)})\initial = 0, \hspace{20pt} (\psi^{(3)} -\i \psi^{(4)})\initial = 0.
\ee
Here we are suppressing the flavor index $a$. Focusing on the first of these equations, one finds
\be
\label{wkjh}
\psi^{(1)}\psi^{(2)} \initial = -\i \psi^{(1)}\psi^{(1)}\initial = -\tfrac{\i}{2}\initial,
\ee
where we used that $\psi\psi = \frac{1}{2}$. Eq.~(\ref{wkjh}) implies that the correlator $g_{12}(0) = -\frac{\i}{2}$. Remembering that $g_{12} = -\frac{\i}{2}x$, this means that $x(0) = 1$. Arguing similarly, the second equation in (\ref{initialConse}) turns out to imply  that $g_{13} = \i g_{14}$, which translates to $y(0) = z(0)$. The final boundary conditions can be analyzed in the same way, and all together we find that the complete set of boundary conditions without operator insertions are
\begin{align}\label{initialcond1}
x(0) &= 1, \hspace{20pt} y(0) = z(0) \\
z(T) &= 1, \hspace{20pt} y(T) = x(T).\label{finalcond1}
\end{align}
The unique solution that connects these two together is $x(t) = y(t) = z(t) = 1$. This can be understood as a configuration in which all of the contours have simultaneous maximal correlation with each other. Note that simultaneous maximal correlation with multiple systems is not normally possible in quantum mechanics, but it is possible in postselected quantum mechanics. The OTOC contour, with different initial and final states, is effectively a postselected problem.

To make a nontrivial OTOC, we need to include operator insertions. Instead of inserting a product of operators in the initial state, it is technically convenient to include an ``entangled'' operator insertion $\exp(\i\gamma \psi^{(1)}_a\psi^{(4)}_a)$, so that the new initial state is
\be\label{defInitialOp}
|\text{initial}\rangle = \exp\left(\i\gamma \psi^{(1)}_a\psi^{(4)}_a\right)\initial.
\ee
We will choose $\gamma > 0$ so that this operator is large in the highly correlated state with no insertions; for this sign, the insertion behaves similarly to a product of operators.

What are the initial conditions for $x,y,z$ in this state? To work this out (continuing to suppress the flavor index) one can write $x,y,z$ out in terms of the Majorana fermions, and then use the Majorana algebra to act with them on this initial state. Concretely, we write $x = 2\i \psi^{(1)}\psi^{(2)}$, and $y = 2\psi^{(1)}\psi^{(3)}$ and $z = 2\i\psi^{(1)}\psi^{(4)}$ and then use $\{\psi^{(i)},\psi^{(j)}\} = \delta_{ij}$ to evaluate
\begin{align}
\left(\begin{array}{c}x \\ y \\ z\end{array}\right) |\text{initial}\rangle &= \exp\left(\i\gamma \psi^{(1)}\psi^{(4)}\right) \exp\left(-\i\gamma \psi^{(1)}\psi^{(4)}\right)\left(\begin{array}{c}x \\ y \\ z\end{array}\right)\exp\left(\i\gamma \psi^{(1)}\psi^{(4)}\right)\initial\\&= \exp\left(\i\gamma \psi^{(1)}\psi^{(4)}\right)\left(\begin{array}{ccc} \cosh(\gamma) & -\sinh(\gamma) & 0 \\ -\sinh(\gamma) & \cosh(\gamma) & 0 \\ 0 & 0 & 1\end{array}\right)\left(\begin{array}{c}x \\ y \\ z\end{array}\right)\initial.
\end{align}
So the effect of the operator insertions is a hyperbolic rotation in the space of the $x,y,z$ variables. Propagating the initial conditions (\ref{initialcond1}) through this rotation, we find 
\be\label{bcINITIAL}
x(0) = \cosh(\gamma) - \sinh(\gamma)z(0), \hspace{20pt} y(0) = \cosh(\gamma)z(0) - \sinh(\gamma).
\ee
Similarly, after including an operator insertion $\exp(\i\gamma \psi^{(1)}_a\psi^{(2)}_a)$ before applying the final state, we end up with the final boundary conditions 
\be\label{bcFINAL}
z(T) = \cosh(\gamma) - \sinh(\gamma)x(T), \hspace{20pt} y(T) = \cosh(\gamma)x(T) - \sinh(\gamma).
\ee
We have a total of four conditions, and only three parameters to vary in the solution, so this problem sounds overconstrained. In fact, the conserved quantities ensure that solutions generically exist. In particular, the initial and final boundary conditions both imply
\be
x^2 - y^2 + z^2 = 1.
\ee
Conservation of this quantity implies that we really only have to solve one of the equations (\ref{bcFINAL}) at the final boundary. The other will automatically be satisfied, up to a possible sign.

\subsection{Solutions}\label{sec:solns}
We will now discuss some solutions to the equations (\ref{simplifiedEOM}) with the boundary conditions (\ref{bcINITIAL}) and (\ref{bcFINAL}). We already saw that without insertions ($\gamma = 0$), there is a solution $x = y = z = 1$.

The next simplest case is where there is an operator insertion at $t = 0$, but no insertion at $t = T$. To desribe this situation we will use notation $\gamma_i$ and $\gamma_f$ for the parameters of the initial and final operator insertions, and consider the case $\gamma_i \neq 0$ and $\gamma_f = 0$. In this situation, both conserved quantities are equal to one. In particular, $r = 1$, so the solution remains with the space (\ref{r1}), and we conclude that the solution has $x(t) = y(t)$ and $z(t) = 1$ for all times. The specific solution can be found by writing the equation of motion for $y(t)$ given these constraints:
\be\label{yeq}
\dot{y} = \frac{J}{2^{q-2}}(y^{q-1}-y).
\ee
The solution is
\be\label{soly}
x(t) = y(t) = \left(\frac{1}{1 + e^{\lambda_L (t-t_0)}}\right)^{\frac{1}{q-2}}, \hspace{20pt} z(t) = 1.
\ee
The parameter $\lambda_L$ is the Lyapunov exponent of the Brownian SYK theory, and it has the value
\be\label{LyapunovExp}
\lambda_L = \frac{(q-2)J}{2^{q-2}}.
\ee
This value can be obtained by linearizing (\ref{yeq}) near the unstable equilibrium $y = 1$. To solve the initial conditions, we need to take $a = 1$, and
\be
e^{-\gamma_i} = y(0) = \left(\frac{1}{1 + e^{-\lambda_L t_0}}\right)^{\frac{1}{q-2}}.
\ee
This solution can be interpreted as follows. We have inserted operators in the initial state, but not in the final state. The chaos-fueled amplification of this operator insertion destroys the $x$ and $y$ correlation as time goes by, but the $z$ correlation remains. The solution $y(t)$ can be interpreted as an OTOC where the insertions at time $t$ are treated in the probe approximation.

Finally, we consider the case with both initial and final operator insertions, and for simplicity, we take $\gamma_i = \gamma_f = \gamma$. First, we plot an example solution of the ODEs, for a value of $r$ that is fairly close to one ($r \approx 1 - 4\cdot 10^{-13}, J = 4, q = 4$):
\be
\includegraphics[valign = c, width = .91\textwidth]{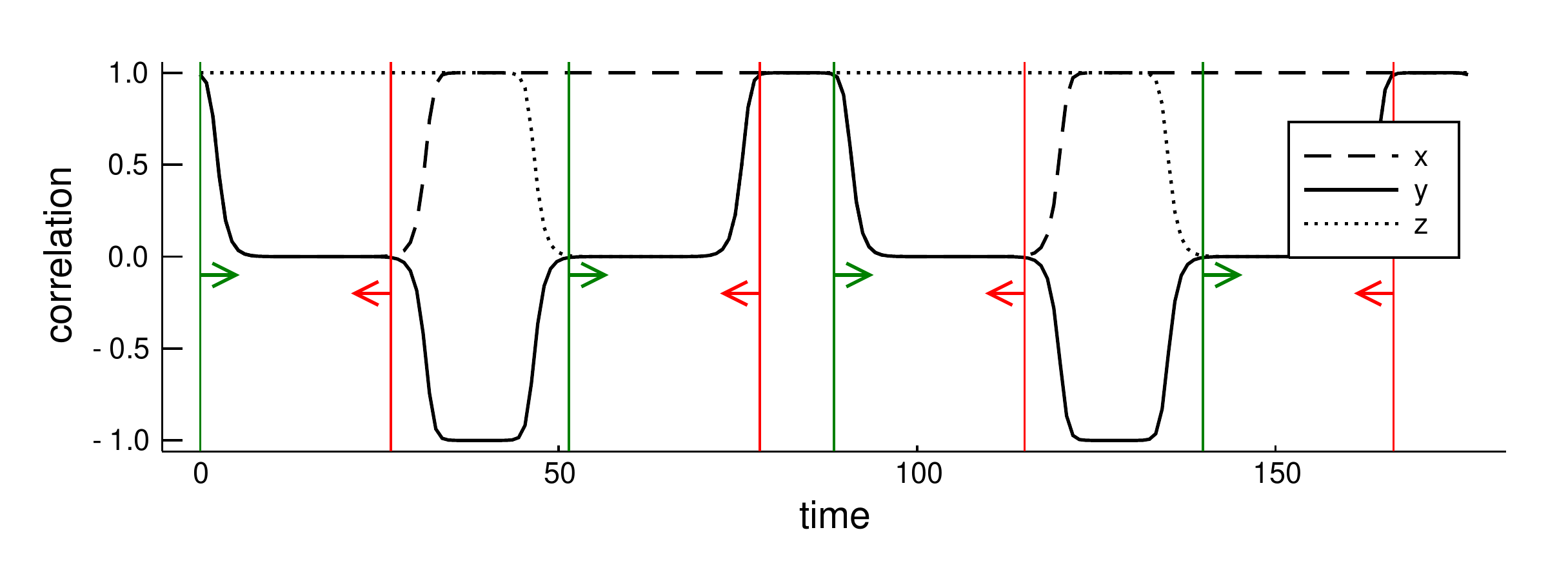}\label{exampleSol}
\ee
Here we plotted two periods of the solution. The vertical lines indicate locations where the initial conditions and final conditions are satisfied, decorated with rightward and leftward arrows respectively. We will now make several comments about this solution.
\begin{enumerate}
\item The full periodic solution consists of approximately constant regions (where it approaches close to a fixed point) separated by transitions. As $r$ approaches one, the solutions approach the fixed points more closely, and the constant regions last longer. So, roughly, the solution can be ``stretched'' or ``compressed'' in time, although the timescale of the transition regions remains fixed.
\item To make a solution that satisfies the initial conditions at time zero and satisfies the final conditions at time $T$, one can pick a pair of initial and final lines in (\ref{exampleSol}) and adjust $r$ so that the final line coincides with time $T$.
\item There are two different types of initial line to choose from: one where $\{x,y,z\}\approx \{1,1,1\}$ and one where  $\{x,y,z\}\approx \{1,0,0\}$. Similarly, there are two different types of final line: one where $\{x,y,z\}\approx \{1,1,1\}$ and one where $\{x,y,z\} \approx \{0,0,1\}$. So there are $2\times 2 = 4$ qualitative classes of solution. 
\item Within each of these four qualitative classes, one has further freedom to insert some number of periods in the ``middle'' of the solution. For large $T$, this number is arbitrary, but for finite $T$, the number of periods cannot be larger than $\text{const.}\times T$, because of the need to accommodate the transition regions, which have fixed width.
	\end{enumerate}
So, displaying explicitly the first two solutions within each of the four qualitative classes, we have the following contributions to the OTOC:
	\begin{align}\label{qualitativeClasses}
&\left(\includegraphics[width = .12\textwidth,valign = c]{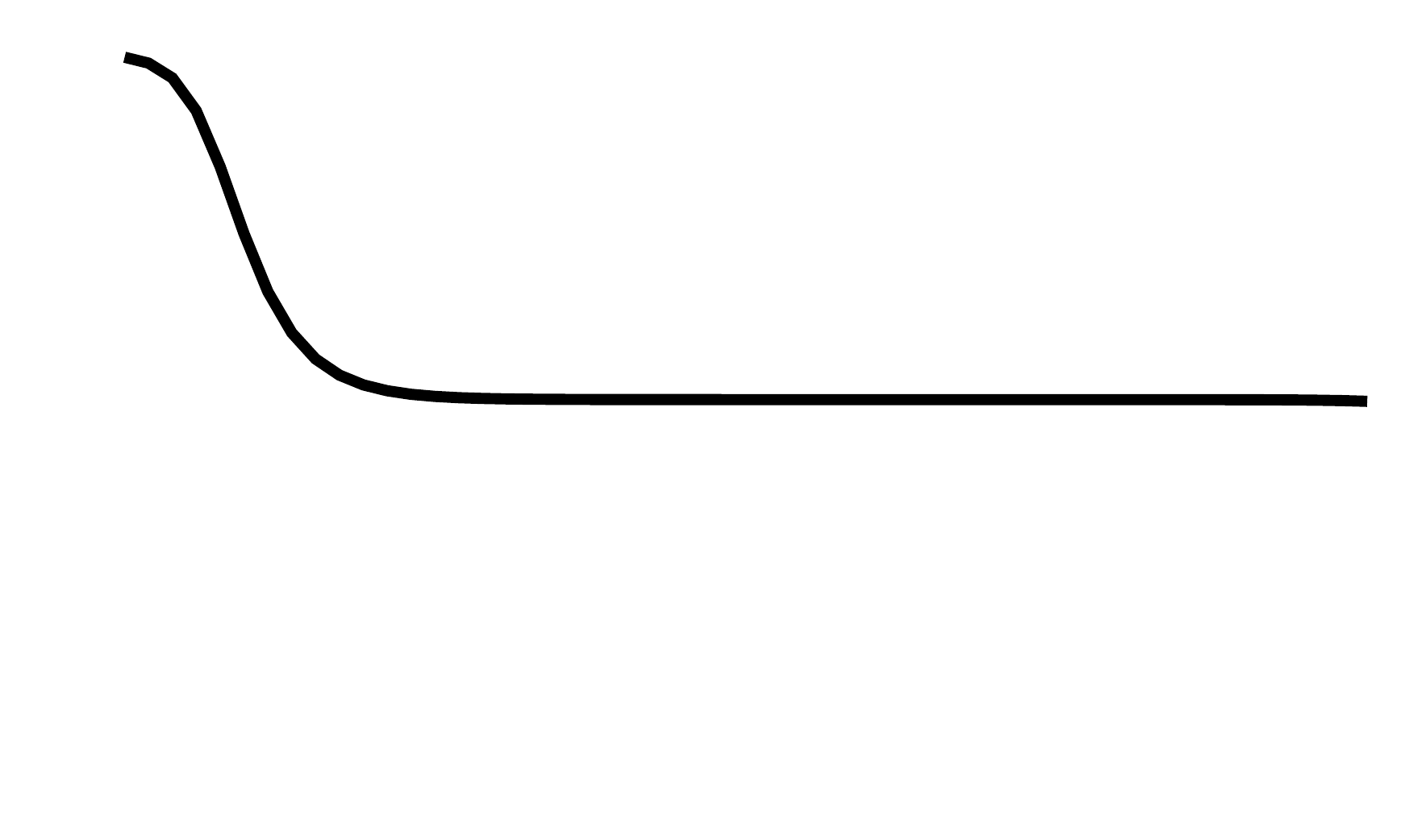} + \includegraphics[width = .12\textwidth,valign = c]{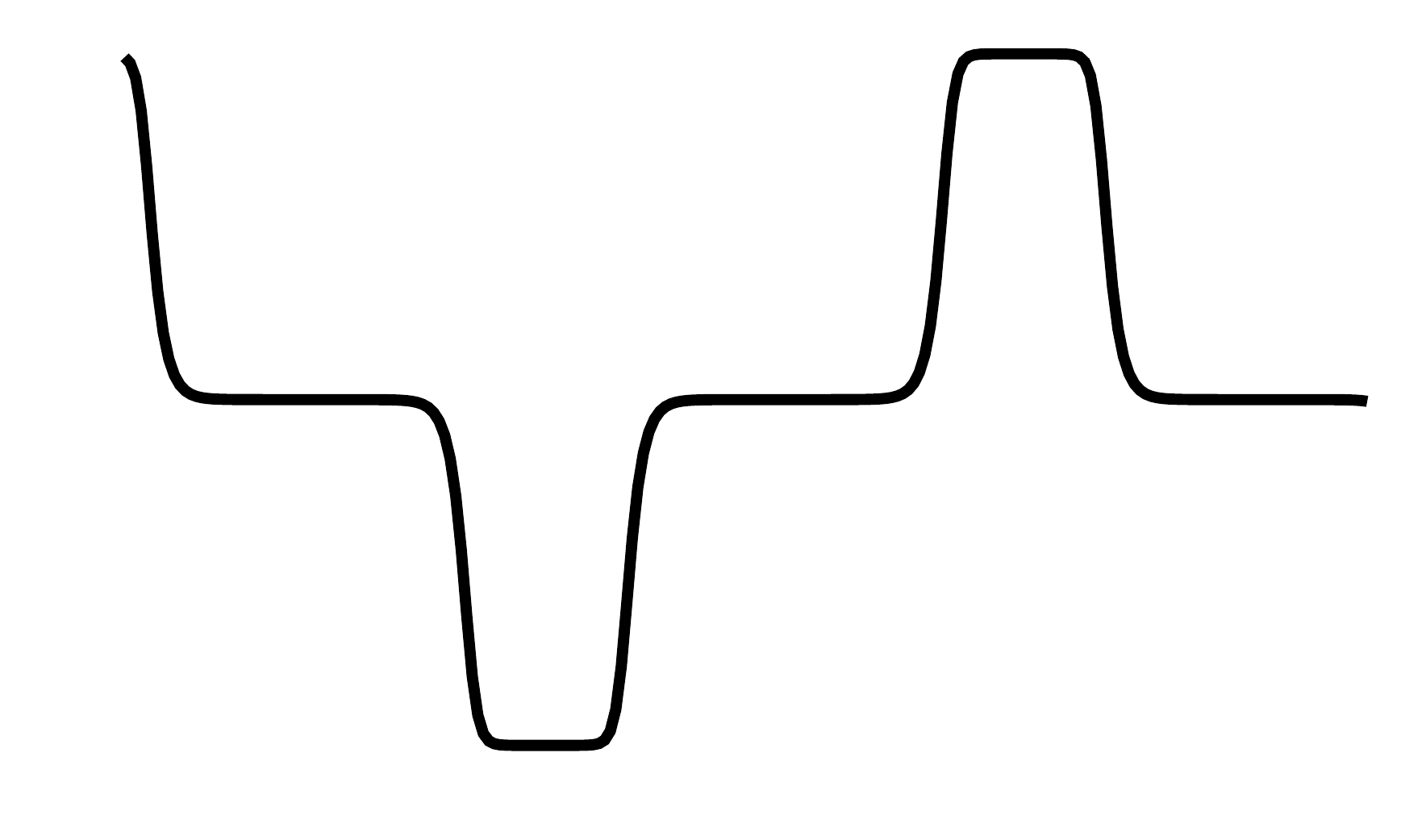} + \dots\right) + \left(\includegraphics[width = .12\textwidth,valign = c]{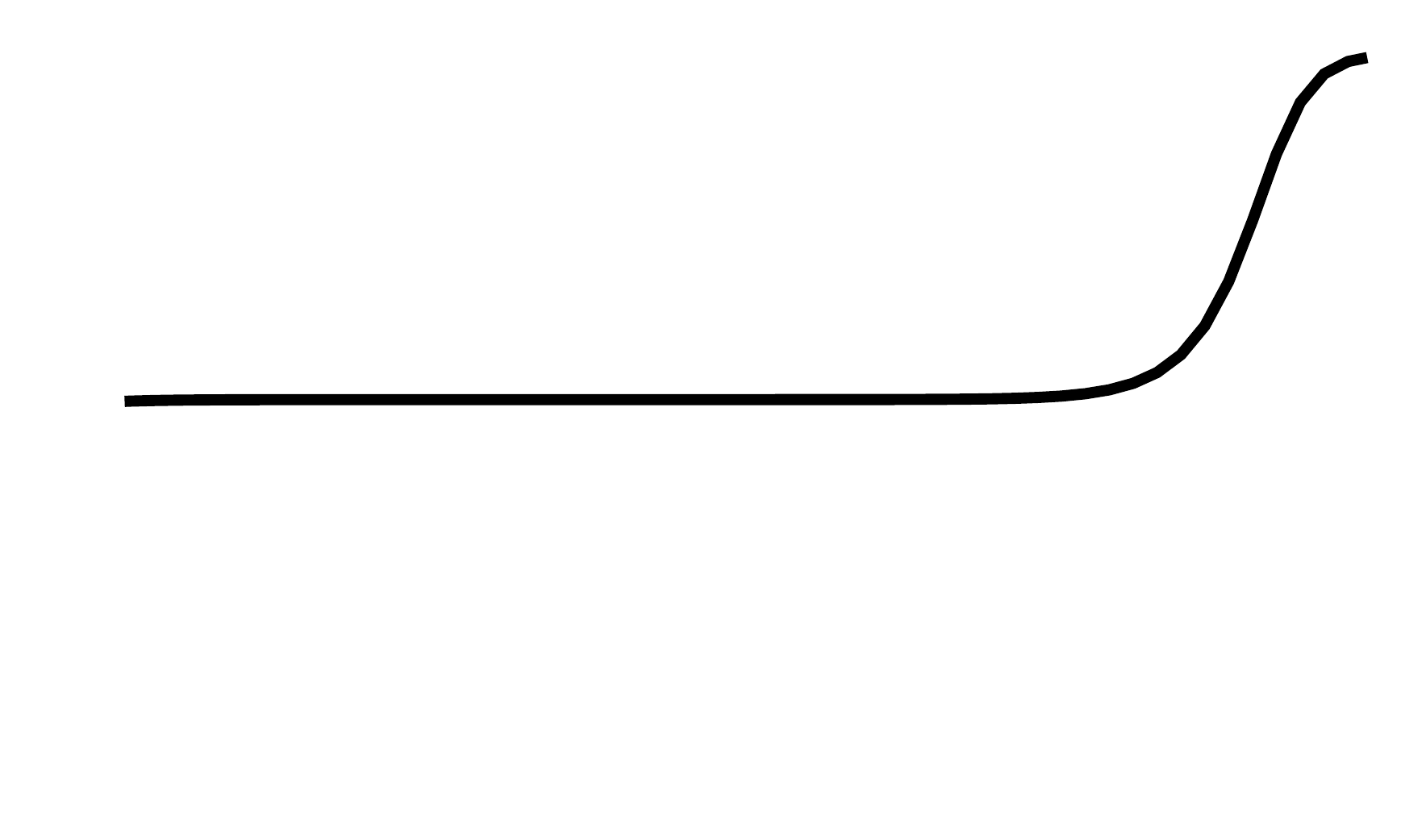} + \includegraphics[width = .12\textwidth,valign = c]{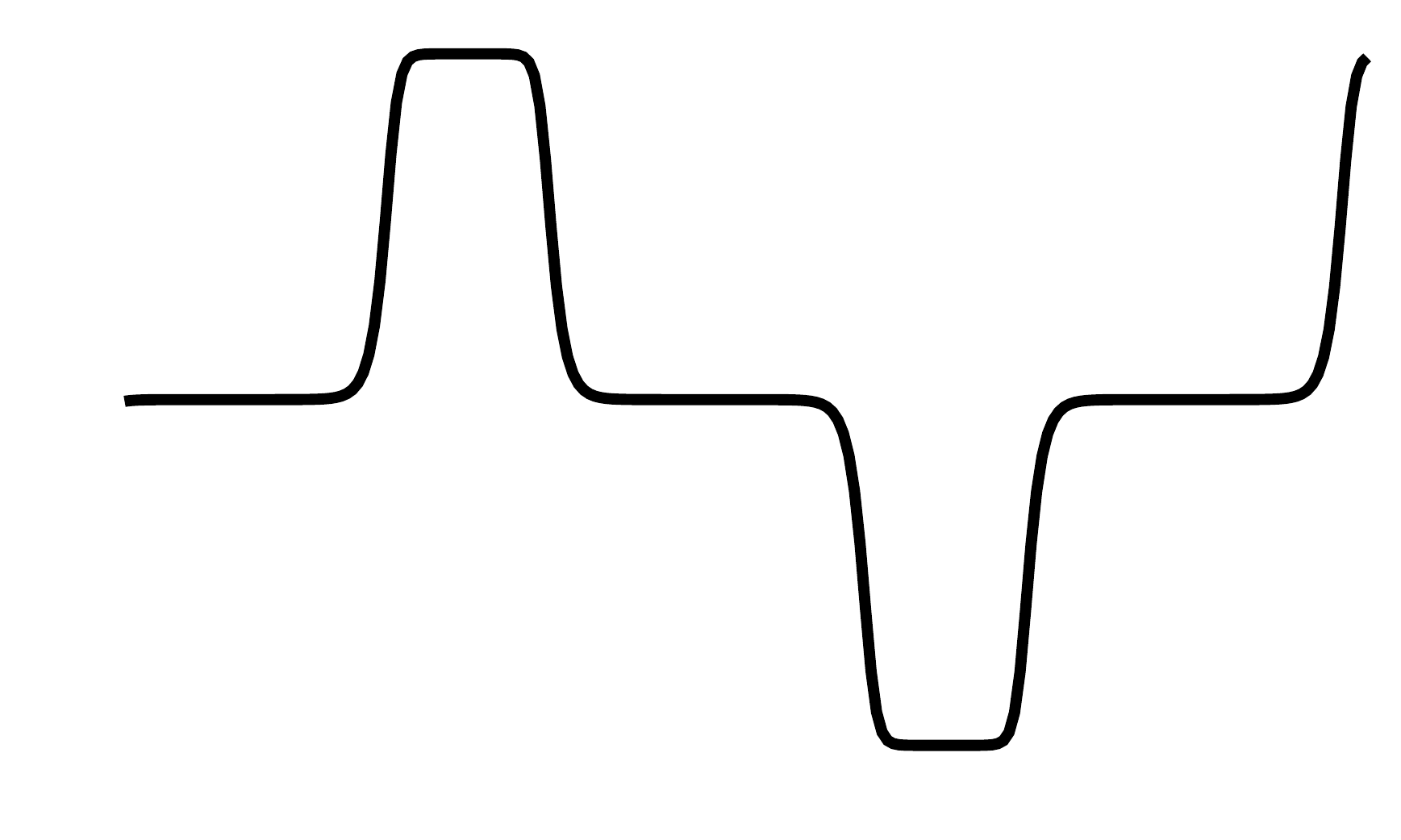} + \dots\right)\\
&\hspace{20pt}+\left(\includegraphics[width = .12\textwidth,valign = c]{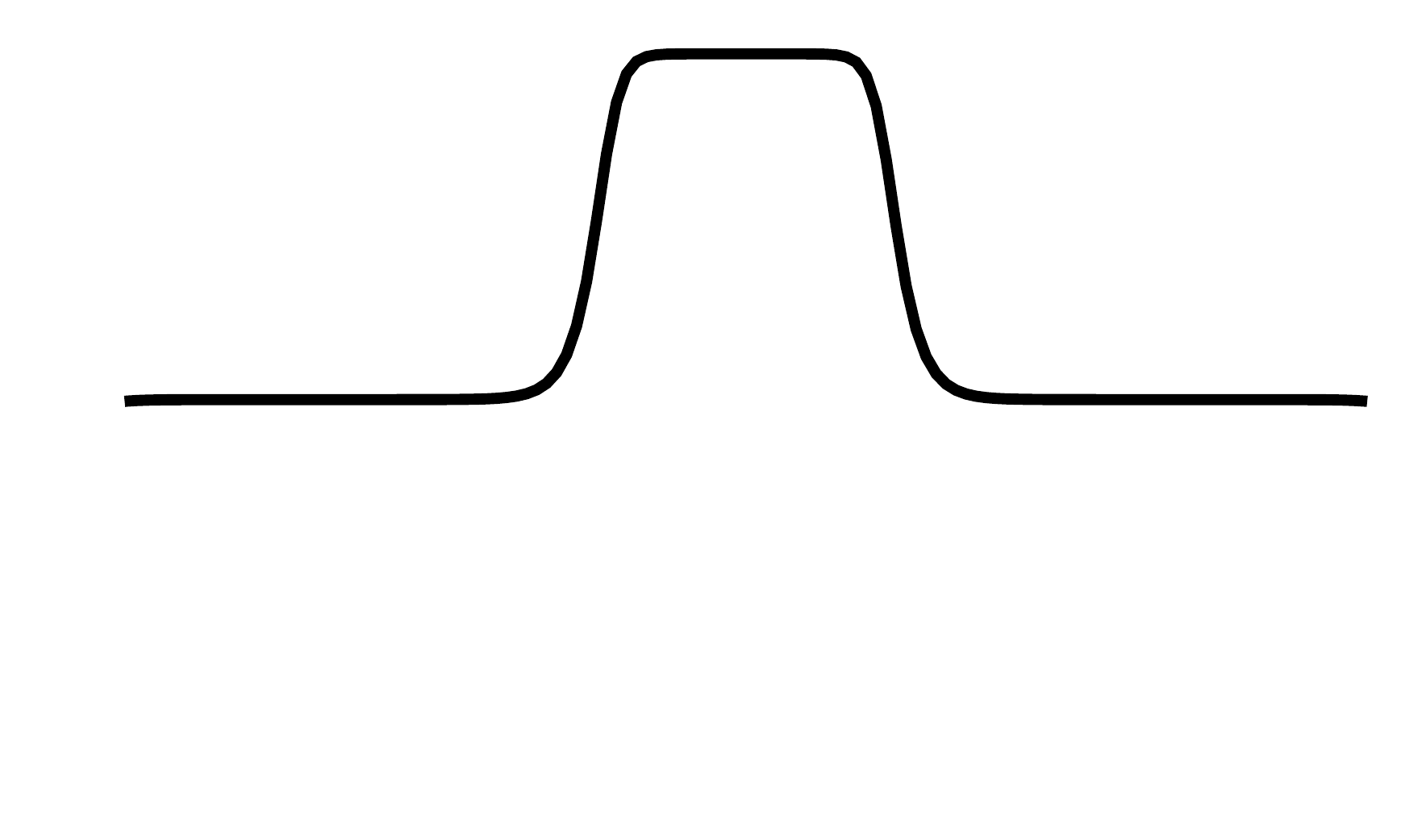} + \includegraphics[width = .12\textwidth,valign = c]{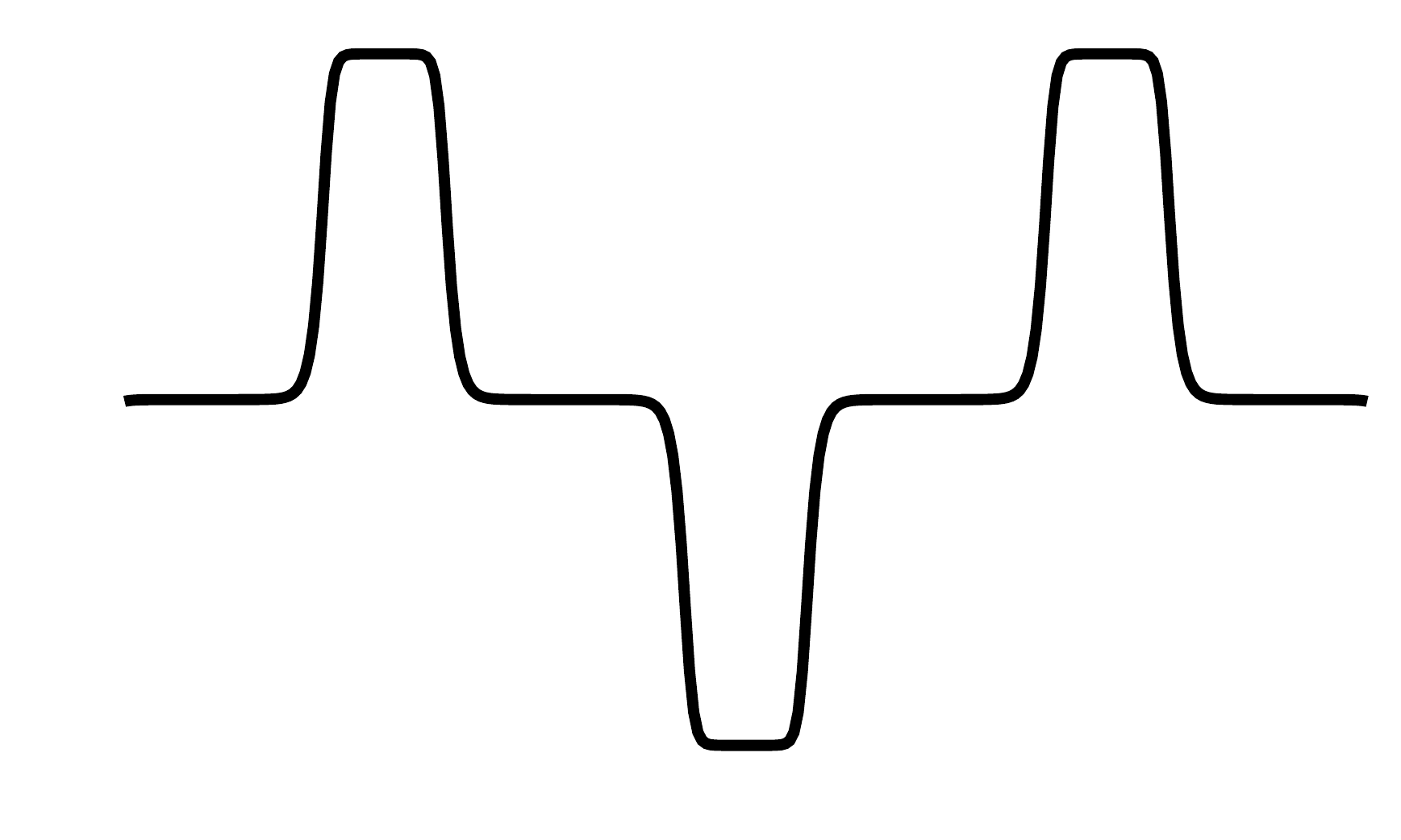} + \dots\right) + \left(\includegraphics[width = .12\textwidth,valign = c]{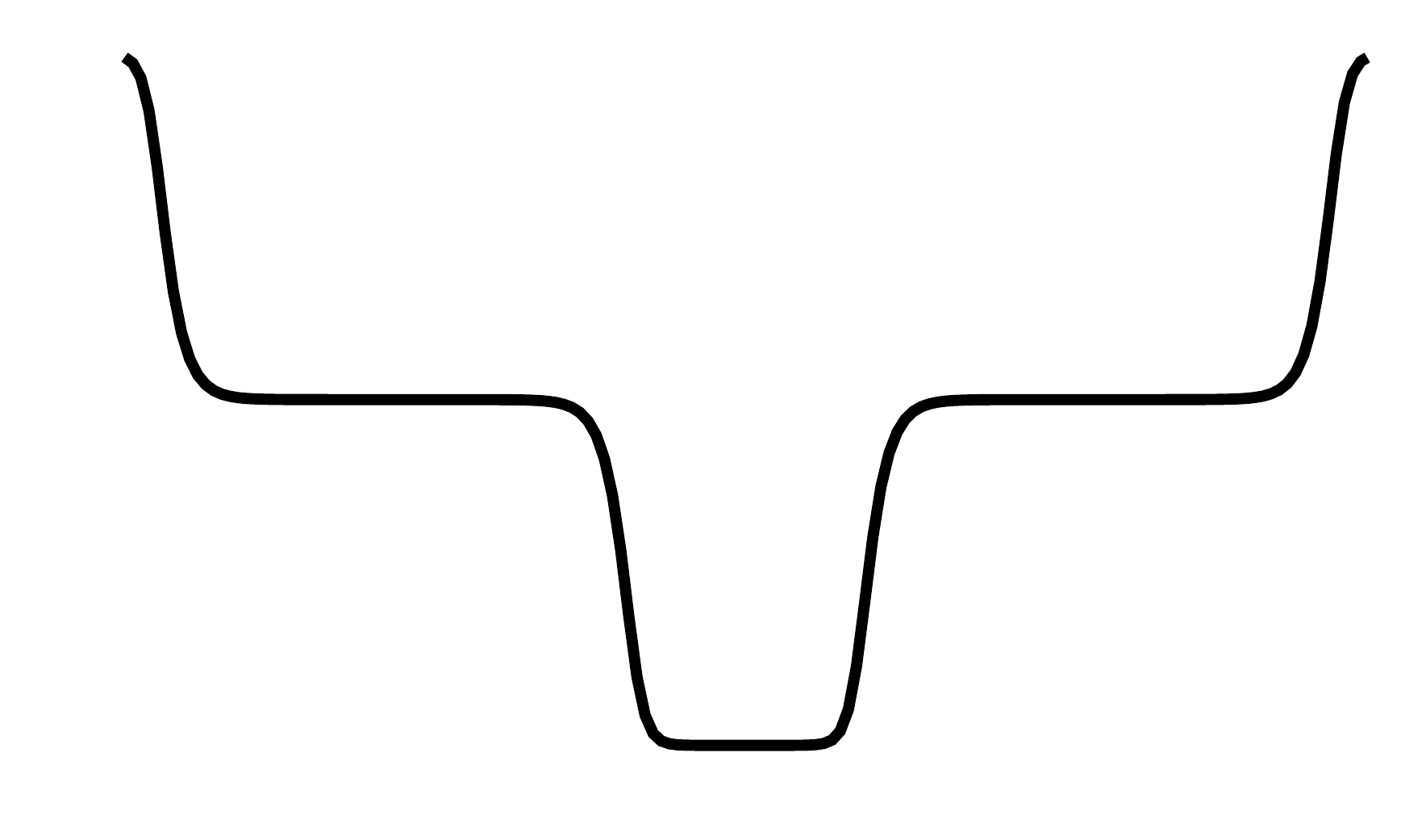} + \includegraphics[width = .12\textwidth,valign = c]{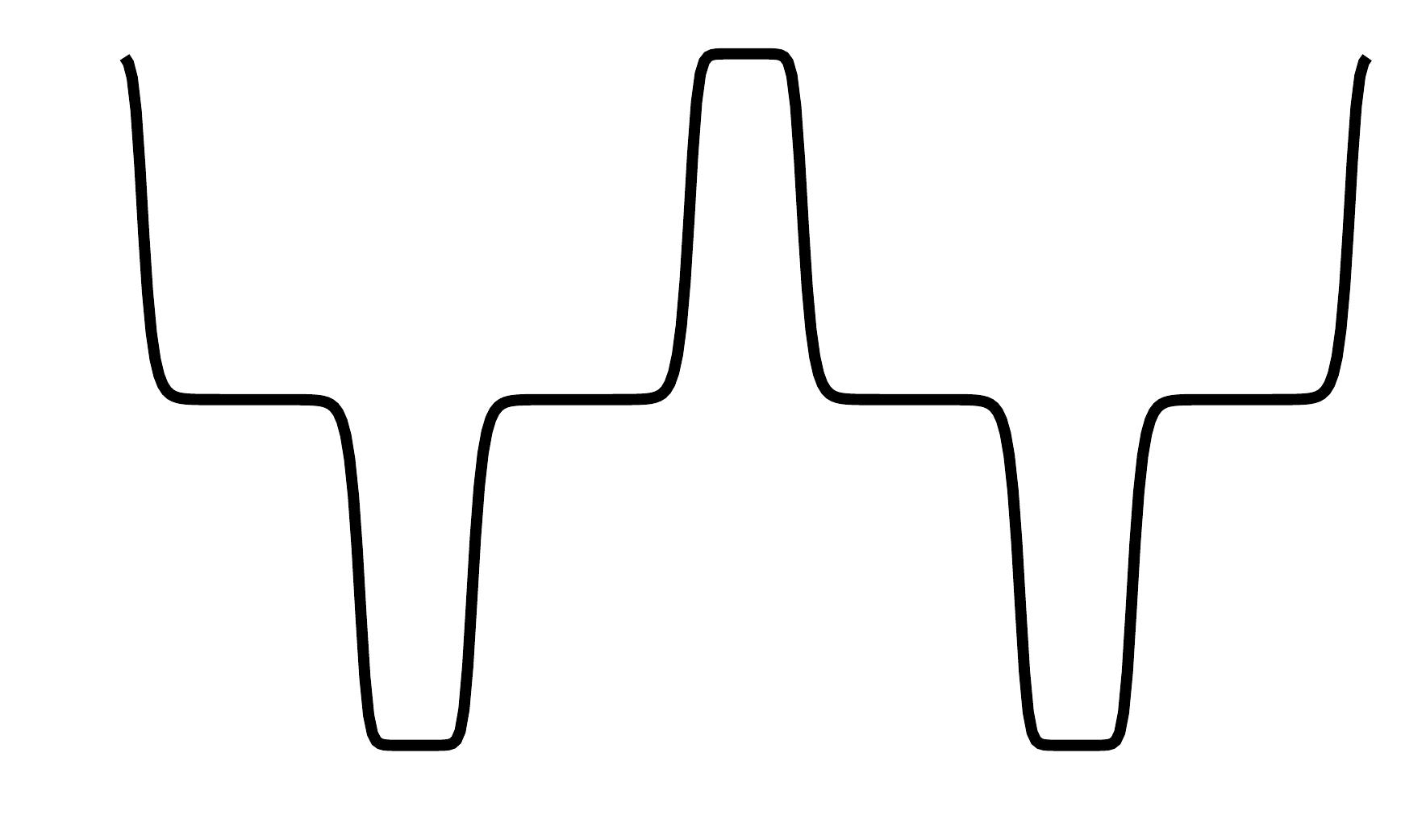} + \dots\right).\notag
	\end{align}
In these figures, the $y(t)$ function is plotted explicitly; $x(t)$ and $z(t)$ can be restored by comparing to (\ref{exampleSol}). Below, we will see that these four contributions correpsond to the four different Weingarten coefficients, and the sums within each parentheses become (for large $T$) a geometric series that can reproduce the specific Weingarten coefficients.

\subsection{The classical action of the solutions}\label{Sec:classicalAc}
Let's now estimate the on-shell action (\ref{actionGSigma}) of these solutions, using the approximation of large $T$. The terms other than the Pfaffian simplify nicely on shell:
\begin{align}
\frac{1}{2}\int\hspace{-8pt}\int_0^T \d t \d t'&\left[\Sigma_{jj'}(t,t')G_{jj'}(t,t') - \frac{J^2(t,t')}{q}s_js_{j'}G_{jj'}(t,t')^q\right]\notag \\
&\hspace{70pt}=\frac{JT}{2^{q-1}q} + \frac{1}{2}\sum_{j\neq j'}\int_0^T \d t\left[\sigma_{jj'}(t)g_{jj'}(t) - \frac{J}{q}s_js_{j'}g_{jj'}(t)^q\right]\\
&\hspace{70pt}=\frac{JT}{2^{q-1}q}+J\frac{q-1}{2q}\int_0^T \d ts_js_{j'}g_{jj'}(t)^q\\
&\hspace{70pt}=\frac{JT}{2^{q-1}q} + \frac{JT}{2^{q-1}}\frac{q-1}{q}r\\
&\hspace{70pt}\approx \frac{JT}{2^{q-1}}.\label{GSRE}
\end{align}
In the first step, we separated off the terms that arise from the coincident point correlator $G_{jj}(t,t) = \frac{1}{2}$, and then took the Brownian limit in the off-diagonal terms. In the second step, we used the equation of motion (\ref{SigmaEOM}). In the third step we used (\ref{gForm}) to write the answer in terms of $x,y,z$, and then used the conservation condition (\ref{defr}). In the final step, used that for late-time solutions, $r\approx 1$.

The Pfaffian term takes a little more work. First of all, we can write it as a fermion integral
\be
\text{Pf}(\partial_\tau - \sigma) = \int_{\psi^{(1)} = \i\psi^{(2)},\psi^{(3)}=\i\psi^{(4)}}^{\psi^{(1)} = -\i\psi^{(4)}, \psi^{(2)} = -\i\psi^{(3)}} \hspace{-50pt}\mathcal{D}\psi^{(1)}\dots \mathcal{D}\psi^{(4)} \exp\left(-\frac{1}{2}\int_0^T \mathrm{d} t\left[\psi^{(j)}\partial_\tau\psi^{(j)} - \sigma_{jj'}(t)\psi^{(j)}\psi^{(j')}\right]\right)
\ee
This is a path integral over four Majorana fermions. Normally, we would need to represent this in a $2^{4/2} = 4$-dimensional Hilbert space. But because the entire evolution remains in the bosonic sector, we can get by with a representation in a two-dimensional Hilbert space, where we write the fermion bilinears in terms of the Pauli $X,Y,Z$ operators
\be
\psi^{(1)}\psi^{(2)} = -\tfrac{\i}{2} X, \hspace{20pt} \psi^{(1)}\psi^{(3)} = \tfrac{\i}{2}Y, \hspace{20pt} \psi^{(1)}\psi^{(4)} = -\tfrac{\i}{2} Z.
\ee
Then the Pfaffian can be written as a Hilbert space quantity
\be\label{expham}
\text{Pf}(\partial_\tau - \sigma) = 2\times \langle 0|\mathcal{T}\exp\left\{-\int_0^T \d t\, h(t)\right\}|+\rangle,
\ee
where the time dependent ``Hamiltonian'' is
\be\label{ham}
h(t) = \frac{J}{2^{q-1}}\left(-x^{q-1}(t) X + \i y^{q-1}(t)Y - z^{q-1}(t)Z\right).
\ee
Here we are using the standard notation that $|\pm\rangle$ are the plus and minus eigenstates of $X$, and $|0\rangle$ and $|1\rangle$ are the plus and minus eigenstates of $Z$.

Let's consider evaluating this for one of the solutions described above. When $T$ is large, the solutions have long regions in which $\{x,y,z\} \approx  \{1,0,0\}$ or $\{x,y,z\} \approx  \{0,0,1\}$. Along these regions, the $h(t)$ operator in (\ref{ham}) is proportional to $X$ or $Z$, and the exponential of $-h(t)$ along these long intervals of time projects the state onto $|+\rangle$ or $|0\rangle$, respectively. The eigenvalue of the corresponding $h(t)$ on these states is $-\frac{J}{2^{q-1}}$, so the contribution of the Pfaffian term cancels the contribution of (\ref{GSRE}) along these portions, so the total effect of these portions of the solution is just a projection onto either $|+\rangle$ or $|0\rangle$.

In addition, there are long regions where $\{x,y,z\} \approx \{1,1,1\}$ or $\{x,y,z\} \approx \{1,-1,1\}$, and there are also transition regions. How do these contribute? First we consider a transition region like the one at the very beginning of the solution (\ref{exampleSol}), where we transition from $\{1,1,1\}$ to $\{0,0,1\}$. There are two facts that simplify the analysis of this region. First, this region is adjacent to a long portion of the solution in the future, where $\{x,y,z\} \approx \{0,0,1\}$. By the logic described above, this effectively gives a future projector onto the state $\langle 0|$. Second, during the transition, we have $x(t)\approx y(t)$ and $z(t)\approx 1$. This means that $h$ is approximately
\be
h(t) \approx \frac{J}{2^{q-1}}\Big[f(t)(-X+\i Y) + Z\Big].
\ee
Here the $f(t)$ depends on the detailed shape of the transition region. But because the $(-X+\i Y)$ operator annihilates the state $\langle 0|$, and because $Z$ is equal to $+1$ on this state, we have
\be\label{Argnear}
\langle 0| \mathcal{T}\exp\left\{-\int \d t\, h(t)\right\} \approx \langle 0| \exp\left\{\int \d t \frac{J}{2^{q-1}}\right\}.
\ee
So the state $\langle 0|$ is unchanged, and the multiplicative factors cancels against the local contribution in (\ref{GSRE}). A similar argument also applies to the other transition regions and to the extended portions of the solution with $\{x,y,z\} \approx \{1,\pm 1 ,1\}$. 

The upshot is that up to a term exponential in $T$ that cancels against (\ref{GSRE}), the Pfaffian reduces to a $\langle 0|\cdot |+\rangle$ matrix element of a sequence of projection operators. We insert the projector $\Pi_0$ for regions in which $\{x,y,z\} = \{0,0,1\}$ and we insert $\Pi_+$ for regions in which $\{x,y,z\} = \{1,0,0\}$. To illustrate this, in the following table we show the calculation of the action for the leading terms in each of the four qualitative classes of solutions (\ref{qualitativeClasses}). 
\be
\begin{tabular}{c|c|c}
 solution & $e^{-\frac{JT}{2^{q-1}}}\cdot$Pf & $e^{-I}$\\
\hline
\includegraphics[width = .12\textwidth,valign = c]{figures/sol100.pdf}  \rule{0pt}{5ex} & $2\langle 0|\Pi_{0}|+\rangle = 2^{1/2}$ & $2^{N/2}$\\
\includegraphics[width = .12\textwidth,valign = c]{figures/sol010.pdf}  & $2\langle 0|\Pi_+|+\rangle = 2^{1/2}$ & $2^{N/2}$\\
\includegraphics[width = .12\textwidth,valign = c]{figures/sol000.pdf}  & $2\langle 0|\Pi_0\Pi_+|+\rangle = 2^{1/2}$ & $2^{N/2}$\\
\includegraphics[width = .12\textwidth,valign = c]{figures/sol110.pdf}  & $2\langle 0|\Pi_+\Pi_0|+\rangle = 2^{-1/2}$ & $2^{-N/2}$
\end{tabular}
\ee
The general solution is obtained from one of these four by inserting additional periods of the periodic solution in the ``middle.'' This introduces additional projectors, and one can check that for each additional period that is inserted in the middle of such a solution, we make one or the other of the replacements
\be
\Pi_+\rightarrow \Pi_+\Pi_0\Pi_+ = \frac{1}{2}\Pi_+,\hspace{20pt} \text{or}\hspace{20pt} \Pi_0\rightarrow \Pi_0\Pi_+\Pi_0 = \frac{1}{2}\Pi_0.
\ee
This means that the Pfaffian decreases by a factor of two for each inserted period, and $e^{-I}$ decreases by a factor of $2^{-N} = L^{-2}$.

\subsection{The OTOC}\label{Sec:BSYKOTOC}
At the level of the classical action, we can now write the answers for the different terms that contribute to the OTOC in (\ref{qualitativeClasses}):
  \begin{align}\label{qualitativeClasses1}
\left(\includegraphics[width = .12\textwidth,valign = c]{figures/sol100.pdf} + \includegraphics[width = .12\textwidth,valign = c]{figures/sol101.pdf} + \dots\right) &\sim e^{N\gamma_i/2}\Big(1 + \frac{1}{L^2} +\dots\Big)\\
\left(\includegraphics[width = .12\textwidth,valign = c]{figures/sol010.pdf} + \includegraphics[width = .12\textwidth,valign = c]{figures/sol011.pdf} + \dots\right) &\sim e^{N\gamma_f/2}\Big(1 + \frac{1}{L^2} +\dots\Big)\label{qualitativeClasses2}\\
\left(\includegraphics[width = .12\textwidth,valign = c]{figures/sol000.pdf} + \includegraphics[width = .12\textwidth,valign = c]{figures/sol001.pdf} + \dots\right) &\sim  \Big(1 + \frac{1}{L^2} +\dots\Big)\label{qualitativeClasses3}\\
\left(\includegraphics[width = .12\textwidth,valign = c]{figures/sol110.pdf} + \includegraphics[width = .12\textwidth,valign = c]{figures/sol111.pdf} + \dots\right)&\sim \frac{e^{N(\gamma_i+\gamma_f)/2}}{L^2}\Big(1 + \frac{1}{L^2} +\dots\Big)\label{qualitativeClasses4}
\end{align}
On the LHS, the dots contain configurations with additional periods (pairs of ``humps'') added, and on the RHS, the dots contain terms with further powers of $1/L^2$. The factors of $e^{N \gamma_i/2}$ and $e^{\gamma_f N/2}$ are present or absent depending on whether the correlation is large (present) or small (absent) at the early and final times. The factors of $L = 2^{N/2}$ are taken from the evaluation of the action in the last subsection. 

At the classical level (meaning at the level of the coefficient of $N$ in the exponent) these terms have precisely the same form as (\ref{OTOC}), once we expand the prefactor $1/(1-L^{-2})$ in powers of $L^{-2}$. However, there will also be order-one coefficients multiplying all terms in the Brownian SYK expression, coming from the integral over quantum fluctuations, and we will need specific values of these coefficients to truly match to (\ref{OTOC}). Naively, the way to compute these factors is to do a one-loop determinant around each of the saddles. However, a complication is that associated to each ``hump'' in the solution (region in the bulk of the solution where $y$ is close to $\pm 1$) there are a pair of nearly-zero modes that have to be treated specially.\footnote{This sounds similar to ``kink'' instantons in the double-well problem in quantum mechanics, where the center-of-mass time of the kink is a nearly-zero mode that has to be treated specially. However, the details are different: in the double-well, the zero-mode integral gives $T^{\text{\# kinks}}/(\text{\# kinks})!$, while we will get $(-1)^{\text{\# humps}}$.}

To better understand what the humps represent, it is helpful to take a step back and think about the random unitary integral. The operator
\be
\int \d U \ U \otimes U^* \otimes U \otimes U^*
\ee
is a projection operator onto the subspace of states that are invariant under $U$. For the case where $U$ is integrated over the entire unitary group, the only states that are invariant are superpositions of the maximally entangled states
\be
|x\rangle = \frac{1}{L}\sum_{i,j} |i\rangle\otimes |\bar{i}\rangle \otimes |j\rangle\otimes |\bar{j}\rangle, \hspace{20pt} |z\rangle = \frac{1}{L}\sum_{i,j} |i\rangle\otimes |\bar{j}\rangle \otimes |j\rangle\otimes |\bar{i}\rangle
\ee
These are normalized states with inner product $\langle x|z\rangle = \frac{1}{L}$, and the projector onto the subspace they span is given by 
\be\label{projector}
\left(\begin{array}{cc} |x\rangle & |z\rangle \end{array}\right) \left(\begin{array}{cc} 1 & \frac{1}{L} \\ \frac{1}{L} & 1 \end{array}\right)^{-1}\left(\begin{array}{c} \langle x| \\ \langle z| \end{array}\right) = \sum_{n = 0}^\infty \left(\begin{array}{cc} |x\rangle & |z\rangle \end{array}\right) \left(\begin{array}{cc} 0 & -\frac{1}{L} \\ -\frac{1}{L} & 0 \end{array}\right)^{n}\left(\begin{array}{c} \langle x| \\ \langle z| \end{array}\right).
\ee

In Brownian SYK, these states $|x\rangle$ and $|z\rangle$ are zero-energy ground states of the theory (\ref{disorderAvgAction}) propagating on the four contours. In the large $N$ collective-field description, $|x\rangle$ and $|z\rangle$ correspond to static solutions to the equations of motion, with ``$x$-type'' correlation $\{x,y,z\} = \{1,0,0\}$ or with ``$z$-type'' correlation $\{x,y,z\} = \{0,0,1\}$. The general solution to the equations of motion that we discussed above is constructed out of long periods of these stationary configurations, connected together by ``hump'' transitions between them. For example, consider the simplest solution with a hump in it, from the leftmost diagram in (\ref{qualitativeClasses3}):
\be\label{contributionlinethree}
\includegraphics[width = .37\textwidth,valign = c]{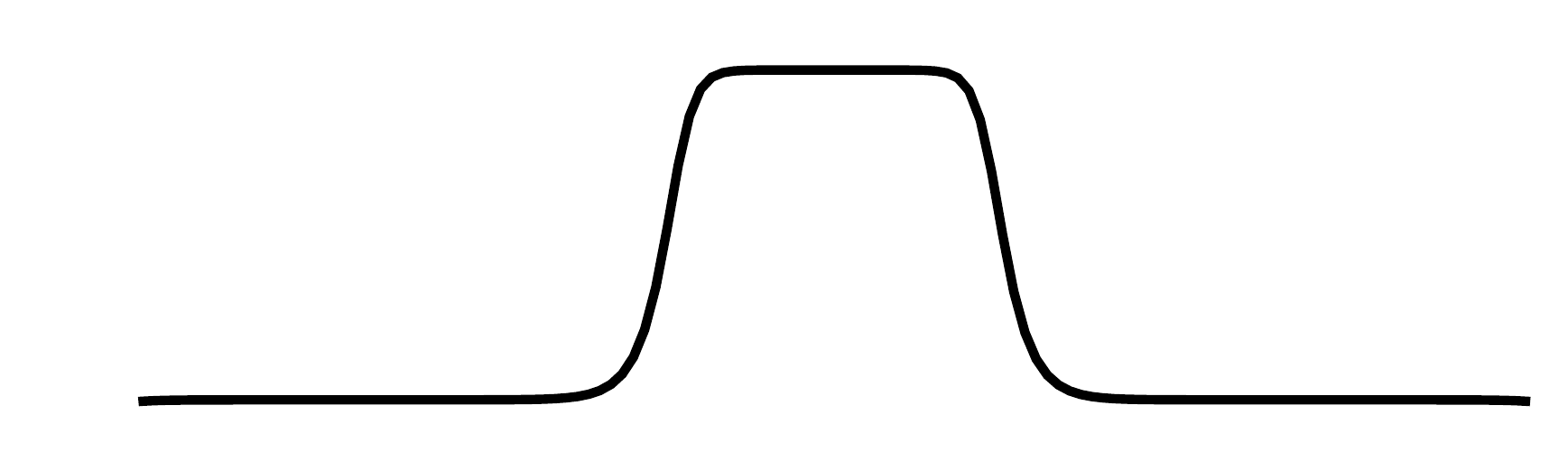}  \ \ = \ \ \includegraphics[width = .4\textwidth,valign = c]{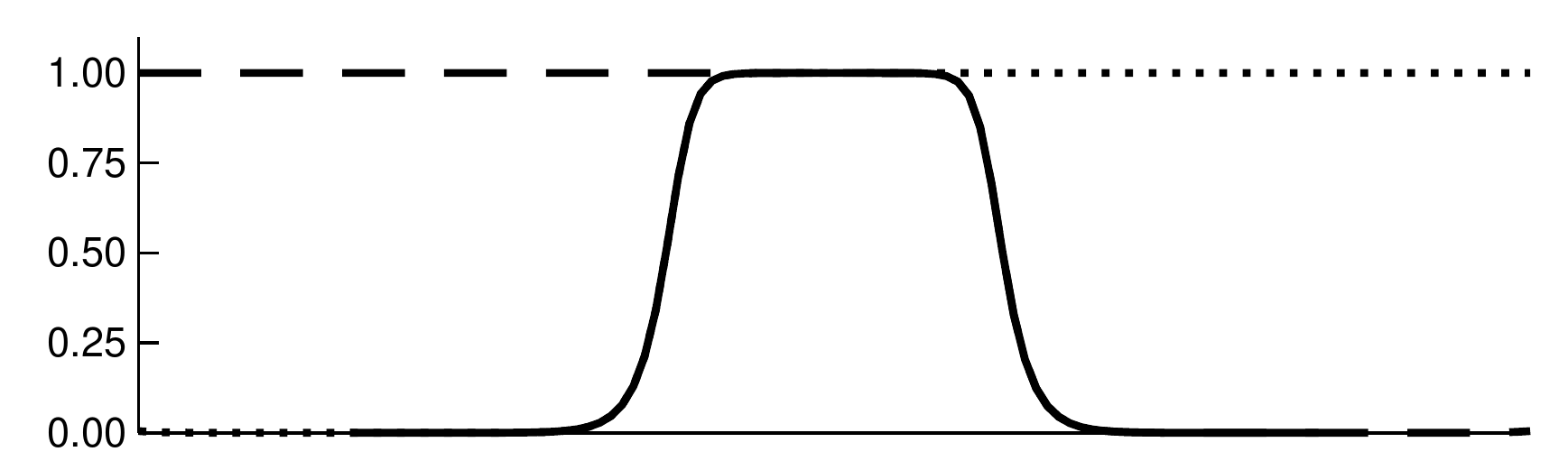}
\ee
On the right, we restored the axes and the $x$ (dashed) and $z$ (dotted) correlations. It is clear that this solution represents a transition from $x$-type correlation at early time to $z$-type correlation at late time. The sum over such transitions in the full set of solutions above corresponds to the sum on the RHS of (\ref{projector}). The action of each transition gives the correct factor of $1/L$, but to understand the minus sign we need to examine the integral over fluctuations.

For this ``one hump'' case, the nearly-zero modes can be parametrized as the times $t_l$ and $t_r$ that mark the beginning and end of the hump. We claim that these can be identified with the variables $x^+$ and $x^-$ of section \ref{sec:qualitativediscussion}, with the relationship
\be\label{xpm}
x^+ \sim e^{\lambda_L (T-t_r)}, \hspace{20pt} x^- \sim e^{\lambda_L t_l}.
\ee
To see this, recall that $x^+$ is the coefficient of a decorrelating ``scramblon'' mode that grows exponentially towards the future, proportional to $x^+ e^{\lambda_L(t-T)}$. When this mode becomes of order one, it has the effect of destroying the $x$-type correlation. In the one-hump solution, the $x$-type correlation gets destroyed at the right boundary of the hump, at time $t_r$. This implies $x^+  e^{\lambda_L (t_r-T)}\sim 1$. The argument is similar for the $x^-$ mode.\footnote{We give two checks of this correspondence in appendix \ref{app:kernel}. First, we show that the action (\ref{actionConjecture}) accurately describes small fluctuations around the disk solution in Brownian SYK, to quadratic order in $x^+$ and $x^-$. Second, we check that the correct integration measure is $\d x^+ \d x^-\propto e^{-\lambda_L(t_r-t_l)} \d t_l\d t_r$. The factor $e^{-\lambda_L(t_r-t_l)}$ arises from the one-loop determinant over the nonzero modes, as explained at the end of appendix \ref{app:kernel}.}

In the effective theory of section \ref{sec:qualitativediscussion}, the full integral over $x^+$ and $x^-$ is\footnote{A minor detail is that if we define $x^\pm$ as in (\ref{xpm}), the $-\i$ is actually not present in the action. In writing this formula, we are implicitly rotating the contour as discussed in appendix \ref{app:kernel}. More generally, the phase of the action depends on the spacing of the operator insertions around th Euclidean circle, and the Brownian SYK case corresponds most closely to equal $\beta/4$ spacing.}
\be
\int \frac{\d x^+\d x^-}{2\pi/a} e^{-\i a x^+ x^-} = +1
\ee
but as we explained, this integral can be decomposed into three pieces $+1 = +1 + 1 - 1$, where the two positive contributions arise from the part of the integral where one of $x^\pm$ is small and the other is large, and the $-1$ contribution arises from the region where both are large. The configurations where one of $x^\pm$ is large and one is small correspond to the two leading terms in (\ref{qualitativeClasses1}) and (\ref{qualitativeClasses2}), which should therefore have coefficient $+1$. The one-hump solution corresponds to a region where $x^\pm$ are both large, and should therefore have a coefficient $-1$.\footnote{There is an interesting feature of this integral: because the semiclassical approximation breaks down for these $x^\pm$ modes, the actual solution is less significant than the integration region that it is part of, which we take to be the region where neither of $x^\pm$ are small, so the $x$-type correlation is small at $t = 0$ and the $z$-type correlation is small at $t = T$. The solution is at a somewhat arbitrary point within this integration region, and some of its featrues are not representative. For example, the integral is dominated by the region where $N e^{\lambda_L T}x^+x^-\sim 1$, so $t_r-t_l\propto \log(N)$, whereas in the solution, $t_r-t_L \propto T$.}

For solutions with more humps, we expect that the integral over fluctuations can be well approximated by a product of the separate integrals for each hump, giving a factor $(-1)^{\# \text{humps}}$. This gives precisely the right set of coefficients (and in particular minus signs!) to explain (\ref{OTOC}). The terms on the first line (\ref{qualitativeClasses1}) and second line (\ref{qualitativeClasses2}) have an even number of humps, and contribute with coefficient $+1$. The terms on the third line (\ref{qualitativeClasses3}) and fourth line (\ref{qualitativeClasses4}) have an odd number of humps, and contribute with coefficient $-1$.

\subsection{Discrete symmetries}
The above picture would be correct for a Brownian SYK-like theory with no global symmetries, for which the time-evolution operator could really converge to a random unitary at late times. However, the Brownian SYK theory has a $(-1)^F$ symmetry (which is anomalous for odd $N$). In cases where $q \equiv 2 $ (mod 4), it also has a time reversal symmetry as explained in appendix A.1 of \cite{Saad:2018bqo}. Due to these symmetries, the late-time ensemble should converge to a subgroup of the unitary group, and the answer for the OTOC should be modified. So we must be missing something in our analysis above.

To explain what is missing, let's consider the random unitary ensemble that preserves the $(-1)^F$ symmetry. The integral 
\be\label{dududu}
\int \d U \ U\otimes U^* \otimes U \otimes U^*
\ee
over such an ensemble is again a projector onto the subspace of invariant states, but this subspace is now larger, spanned by
\be\label{newstates}
\sum_{i,j} |i\rangle\otimes (-1)^{\eta_1 F}|\bar{i}\rangle \otimes |j\rangle\otimes (-1)^{\eta_2 F}|\bar{j}\rangle, \hspace{20pt} \sum_{i,j} |i\rangle\otimes (-1)^{\eta_1 F}|\bar{j}\rangle \otimes |j\rangle\otimes (-1)^{\eta_2 F}|\bar{i}\rangle
\ee
where $\eta_1,\eta_2$ can be either zero or one. This splits into two orthogonal subspaces with $\eta_1 = \eta_2$ and with $\eta_1 \neq \eta_2$, and the initial and final conditions for the OTOC are orthogonal to the $\eta_1\neq \eta_2$ subspace, so it plays no role. So for the purposes of the OTOC computation, the upshot is that we have two additional ground states, namely (\ref{newstates}) with $\eta_1 = \eta_2 = 1$.

In Brownian SYK, these extra states correspond to fact that in the large $N$ theory, there are further static solutions $\{x,y,z\} = \{-1,0,0\}$ and $\{ x,y,z\} = \{ 0,0,-1\}$ for which the action is independent of time. For late time, one can then have approximate solutions involving transitions between these patterns of correlation and also the patterns $\{x,y,z\} = \{1,0,0\}$ and $\{0,0,1\}$ that we discussed previously. A subtlety is that strict classical solutions will not connect $x>0$ to $x<0$, as discussed in section \ref{sec:conservation}. However, there are configurations that will accomplish this with a violation of the equations of motion that is exponentially small in $T$, since we can ``by hand'' change the sign of $x$ and $y$ during a period when they are exponentially small. 

For cases with time-reversal symmetry, there are new invariant states where the subsystems with $U$ operators in (\ref{dududu}) are entangled with each other, and likewise the subsystems with $U^\dagger$ operators. In Brownian SYK, these correspond to stationary configurations with ``$y$-type'' correlation $\{x,y,z\} = \{0,\pm \i,0\}$, which can again transition by slightly off-shell configurations to the configurations with $x$-type or $z$-type correlation. We have checked that the contribution of the Pfaffian to the action for such transitions gives the $N$ mod 8 periodicity expected for this type of system with both $(-1)^F$ symmetry and time-reversal symmetry, see e.g.~\cite{Stanford:2019vob}.

\subsection{Comparison to the handle-disk}
For the subleading correction to the OTOC
\be\label{comparisonWWVV}
-\frac{1}{L^2}\langle VV\rangle\langle WW\rangle
\ee
it is interesting to relate the JT and Brownian SYK results by comparing the pattern of correlation on the handle-disk with the Brownian SYK saddle point. To do so, consider the following two-point function of probe $\mathcal{O}$ operators on the handle-disk:
\be
\includegraphics[width = .25\textwidth,valign = c]{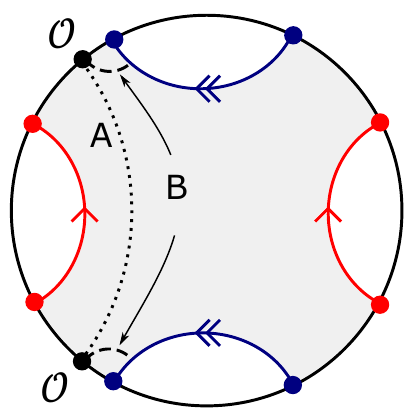} 
\ee
Here we use the same notation and color as in section \ref{sec:JT}. The $W$ insertions at $t=0$ are represented by the blue dots and the $V$ insertions at $t=T$ are red dots. The two black dots probe the pattern of correlation, and by placing them at corresponding locations on the two segments of the boundary indicated, we are probing the ``$x$-type'' correlator analogous to $x(t)$ in the Brownian SYK solutions.

There are two leading geodesics contributing to the correlator, labelled A and B.
When the probe operators are inserted close to the $W$ operator at $t=0$, both types of geodesic contribute significantly to the correlator. However, the B geodesic intersects the blue geodesic associated to the $W$ operators, and as we increase the time of the $\mathcal{O}$ insertions, the relative boost of this intersection increases, and the B geodesic becomes longer, contributing little to the correlator. This represents the scrambling effect of the $W$ operator insertion.

What about the contribution of A? This geodesic doesn't intersect with anything, so it seems to be immune to scrambling effects. And indeed, in the classical solution in pure JT gravity the length of A is independent of time. (This is not true in the above Euclidean geometry, but it is true in the Lorentzian geometry appropriate to the OTOC, where the time is moving in opposite directions on the two boundaries, similar to the boost isometry of the two-sided black hole. The Lorentzian geometry is hard to draw so we will continue to use the Euclidean picture.)

However, at late times there are important fluctuations around the classical solution, where slightly off-shell ``scramblon'' modes are activated. Concretely, these modes are the modes that would be sourced by a very small amount of stress energy propagating around the ``horizontal'' or ``vertical'' cycles in the handle-disk. The horizontal one is pictured below:
\be
\includegraphics[width = .25\textwidth,valign = c]{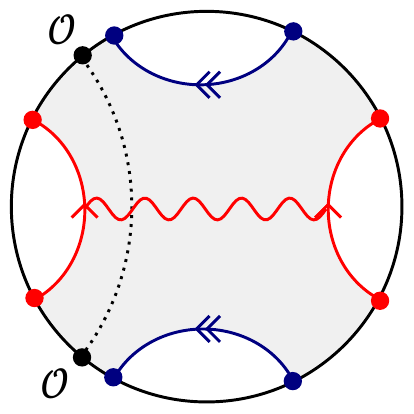} 
\ee
The part of the bulk path integral that is responsible for the contribution (\ref{comparisonWWVV}) is a slightly off-shell one where both the ``horizontal'' and ``vertical'' scramblon modes are turned on (the analogous point was explained in detail for the disk topology in section \ref{sec:qualitativediscussion} near (\ref{intanswer2})). For such a configuration, the length of the A geodesic is time-dependent, because it depends on the relative boost between the off-shell scramblon mode at time $T$ and the A geodesic at time $t$. This relative boost is determined by the time difference, and at early times, the boost is very large, and the length of the A geodesic is correspondingly large, leading to very little $x(t)$ correlation. At late times, the relative boost is small, and $x(t)$ is large. 

We can now compare this to the handle-disk solution in Brownian SYK that gives (\ref{comparisonWWVV}). In the plot below, the dashed line represents the $x(t)$ correlator for this solution:
\be
\includegraphics[width = .4\textwidth, valign = c]{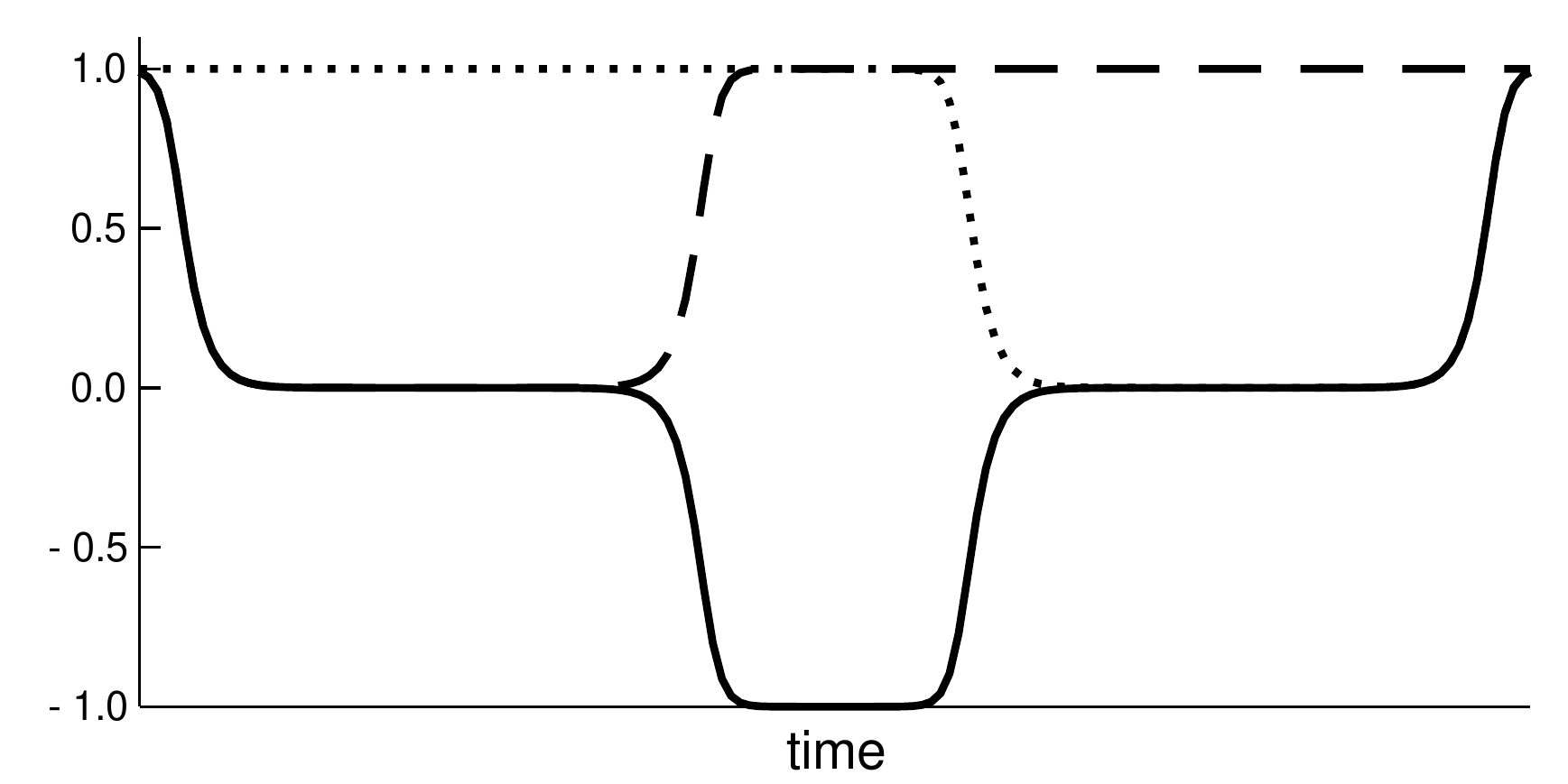}
\ee
The initial value of $x(t)$ is large, reflecting the contribution of the B geodesic. Then the value becomes small for a long time, before eventually rising to an $O(1)$ value. This is qualitatively consistent with the pattern we expected for the off-shell handle disk.

But there is one important subtlety. We argued that this pattern of correlation represents the contribution of a slightly off-shell handle disk, but here we see it arising from an on-shell solution in Brownian SYK. The explanation for this discrepancy is that Brownian SYK is analogous to JT gravity with dynamical matter fields, and on a handle-disk geometry, the scramblon modes can have small but nonzero on-shell values sourced by these matter fields propagating around the closed cycles of the handle-disk. So it happens that there is an actual solution to the Brownian SYK equations with the scramblon modes turned on, although to somewhat arbitrary values determined by the bulk matter. These modes are only very weakly stabilized at these values, and we need to integrate over their fluctuations anyhow, so the details of these classical solutions are not particularly significant, but the qualitative behavior of the correlation is right.

\section{Discussion}
In this paper, we studied how the time evolution $U(t)$ in two quantum chaotic systems (JT gravity with matter and Brownian SYK) can reproduce small but important effects present in the average over unitary matrices. In JT gravity, we matched the first subleading terms, and the most interesting effect came from the topology of a disk with a handle inserted. In Brownian SYK, we found path integral configurations that can reproduce the exact answer for the ensemble of random unitary matrices. In both systems, an important minus sign arises from large fluctuations in modes for which the large $N$ approximation breaks down at late times. We explained these modes using an effective action associated to the chaos of the theory.

The full answers for the unitary integral involve a convergent series in $(\text{dim Hilbert space})^{-2}$. In Brownian SYK, this series is reproduced by a set of path integral configurations that involve transitions between two different basic patterns of correlation. This seems to be an interesting concept with some similarity to the genus expansion in two-dimensional gravity, but which might make sense in a wider class of systems. Perhaps the Brownian SYK patterns are the universal ``skeleton'' of an expansion present in chaotic theories, which in 2d gravity it is provided by a part of the integration over moduli space. It would be interesting to understand this better, and in particular, to generalize the effective theory of $x^\pm$ to describe the space of these configurations. 

Some other future directions include
\begin{itemize}
\item Can one find the $x^+$ and $x^-$ action in other systems, such as large $q$ SYK at different temperatures?\footnote{Added in v2: this problem has been solved in \cite{Gu:2021xaj}.}
\item For an evaporating black hole, can one find the third and fourth terms in (\ref{rhoarhob})? In a simple two-dimensional gravity setup with ``one-sided'' black holes, the first term corresponds to two disks, and the second term corresponds to one disk. The third term should correspond to two disks connected by a wormhole. We expect that this topology contains a region that describes a wormhole contribution to $|\langle a|b\rangle|^2$, and also a region that descibes the third term in (\ref{rhoarhob}). However, the bulk analysis looks tricky, and comparison to the random unitary formula may not be precise, due to the non-equilibrium nature of evaporation.
\item What is the boundary dual of JT gravity with operator insertions but without dynamical matter loops? Perhaps this can give a hint to some generalization of random matrix theory. To what extent can one make sense of the theory with matter loops included?

\end{itemize}

\section*{Acknowledgements} 

We thank Andreas Blommaert and Geoff Penington for initial collaboration, and Phil Saad, Arvin Shahbazi-Moghaddam, and Stephen Shenker for discussions. DS was supported in part by DOE grant DE-SC0021085. ZY is supported in part by the Simons Foundation.  SY was supported in part by NSF grant PHY-1720397. 

\appendix

\section{Random unitary evolution vs.~random energy basis}\label{App:REB}
In the main text, our goal was to match gravity and SYK computations to a formula for the OTOC in a model where the time evolution operator is a random unitary. This is an accurate description of Brownian SYK at late times, but it isn't such a good model of gravity --- it neglects, for example, the conservation of energy. In this appendix we will discuss a more complicated model that conserves energy.

In a system with a time-independent Hamiltonian, the time evolution operator is
\be
U(t) = u^\dagger e^{-\i h t} u
\ee
where $h$ is diagonal and $u$ is the change of basis matrix between some reference basis (e.g. the local basis where simple operators are simple) and the energy eigenbasis. Here a simple model would be to take the matrix $u$ to be Haar random. Of course, this is still not a great model for a real quantum system, since it fails to encode the fact that simple operators do not change the energy by a huge amount. But we believe it is useful at a qualitative level.

With $V$ insertions at times $\pm \frac{\delta V}{2}$ and $W$ insertions at times $t \pm \frac{\delta W}{2}$, the averaged OTOC in this model is
\begin{align}\label{complicated}
\int \d u \left\langle U(t + \delta_V + \delta_W)^\dagger \ W \ U(t + \delta_W) \ V \ U(t)^\dagger \ W \ U(t+\delta_V) \  V \right\rangle.
\end{align}
We will mainly set $\delta_W = \delta_V = 0$, but in some formulas below we will retain the dependence on these parameters. Because each $U$ operator involves one $u$ and one $u^\dagger$, (\ref{complicated}) involves an average of a product of four $u$ and four $u^\dagger$ operators. The full Weingarten expression for this includes $(4!)^2 = 576$ different contractions. It is not possible to write the whole expression, but one can explore parts of the answer. So for example, here are all of the terms that appear at leading order $L^0$:
\begin{align}
&\langle W\rangle^2\langle V\rangle^2\Big(4|\langle e^{-\i h t}\rangle|^2 + \langle e^{2\i h t}\rangle\langle e^{-\i h t}\rangle^2 +\langle e^{-2\i h t}\rangle\langle e^{\i h t}\rangle^2  -5|\langle e^{-\i h t}\rangle|^4 -1 \Big) \\
&+\langle WV\rangle \langle W\rangle\langle V\rangle\Big(-4|\langle e^{\i h t}\rangle|^2 - 2\langle e^{2\i h t}\rangle\langle e^{-\i h t}\rangle^2 -2\langle e^{-2\i h t}\rangle\langle e^{\i h t}\rangle^2 + 8|\langle e^{-\i h t}\rangle|^4 \Big)\\
&+\Big(\langle W\rangle\langle WV^2\rangle + \langle W^2V\rangle\langle V\rangle\Big) \Big(2|\langle e^{\i h t}\rangle|^2-2|\langle e^{-\i h t}\rangle|^4\Big)\\
&+\Big(\langle W\rangle^2\langle V^2\rangle + \langle W^2\rangle \langle V\rangle^2\Big)\Big(-2|\langle e^{\i h t}\rangle|^2 +|\langle e^{-\i h t}\rangle|^4 +1 \Big)\\
&+\langle W V\rangle^2 \Big(-2|\langle e^{-\i h t}\rangle|^4 +\langle e^{2\i h t}\rangle\langle e^{-\i h t}\rangle^2 +\langle e^{-2\i h t}\rangle\langle e^{\i h t}\rangle^2\Big)\\
&+\langle W V W V\rangle |\langle e^{-\i h t}\rangle|^4.
\end{align}
More precisely, these are the terms that appear at leading order for small times $t$. For large $t$, the terms with time dependence become subleading, and only three terms from the above list survive:
\be
\langle W^2\rangle \langle V\rangle^2 + \langle W\rangle^2\langle V^2\rangle - \langle W\rangle^2\langle V\rangle^2.
\ee
These are the same terms that appear at leading order in the simpler model where $U(t)$ itself is a random unitary matrix. Putting back the more refined time dependence, these three terms split up into several terms that can be written as
\begin{align}
\langle V\rangle^2\Big[\langle W^2\rangle_c |\langle e^{\i \delta_W h}\rangle|^2  +\langle W\rangle^2\Big]+ \langle W\rangle^2\Big[\langle V^2\rangle_c|\langle e^{\i \delta_V h}\rangle|^2 + \langle V\rangle^2\Big] - \langle W\rangle^2\langle V\rangle^2
\end{align}
where the $\langle\rangle_c$ means the connected correlator. The expressions in brackets are simply the time-dependent two point functions e.g.~$\langle W(\delta_W)W(0)\rangle$ in this model.

What about the terms at order $L^{-2}$? This rapidly gets complicated, so we will consider the special case (which corresponds to what we studied in the JT gravity section) in which (i) one point functions are small, (ii) correlation between $V$ and $W$ operators is small, so that all of the following correlators are of order $L^{-2}$:
\be
\langle W\rangle^2, \hspace{10pt} \langle V\rangle^2, \hspace{10pt} \langle W\rangle\langle W V^2\rangle,\hspace{10pt} \langle W^2V\rangle\langle V\rangle, \hspace{10pt} \langle W V\rangle^2.
\ee
We will further assume that the time $t$ is large. Then one finds that the only terms at order $L^{-2}$ are
\be
\langle W^2\rangle\langle V\rangle^2  + \langle W\rangle^2\langle V^2\rangle   -\frac{1}{L^2}\langle W^2\rangle\langle V^2\rangle.
\ee
Again, these are the same terms that appeared at this order under the assumption that $U(t)$ was a random unitary. The third term is obviously of order $L^{-2}$, and the first two are of this order by our assumption above. Putting the dependence on $\delta_W$ and $\delta_V$ back in, one finds
\be\label{eqn:RMOTOC}
\langle W^2\rangle \left\{\Big[\langle V\rangle^2 {-} \frac{\langle V^2\rangle}{L^2}\Big]|\langle e^{\i \delta_W h}\rangle|^2 + \frac{\langle V^2\rangle}{L^2}\right\} +\langle V^2\rangle\left\{\Big[\langle W\rangle^2 {-} \frac{\langle W^2\rangle}{L^2}\Big]|\langle e^{\i \delta_Vh}\rangle|^2 + \frac{\langle W^2\rangle}{L^2}\right\}   - \frac{1}{L^2}\langle W^2\rangle\langle V^2\rangle.
\ee
This is the random matrix analog of the bulk handle-disk result (\ref{finalhandledisk}).
Recall that in pure JT, we have $\langle \mathcal{O}\rangle^2={\langle \mathcal{O}^2\rangle\over L^2}$, which eliminates the time dependence in the above. When one adds bulk matter fields to JT, the squared one point function can be separated into the pure JT piece $\frac{\langle \mathcal{O}^2\rangle}{L^2}$, and a piece $\langle \mathcal{O}\rangle^2 - \frac{\langle \mathcal{O}^2\rangle}{L^2}$ that represents the effects of bulk matter loops. The Gauss-law constraint implies that only the part involving matter loops has time dependence, and the $|\langle e^{\i \delta h}\rangle|^2$ pieces in the above formula are in the right places and depend on the right combination of times to match (\ref{finalhandledisk}).

\section{The kernel for fluctuations in Brownian SYK}\label{app:kernel}
In this appendix we will discuss the integral over fluctuations around a saddle point of the Brownian SYK theory. We will work out the details explicitly for the ``trivial'' saddle point $x = y = z = 1$ in the case with no operator insertions.

For this case, the forwards and backwards time evolution should cancel exactly, so the full answer for the path integral should be simply $2^{N/2}$, independent of the length of the timefold. This is obvious in the Hamiltonian formulation, but of course one can also establish it in the path integral formulation by thinking about doing the path integral slice by slice. For example, suppose we start from the final boundary condition. By the argument near (\ref{Argnear}), the Pfaffian is independent of one linear combination of $\sigma$ fluctuations that correspond to the arbitrary $f(t)$ in (\ref{Argnear}). This linear combination therefore enters the action only through the $g \sigma$ term, and it imposes a delta function constraint setting to zero the corresponding combination of the $g$ variables. In turn, this is enough to ensure that the $g^q$ interaction term vanishes, which makes the action independent of some further $g$ variables, which then function as Lagrange multipliers, setting the final components of $\sigma$ at that time slice to zero. This argument can then be repeated at the next time slice.

We would like to set this argument aside, however, and study the problem from a more general-purpose traditional viewpoint, where we expand the action around the saddle and evaluate the integral over small fluctuations. The action is 
\begin{align}
-\frac{I}{N} &= \log \text{Pf}\left(\partial_t - \sigma\right) - \frac{1}{2}\int_0^T \d t \left[\sigma_{jj'}(t)g_{jj'}(t) - \frac{J}{q}s_js_{j'}g_{jj'}(t)^q\right]
\end{align}
Using the exact reduction in section \ref{sec:reducing} to variables 
\be
g = \delta_{ij}\frac{\text{sgn}(0)}{2} + \frac{1}{2}\left(\begin{array}{cccc} 0 & -\i x & y & -\i z \\
\i x & 0 & -\i z & -y \\
-y & \i z  & 0 & -\i x\\
\i z  & y & \i x & 0  \end{array}\right), \hspace{20pt} \sigma = \left(\begin{array}{cccc} 0 & -\i \sigma_x & \sigma_y & -\i \sigma_z \\
\i \sigma_x & 0 & -\i \sigma_z & -\sigma_y \\
-\sigma_y & \i \sigma_z  & 0 & -\i \sigma_x\\
\i \sigma_z  & \sigma_y & \i \sigma_x & 0  \end{array}\right)
\ee
and writing the Pfaffian as in section \ref{Sec:classicalAc}, this action can be written as
\begin{align}
-\frac{I}{N}&= \log\left\{2\times \langle 0|\mathcal{T} \exp\left[\int_0^T \d t \Big(-\sigma_x(t)X +\i \sigma_y(t)Y - \sigma_z(t) Z\Big)\right]|+\rangle\right\}\\
&\hspace{20pt}+ \int_0^T\d t\left[\Big(\sigma_x(t) x(t) - \sigma_y(t) y(t) + \sigma_z(t) z(t)\Big) - \frac{J}{2^{q-1}q}\Big(1-x^q(t) + y^q(t) - z^q(t)\Big)\right]\notag.
\end{align}

At this point we specialize to expanding around the solution for the disk
\be
\{x,y,z\} = \{1,1,1\} + 2\underbrace{\{\hat{x}, \hat{y}, \hat{z}\}}_{\hat{g}}, \hspace{20pt} \{\sigma_x,\sigma_y,\sigma_z\} = -\frac{J}{2^{q-1}}\{1,1,1\} + \underbrace{\{\hat{\sigma}_x,\hat{\sigma}_y,\hat{\sigma}_z\}}_{\hat{\sigma}}.
\ee
The action, to quadratic order in the fluctuations, is
\be\label{eqn:K}
-\frac{I}{N} = \log(\sqrt{2}) + \left(\begin{array}{c}\hat{\sigma} \\ \hat{g}\end{array}\right) \cdot \left(\begin{array}{cc} \frac{2^{q-2}}{(q-1)J}K & S \\ S & \frac{(q-1)J}{2^{q-2}}S\end{array}\right)\left(\begin{array}{c}\hat{\sigma} \\ \hat{g}\end{array}\right).
\ee
Here the $K$ and $S$ operators act on a space indexed by a time and a choice of $x,y,z$ component, and are explicitly:
\be
K = \frac{(q{-}1)J}{2^{q-2}}e^{-\frac{J|t_{12}|}{2^{q-2}}}\left(\begin{array}{ccc} 0  & \theta(t_{12}) & -\theta(t_{12}) \\ \theta(t_{21}) & -1 & \theta(t_{12}) \\ -\theta(t_{21}) & \theta(t_{21}) & 0\end{array}\right), \hspace{20pt} S = \delta(t_{12})\left(\begin{array}{ccc} 1  & 0 & 0 \\ 0 & -1 & 0 \\ 0 & 0 & 1\end{array}\right)
\ee
In expanding around a more general solution, the $K$ matrix would change to a new form determined by the correlation functions of the new saddle point. Using $S^{-1} = S$ and the formula for a determinant of a block matrix, one finds that the one-loop determinant is
\be
\frac{2^{\frac{N}{2}}}{\sqrt{\text{det}(S)\text{det}(S-K)}} = \frac{2^{\frac{N}{2}}}{\sqrt{\text{det}(1-SK)}}.
\ee
In principle, one can compute the determinant using the formula
\be
\text{det}(1-k) = \exp\left(\text{Tr}\log(1-k)\right) = \exp\left(-\sum_{p = 1}^\infty \frac{1}{p}\text{Tr}(k^p)\right).
\ee
One can show that traces of $(SK)^p$ for powers $p \ge 2$ vanish, while as written, $\text{Tr}(SK) = \frac{(q{-}1)JT}{2^{q-2}}$. This suggests that the determinant decays exponentially. However, the diagonal entries in $K$ come from coincident point correlators, which are sensitive to regularization. It seems that the regularization that preserves manifest unitarity is to set the diagonal entries at concident times to zero, so $\text{Tr}(SK) = 0$ and $\text{det}(1-k) = 1$. If we choose a different regularization, then we need to include a counterterm that cancels the $\text{Tr}(SK)$ term.

Another approach to computing the determinant is to diagonalize the operator $K-S$. This is a real-symmetric operator and therefore has complete set of eigenvectors. The equations for an eigenvalue of of $K - S$ are 
\begin{align}
(\lambda+1)\hat{\sigma}_x(t_1) &= \frac{(q{-}1)J}{2^{q-2}}\int_0^{t_1}\d t_2\, e^{-\frac{J}{2^{q-2}}t_{12}}\left(\hat{\sigma}_y(t_2) - \hat{\sigma}_z(t_2)\right)\\
(\lambda-1)\hat{\sigma}_y(t_1) &= -(\lambda+1)\left(\hat{\sigma}_x(t_1)+\hat{\sigma}_z(t_1)\right)\\
(\lambda+1)\hat{\sigma}_z(t_1) &= \frac{(q{-}1)J}{2^{q-2}}\int_{t_1}^{T}\d t_2\, e^{-\frac{J}{2^{q-2}}t_{21}}\left(\hat{\sigma}_y(t_2) - \hat{\sigma}_x(t_2)\right).
\end{align}
These equations imply the differential equations
\be\label{diffVersion}
\frac{\d}{\d t}\left(\begin{array}{c}\hat{\sigma}_x \\ \hat{\sigma}_z\end{array}\right) = \frac{J}{2^{q-2}}\left(\begin{array}{cc}-\frac{\lambda+q-2}{\lambda-1} & -\frac{2(q{-}1)\lambda}{\lambda^2-1} \\ \frac{2(q{-}1)\lambda}{\lambda^2-1}  & \frac{\lambda+q-2}{\lambda-1} \end{array}\right)\left(\begin{array}{c}\hat{\sigma}_x \\ \hat{\sigma}_z\end{array}\right)
\ee
together with the boundary conditions that $\hat{\sigma}_x(0) = 0$ and $\hat{\sigma}_z(T) = 0$.

The differential equation implies
\be
\hat{\sigma}_x'' = -\omega^2\hat{\sigma}_x, \hspace{20pt} \omega = \frac{J}{2^{q-2}}\sqrt{\left(\frac{2(q{-1})\lambda}{\lambda^2-1}\right)^2-\left(\frac{\lambda+q-2}{\lambda-1}\right)^2 }.
\ee
We can solve this equation and one of the boundary conditions by setting $\hat{\sigma}_x = \sin(\omega t)$. Using (\ref{diffVersion}), we will also satisfy the other boundary condition if 
\be
\hat{\sigma}_x'(T) + \frac{J}{2^{q-2}}\frac{\lambda+q-2}{\lambda-1}\hat{\sigma}_x(T) = 0.
\ee
This gives the quantization condition for $\lambda$
\be
\omega \cos(\omega T)+ \frac{J}{2^{q-2}}\frac{\lambda+q-2}{\lambda-1}\sin(\omega T) = 0.
\ee
To find the determinant, one multiplies together all of the values of $\lambda$ that solve this equation. Numerically, this reproduces the formula with the regularization where $\text{det}(1-KS) = e^{-(q-1)JT/2^{q-2}}$. However, this approach makes it possible to understand an important feature of the spectrum of fluctuations, which are the nearly-zero-modes that were discussed in \ref{sec:qualitativediscussion}.

For large $T$, the eigenvalue equation has two special solutions that are exponentially small
\be\label{knowneigs}
\lambda^{(S)} = \frac{q-2}{q-1} e^{-\frac{(q-2)}{2^{q-2}}JT}, \hspace{20pt} \lambda^{(A)} = - \frac{q-2}{q-1} e^{-\frac{(q-2)}{2^{q-2}}JT}
\ee
The corresponding eigenvectors of $K-S$ are (recognizing the exponent as $\lambda_L$ see (\ref{LyapunovExp}))
\be\label{knownvec}
\hat\sigma^{(S)}\propto \left(\begin{array}{c} e^{\lambda_L (t-T)} \\ e^{\lambda_L (t-T)}+ e^{-\lambda_L t} \\ e^{-\lambda_L t}\end{array}\right),\hspace{20pt} \hat\sigma^{(A)}\propto \left(\begin{array}{c} e^{\lambda_L (t-T)} \\ e^{\lambda_L (t-T)}- e^{-\lambda_L t} \\ -e^{-\lambda_L t}\end{array}\right).
\ee
Here we have labeled the eigenvectors $S$ and $A$ to indicate whether they are symmetric or antisymmetric under a time-reversal symmetry that also interchanges the $x$ and $z$ components. 

These two eigenvectors represent the nearly-zero modes. Within this subspace, the corresponding $\hat{g}$ component is given, to leading order, given by stationarizing the action (\ref{eqn:K}). This is equivalent to integrating out $\hat{g}$, and one finds that the resulting action for $\hat{\sigma}$ is
\be\label{knownaction}
-\frac{I}{N} = \frac{2^{q-2}}{(q-1)J}\hat{\sigma}\cdot (K-S)\hat{\sigma}.
\ee
This can be related to the effective action from section (\ref{sec:qualitativediscussion}) as follows. We can define the $x^+$ and $x^-$ variables by using them to parametrize the backreacted correlator in the presence of initial or final perturbations, see (\ref{soly}):
\begin{align}
x(t)  &= \left(\frac{1}{1 + e^{\lambda_L (t-T)}x^+}\right)^{\frac{1}{q-2}}, \hspace{20pt}\sigma_x(t)=-{J\over 2^{q-1}}\left(\frac{1}{1 + e^{\lambda_L (t-T)}x^+}\right)^{\frac{q-1}{q-2}}\\
z(t)  &= \left(\frac{1}{1 + e^{-\lambda_L t}x^-}\right)^{\frac{1}{q-2}}, \hspace{34pt}\sigma_z(t)=-{J\over 2^{q-1}}\left(\frac{1}{1 +  e^{-\lambda_L t}x^-}\right)^{\frac{q-1}{q-2}}.
\end{align}
To first order in $x^+$ and $x^-$, these correspond precisely to perturbations in the subspace of our linearized nearly-zero modes (\ref{knownvec}). One can then therefore use the known eigenvalues of those modes (\ref{knowneigs}), and the formula for the action (\ref{knownaction}) to get the action in terms of $x^\pm$. One finds
\be\label{aBSYK}
-I = a x^+x^-, \hspace{20pt} a = \frac{1}{2(q-2)^2}N e^{-\lambda_L T}.
\ee
This is indeed an example of the action (\ref{actionConjecture}), although our derivation here is only valid to quadratic order in the fluctuation around the disk solution. Added in v2: the derivation in \cite{Gu:2021xaj} is valid in a larger region.

We can use this action to compute the OTOC in Brownian SYK including the resummation of effects of order $e^{\lambda_L T}/N$:
\be
\langle \psi_1(T)\psi_2(0)\psi_1(T)\psi_2(0)\rangle=-\frac{1}{4}\int {dx^+ dx^-\over 2\pi \i /a } e^{a x^+ x^-}  \left(\frac{1}{1 +  x^+}\right)^{\frac{1}{q-2}}  \left(\frac{1}{1 +  x^-}\right)^{\frac{1}{q-2}}.
\ee
with $a$ given in (\ref{aBSYK}). Surprisingly, this is structurally identical to the bulk expression (\ref{secondLine}) in AdS$_2$ due to what appears to be a coincidental agreement between the couplings of the two-point functions to the $x^\pm$ modes in the two cases. The factor of $\i$ is different compared with the action in (\ref{secondLine}). This is because the OTOC considered in Brownian SYK is more closely related to OTOC with even Euclidean time separation in JT gravity. Indeed, if one considers JT gravity with even Euclidean time separation one will find the same action without the factor of $\i$. Related to this, notice that the integral is not convergent on the real axis. One valid contour is to integrate say $x^+$ along the imaginary axis,
\begin{align}
\langle \psi_1(T)\psi_2(0)\psi_1(T)\psi_2(0)\rangle  &= -\frac{1}{4}\int_{-\infty}^\infty {dX^+ dx^-\over 2\pi /a } e^{\i a X^+ x^-}  \left(\frac{1}{1 +  \i X^+}\right)^{\frac{1}{q-2}}  \left(\frac{1}{1 +  x^-}\right)^{\frac{1}{q-2}}\\
&= -\frac{a^{\frac{1}{q-2}}}{4} U(\frac{1}{q-2},1,a)\\
&= -\frac{1}{4}\left(1 - 2\frac{e^{\lambda_L T} }{N} + 2(q-1)^2\frac{e^{2\lambda_L T}}{N^2} + \dots\right).
\end{align}
This reproduces the numerical result found in \cite{Sunderhauf:2019djv} to the expected precision.\footnote{We are grateful to the authors of \cite{Sunderhauf:2019djv} for sharing their data.} Note that integral is convergent if we do the $X^+$ integral first, because it vanishes for $x^- < 0$. The $U$ function is the Tricomi confluent hypergeometric function.

One final comment: note that the contribution to the one-loop determinant from the nonzero modes must cancel the contribution from the nearly-zero modes, so that the whole answer can be one. The contribution of the nearly-zero modes is proportional to $1/\sqrt{\lambda_+\lambda_-}$, so the contribution of the nonzero modes must be proportional to $\sqrt{\lambda_+\lambda_-} \propto e^{-\lambda_L T}$. We expect the contribution of the nonzero modes to be approximately local, and we can use this to figure out their contribution on the ``one-hump'' solution shown in (\ref{contributionlinethree}). This solution has a central region of width $t_r-t_l$ that resembles the unperturbed disk solution that we studied in this appendix, so the determinant from the nonzero modes in this region will be proportional to $e^{-\lambda_L(t_r-t_l)}$. In addition, we have regions of the solution where $\{x,y,z\} \approx \{1,0,0\}$ or $\{0,0,1\}$. In these regions, one can check that the kernel matrix $K$ vanishes, so there is no contribution to the determinant. So the contribution to the determinant from everything other than the nearly-zero modes $t_l,t_r$ will be proportional to $e^{-\lambda_L(t_r-t_l)}$, as claimed in section \ref{Sec:BSYKOTOC}.

\section{Soft modes on the disk}\label{app:softModesDisk}

In this appendix, we will derive the Dray-t'Hooft shockwave action from the exact quantization of  JT gravity \cite{Kitaev:2018wpr,Yang:2018gdb}.  Although the Dray-t'Hooft can be derived by directly looking at the dependence of the gravitational action on the shockwave modes \cite{Maldacena:2016upp}, the approach in this appendix examines the effect of other modes and is also useful when we study higher genus corrections in appendix \ref{App:handle-disksoftmode}.

Let's start by considering the JT gravitational path integral on an OTOC contour with equal spacing around the thermal circle, that is an evolution form $t_1=0$ to $t_2={\beta\over 4}+\i T$, back to $t_3={\beta\over 2}$, forward to $t_4={3\beta\over 4}+\i T$ and then back to $t_5=\beta$ which is identified with the initial location at $t_1=0$.
From the boundary point of view, there is nothing mysterious about the path integral on such a contour since the Lorentzian forward and backward time evolutions cancel between each other due to the unitarity property and the final result of the path integral should be simply the euclidean disk path integral $Z(\beta)$.
However, to see this property in the bulk directly using the gravitational variables is a nontrivial task.

On the disk topology, JT gravity contains only boundary dynamics; the bulk is rigid. Therefore the natural gravitational variables are the locations of the boundary as a function of boundary time: $\vec{x}(t)$.
The gravitational path integral can be written as a path integral of the boundary locations on the OTOC contour:
\be
Z=\int_{\text OTOC}\mathcal{D}\vec{x}(t) e^{-I}.
\ee
This has a form of an ordinary quantum mechanical particle path integral, and the gravitational action induces the Schwarzian action for the boundary particle.
$\vec{x}(t)$ on an OTOC contour is an infinite dimensional variable, but we can reduce the path integral into a finite dimensional integral if we know the propagator of the particle.
This is done in \cite{Kitaev:2018wpr,Yang:2018gdb} using the particle in magnetic formalism of JT gravity, and one gets:
\be\label{eqn:disk integral}
Z=\int \frac{\prod_{i=1,2,3,4}\d\vec{x}_i}{\text{vol(PSL(2,R))}} ~K({\beta\over 4}+\i T; \vec{x}_1,\vec{x}_2) K({\beta\over 4}-\i T; \vec{x}_2,\vec{x}_3) K({\beta\over 4}+\i T; \vec{x}_3,\vec{x}_4) K({\beta\over 4}-\i T; \vec{x}_4,\vec{x}_1),
\ee
where $\vec{x}_{1,2,3,4}$ are the locations of the boundary at time $t_{1,2,3,4}$.
Here we use the notation in \cite{Yang:2018gdb} to denote the quantum propagator of the boundary particle from location $\vec{x}_i$ to $\vec{x}_j$ with total time evolution $\delta t$ as $K(\delta t; \vec{x}_i,\vec{x}_j)$.
The integral measure is determined by the hyperbolic metric and the vol(PSL(2,R)) factor means that one needs to quotient by the gauge symmetry corresponding to different embeddings of the physical manifold into the hyperbolic plane.

The propagator $K$ satisfies a composition rule just like a propagator of any ordinary particle:
\be
\int \d\vec{x}_j K(\delta t_1; \vec{x}_i,\vec{x}_j) K(\delta t_2; \vec{x}_j,\vec{x}_l)=K(\delta t_1+\delta t_2;\vec{x}_i,\vec{x}_l).
\ee
This automatically guarantees the cancellation of the Lorentzian evolution as required by the unitarity property. 
However we would like to see how this works out at the level of perturbation theory explicitly,
since when we later add operator insertions, the composition rule can be no longer used, while the perturbation theory may still be a useful tool.
At this stage if there is no Euclidean evolution, we can already give an argument for the existence of the nearly zero modes using the semiclassical approximation of the propagator:
\be\label{Kformula}
K(\delta t; \vec{x}_i,\vec{x}_j)\approx \sqrt{\det{\partial^2 S_{ij}\over \partial \vec{x}_i\partial \vec{x}_j}}e^{\i S_{ij}}.
\ee
Here $S_{ij}$ is the action of the classical trajectory that propagates from $\vec{x}_i$ to $\vec{x}_j$ with time $\delta t$. 
Consider any classical saddle on the OTOC contour (\ref{eqn:disk integral}) with $\beta=0$, the forward and backward propagator have the same trajectory and therefore are complex conjugate to each other.
This means that the quadratic fluctuation around the saddle has generically this form:\footnote{This action has two kinds of zero modes: one is a global translation which is the gauge symmetry; the other is a translation of all the initial or final locations, this is due to the zero euclidean time evolution. In gravity this corresponds to the large energy fluctuations not suppressed by the Boltzmann distribution, and in the ordinary particle case this is due to the existence of an infinite number of classical trajectories.  }
\be
S_{12}-S_{23}+S_{34}-S_{41}\supset {\partial^2 S_{12}\over \partial x^a \partial x'^b}(\delta x_1^a-\delta x_3^a)(\delta x_2^b-\delta x_4^b)
\ee
where we set the initial and final locations to be $\vec{x}_1=\vec{x}_3=\vec{x}, \vec{x}_2=\vec{x}_4=\vec{x}'$ and $a,b$ is the spatial index.
Using the Hamilton-Jacobi equation, the coefficient in front of the action just measures the Lyapunov exponent:
\be
{\partial^2 S_{12}\over \partial x^a \partial x'^b}={\partial p_b'\over \partial x^a}
\ee
with the final momentum $p'_a$. Since the particle moving on the hyperbolic plane is chaotic, this leads to a pair of modes with exponentially small action, as in section \ref{sec:qualitativediscussion}.
The integral over these soft modes will give a large answer, but the small determinant from all of the other modes (the prefactor in (\ref{Kformula})) will cancel this at one loop order, ensuring a time independent answer.

To work this phenomenon out explicitly in JT gravity and also incorporate the correction coming from finite Euclidean time evolution, we would like a semiclassical formula for the propagator. To get this, it is actually convenient to start with the exact propagator \cite{Yang:2018gdb}:
\be\label{eqn:propagator}
\begin{split}
K(\delta t; \vec{x}_i,\vec{x}_j)&=\Theta(\theta_{ij}) e^{-N\phi_{ij}}{2e^{-\rho_i/2-\rho_j/2}\over \pi^2 \sin{\theta_{ij}/2}}\int_0^{\infty} \d s s\sinh(2\pi s) e^{-{s^2\over 2N}\delta t} K_{2\i s}({4Ne^{-\rho_i/2-\rho_j/2}\over \sin{\theta_{ij}/2}}),\\
\phi_{ij}&=2 {e^{-\rho_i} \cos^2{\theta_j\over 2}+e^{-\rho_j}\cos^2{\theta_i\over 2}\over \cos{\theta_i\over 2}\cos{\theta_j\over 2}\sin{\theta_{ij}\over 2}}.
\end{split}
\ee
Here we are using polar coordinates for the hyperbolic space, and we are using a ``regularized'' coordinate $\rho$ that measures the radial distance from a large reference value $\rho_c$, where the holographic renormalization parameter is $\epsilon=e^{-\rho_c}$. Explicitly, the metric is
\be
\d^2s= \d \rho^2 + \sinh^2(\rho + \rho_c) \d\theta^2 \approx \d^2\rho+{e^{2\rho+2\rho_c}\over 4}\d^2\theta.
\ee
In order to go to the semiclassical limit, it will be convenient for us to introduce the integral representation of the Bessel K function:
\be
K_{2\i s}(x)=\int_{-\infty}^{\infty} d\xi e^{-x\cosh \xi+2\i s \xi},
\ee
and integrate over the $s$ variable.
This leads to a leading classical approximation of the propagator:
\be\label{eqn:propatator2}
K(\delta t;\vec{x}_i,\vec{x}_j)\approx \int d\xi \exp\left(-2N{(\xi-\i \pi)^2\over \delta t}-4N{e^{-\rho_i/2-\rho_j/2}\over \sin{\theta_{ij}/2}}\cosh\xi-N\phi_{ij} \right).
\ee
All the boundary fluctuations that leave the end points invariant have been integrated out except this single $\xi$ mode which can be thought of as representing the energy fluctuation, and whole effect of the rest is the finite one-loop piece which we didn't write down explicitly.
All terms in the exponent are proportional to $N$, so in the large $N$ limit one can simply use the saddle point approximation for $\xi$ by solving its equation of motion. This fact will be used later.

Now, let's return to the disk path integral on an OTOC contour (\ref{eqn:disk integral}).
First we can use the $SL(2,R)$ gauge symmetry to reduce the four angular variables on the OTOC contour down to one variable $\theta$:
\be
\theta_1=0,~~~\theta_2=\theta,~~~\theta_3=\pi,~~~\theta_4=\pi+\theta.
\ee
This gives a simple result for the sum of the $\phi_{ij}$ variables:
\be
\phi_{21}+\phi_{32}+\phi_{43}+\phi_{14}=4{e^{-\rho_1}+e^{-\rho_2}+e^{-\rho_3}+e^{-\rho_4}\over  \sin\theta}.
\ee
Second, we can work out the saddle point approximation of the integral (\ref{eqn:disk integral}).
It is a $9$ variable integral over $\rho_{1,2,3,4},\xi_{21,32,43,14}$ and $\theta$.
The reduced action in these variables is\footnote{Here we neglect the topological term proportional to $S_0$.}:
\be
-I=-N\sum_{i=1}^4\left( 2{(\xi_{i+1,i}-\i \pi)^2\over t_{i+1}-t_i}+4{e^{-\rho_i/2-\rho_{i+1}/2}\over \sin{\theta_{i+1}-\theta_i\over 2}}\cosh \xi_{i+1,i}+4 {e^{-\rho_i}\over \sin\theta}\right).
\ee
Our goal now is to show that this action contains two soft scrambling modes at late time.
In particular, this means that the classical equations of motion allow two families of approximate solutions at late time.

\begin{figure}[h]
\begin{center}
\includegraphics[width=0.4\textwidth]{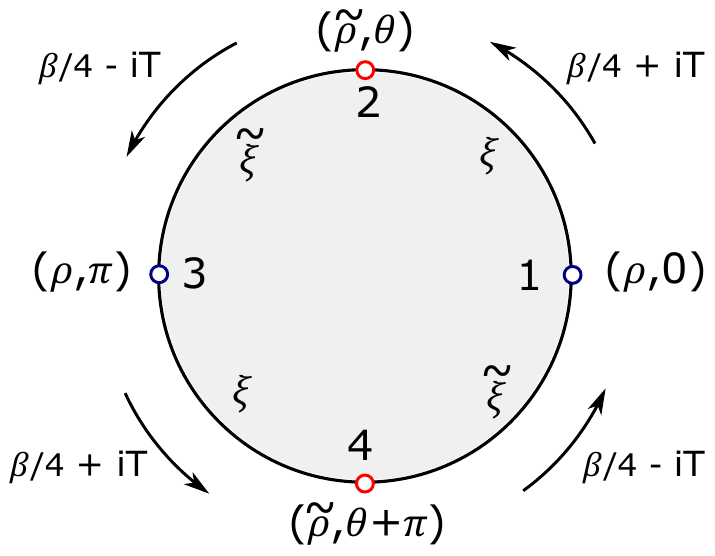}
\caption{Integration parameters on the disk.}
\label{fig:modulidisk}
\end{center}
\end{figure}

The OTOC contour has a permutation symmetry which exchanges $1234$ to $3412$, therefore we can use the following ansatz for the classical saddles:
\be
\rho_1=\rho_3=\rho,~~~\rho_2=\rho_4=\tilde\rho,~~~\xi_{21}=\xi_{43}=\xi,~~~\xi_{32}=\xi_{14}=\tilde\xi.
\ee
These parameters are shown in figure \ref{fig:modulidisk}.
With this ansatz, the action simplifies:
\be
-I=-8N\left({1\over 2}{(\xi-\i \pi)^2\over {\beta\over 4}+\i T}+{1\over 2}{(\tilde\xi-\i \pi)^2\over {\beta\over 4}-\i T}+{e^{-\rho/2-\tilde\rho/2}\over \sin{\theta\over 2}}\cosh\xi+{e^{-\rho/2-\tilde\rho/2}\over \cos{\theta\over 2}}\cosh\tilde\xi+{e^{-\rho}+e^{-\tilde\rho}\over\sin\theta}\right)
\ee
The corresponding equations of motion are now easy to get by taking derivatives of the action:
\be
\begin{split}
&\text{EoM for $\xi$, $\tilde\xi$:}~~~{\xi-\i \pi\over {\beta\over 4}+\i T}+{e^{-\rho/2-\tilde\rho/2}\over \sin{\theta\over 2}}\sinh\xi=0,{\tilde\xi-\i \pi\over {\beta\over 4}-\i T}+{e^{-\rho/2-\tilde\rho/2}\over \cos{\theta\over 2}}\sinh\tilde\xi=0\\
&\text{EoM for $\rho$, $\tilde \rho$:}~~~{1\over 2}e^{-\rho/2-\tilde\rho/2}({\cosh\xi\over \sin{\theta\over 2}}+{\cosh\tilde\xi\over \cos{\theta\over 2}})=-{e^{-\rho}\over \sin\theta}=-{e^{-\tilde\rho}\over \sin\theta}\\
&\text{EoM for $\theta$:}~~~~~~{1\over 2}e^{-\rho/2-\tilde\rho/2}({\cosh\xi\cos{\theta\over 2}\over \sin^2{\theta\over 2}}-{\cosh\tilde\xi\sin{\theta\over 2}\over \cos^2{\theta\over 2}})+(e^{-\rho}+e^{-\tilde\rho}){\cos\theta\over \sin^2\theta}=0
\end{split}
\ee
The exact solution of this set of equations is:
\be
\xi-\i\pi=\i{\theta\over 2};~~~\tilde\xi-\i\pi=\i({\pi\over 2}-{\theta\over 2});~~~\theta={2\pi\over\beta}({\beta\over 4}+\i T);~~~\rho=\tilde\rho=\log{\beta\over \pi},
\ee
which is simply the analytic continuation of the Euclidean disk:
$\theta$ is equal to the amount of time evolution of the $12$ segment  rescaled by a factor of ${2\pi\over \beta}$ in order to keep the period to be $2\pi$.
This also determines the boundary location by the boundary metric condition $e^{\rho}{d\theta\over 2}=dt$, therefore $\rho=\log{\beta\over \pi}$.

Now if we do a perturbative expansion around the saddle, we will find two soft modes with eigenvalues that decays exponentially in time.
Instead of doing that, we will use another trick by directly studying the approximate solution of the equations of motion at late time.
Notice that for large $\xi, \tilde\xi$ and $\theta $ (which is property of the classical saddle at late time) we can simplify the hyperbolic and trigonometric functions by dropping pieces of order $e^{-|\xi|},e^{-|\tilde\xi|}$ and $e^{- |\theta|}$.
This simplifies the equations of motion dramatically:
\be\label{eqn:lateTeoms}
{\xi-\i\pi\over {\beta\over 4}+\i T}+\i e^{-\xi-{\rho\over 2}-{\tilde\rho\over 2}+\i{\theta\over 2}}\approx 0,~~
{\tilde\xi-\i\pi\over {\beta\over 4}-\i T}+ e^{\tilde\xi-{\rho\over 2}-{\tilde\rho\over 2}+\i{\theta\over 2}}\approx 0,~~~
(e^{-\xi}+\i e^{\tilde\xi})e^{-{\rho\over 2}-{\tilde\rho\over 2}+\i{\theta\over 2}}\approx 0.
\ee
Clearly, these equations allow two families of solutions that is related by shifts $\rho\rightarrow \rho+a,\tilde\rho\rightarrow \tilde\rho+b,\theta\rightarrow \theta-\i a-\i b$, and these are the two soft modes.

These two modes do not change the geodesic length between the end points, they are describing the relative angle between the various propagators. 
Or equivalently, they describe the  geodesic lengths of $\vec{x}_{13}$ and $\vec{x}_{24}$.
This gives us two families of  approximate solutions parametrized by $\rho,\tilde\rho$:
\be
\xi=\i{5\pi\over 4}-{T\pi\over \beta};~~~\tilde\xi=\i{5\pi\over 4}+{T\pi\over \beta};~~~\theta={\pi\over 2}+\i {2\pi T\over \beta}+\i (2\log{\beta\over \pi}-\rho-\tilde\rho).
\ee
Evaluating the action along these soft directions, we get:
\be
\begin{split}
-I&={2\pi^2N\over\beta}+{16\pi^3N\over \beta^3} {e^{2\pi T\over \beta}(e^{\rho}-{\beta\over \pi})(e^{\tilde\rho}-{\beta\over \pi})\over e^{4\pi T\over \beta}+{\pi^4\over \beta^4}e^{2(\rho+\tilde\rho)}}\\
&\approx {2\pi^2N\over\beta}+{16\pi^3N\over \beta^3}e^{-2\pi T\over\beta}(e^{\rho}-{\beta\over \pi})(e^{\tilde\rho}-{\beta\over \pi}).
\end{split}
\ee
The first piece ${2\pi^2 N\over \beta}$ is just the disk saddle.
The second piece is the action for the two soft modes. 
 Locally around the classical saddle, the action has two eigenmodes with opposite eigenvalues that decays exponentially in time. We have to rotate the contour for one of these modes, giving a factor of $\i$ to the path integral, which will be cancelled below.
 
 After the scrambling time, the eigenvalues are very small, and the main contribution to the integral will come from a region where $e^{\rho+\tilde\rho}\sim {1\over N}e^{2\pi T\over \beta}$.
In this region, our action is still valid:
corrections coming from the denominator are ${1\over N}$ suppressed in this region, the same order as the violation of the equation of motion.

To recover the disk partition function, we use the fact that the integral of $\rho$ imposes a delta function constraint on $\tilde\rho$ and vice versa.
Therefore we recover the naive saddle point approximation even with the presence of large fluctuations.

This action is promised shockwave action (\ref{actionConjecture})  but expressed in terms of different variables.
 The relation between $\rho,\tilde\rho$ and the shockwave modes $x^{\pm}$ can be identified by comparing their effect on the correlators at late time.
The two side correlator between $t_2$ and $t_4$ is given by the geodesic approximation which is $e^{-2\Delta \rho}$ in our gauge.
Comparing with the backreaction coming from a shockwave excitation in AdS$_2$ which is $(\frac{\beta}{\pi}+(\frac{\beta}{2\pi})\frac{x^+}{\sqrt{2}})^{-2\Delta}$ (\ref{secondLine}), we can identify the relations
\be
\sqrt{2}(\frac{2\pi}{\beta}e^{\rho}-2)\sim x^-,~~~\sqrt{2}(\frac{2\pi}{\beta}e^{\tilde{\rho}}-2)\sim x^+.
\ee
and the soft mode action becomes the Dray-t'Hooft shockwave action:
\be\label{eqn:JTDHaction}
-I\sim \frac{2\pi}{\beta}N e^{-{2\pi T\over \beta}} x^-  x^+.
\ee

Finally let's comment on the contribution coming from other modes. Expanding the action in fluctuations $\delta \xi,\delta\tilde\xi,\delta\theta$ around the family of soft configurations, we have the leading correction\footnote{One can also do a similar analysis for the modes that violate the permutation symmetry, the result is that they are stable and gapped by $N$.}:
\be
\begin{split}
-\delta I&\approx-4N({1\over {\beta\over 4}+\i T}+{\i \pi\over \beta})\delta\xi^2-4N({1\over {\beta\over 4}-\i T}-{\i \pi\over \beta})\delta\tilde\xi^2-{4\pi N\over \beta}(\delta \xi+\delta\tilde\xi)\delta \theta\\
&\approx-{\i 4\pi N\over \beta}(\delta\xi-{\i\over 2}\delta\theta)^2+{\i 4 \pi N \over \beta}(\delta\tilde\xi+{\i\over 2}\delta\theta)^2+N{\beta\over 2T^2}\delta\theta^2.
\end{split}
\ee
None of these modes become exponentially soft at late times. Note that we have to rotate the contour for the $\theta$ direction, which contributes another factor of $\pm\i$ that combines with the unstable direction in the shockwave modes so that the final result is real.

\section{Soft modes on the handle-disk}\label{App:handle-disksoftmode}
\begin{figure}[h]
\begin{center}
\includegraphics[width=0.4\textwidth]{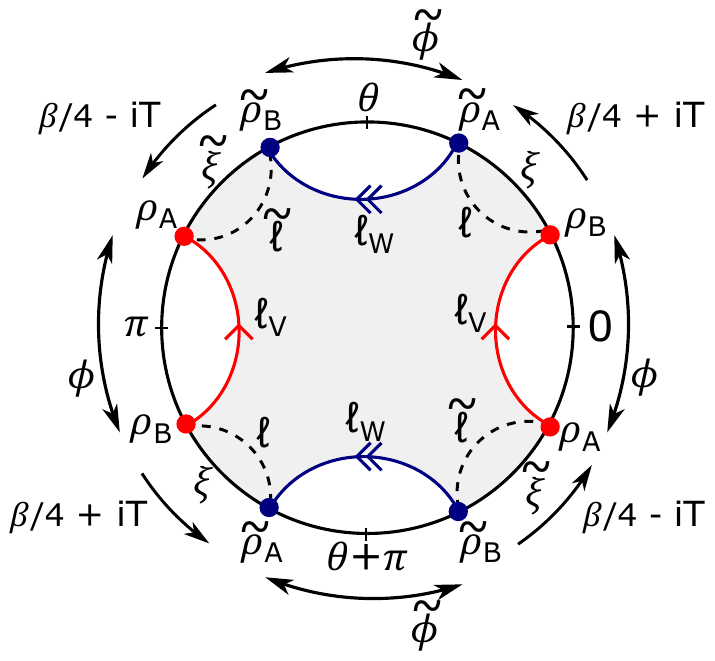}
\caption{Useful variables for discussing the handle disk}
\label{fig:modulihandle-disk}
\end{center}
\end{figure}

In this appendix, we will show that the late time handle-disk path integral is dominated by two soft scramblon modes as in the disk case.
First let's introduce the parameters associated to the moduli space of the handle-disk.  By cutting the handle-disk along the geodesics where the operators follow (figure \ref{fig:handle-disk}), we can embed the geometry into a hyperbolic plane (see figure \ref{fig:modulihandle-disk}).
The moduli space of the handle-disk is given by the locations of the 8 boundary end points mod a SL(2,R) gauge transformation, with the constraint that the geodesic distances between the cut geodesics are the same.
This leads to $8\times2-3-2=11$ parameters in total.
We can do a slight simplification of the problem using the permutation symmetry of the OTOC contour (rotation of $\pi$ in figure \ref{fig:modulihandle-disk}) as in the disk discussion to reduce the moduli space down to $7$ dimensions.
They are shown in figure \ref{fig:modulihandle-disk}.
Each end point is labeled by their location in the coordinates $\lbrace\rho_i,\theta_i\rbrace$. 
Using SL(2,R), one can first fix the gauge so that
\be
\theta_1+\theta_2=0,~~~\theta_3+\theta_4=2\theta,~~~\theta_5+\theta_6=2\pi,~~~\theta_7+\theta_8=2\pi+2\theta.
\ee
We denote the difference of the $\theta$s as $\phi$. Using the assumption of permutation symmetry, we have:
\be
\theta_2-\theta_1=\phi,~~~\theta_4-\theta_1=\tilde\phi,~~~\theta_6-\theta_5=\phi,~~~\theta_8-\theta_7=\tilde\phi.
\ee
Finally, the permutation symmetric configuration for the radial directions are:
\be
\rho_1=\rho_5=\rho_A,~~\rho_2=\rho_6=\rho_B,~~~\rho_3=\rho_7=\tilde\rho_A,~~~\rho_4=\rho_8=\tilde\rho_B.
\ee
To find the soft modes, we use the same procedure as in the disk case: first we write down the action depending on the 7 moduli, work out their equations of motion, find the approximate solutions at late time, and then evaluate the action of these approximate solutions.

So first, let's evaluate the action.  After further cutting along the dashed lines in figure \ref{fig:modulihandle-disk}, we can write the integrand in the moduli-space integral as a product of three types of factors: the path integral between the asymptotic boundaries and the dashed lines (these are sometimes called ``Wheeler de-Witt'' wavefunctions), the path integral in the central hyperbolic polygon enclosed by various geodesics, and finally the weighting from the operator insertions. The final factor is simple, we just need to insert the $e^{-\Delta \ell_{V,W}}\equiv e^{-N \tilde \Delta \ell_{V,W}}$ factors, where $\ell_{V,W}$ is the geodesic length labeled in the figure. The action of the hyperbolic polygon is given by the regularized area of the polygon, and it is equal to the sum of the phase factors $-N\phi_{ij}$ in the propagator (\ref{eqn:propagator}).
Finally, the WdW wavefunction is the boundary particle propagator with the phase factor stripped off \cite{Yang:2018gdb}. The total weighting factor written down this way is somewhat messy, mainly due to the area of the hyperbolic polygon.  Since we are interested in the late time behavior, we can use the assumption that $-\i\theta $ is large (based on our previous disk analysis) to simplify the action up to order $e^{\i \theta}$. One finds:
\be
\begin{split}
{I\over 8N}&\approx{1\over 2}{(\xi-\i \pi)^2\over {\beta\over 4}+\i T}+{1\over 2}{(\tilde\xi-\i \pi)^2\over {\beta\over 4}-\i T}-2\i e^{-{\tilde\rho_A\over 2}-{\rho_B\over 2}+\i {\theta\over 2}-\i{\phi+\tilde\phi\over 4}}\cosh\xi +2e^{-{\rho_A\over 2}-{\tilde\rho_B\over 2}+\i{\theta\over 2}+\i {\phi+\tilde\phi\over 4}}\cosh\tilde\xi\\
&+\i ({e^{-\tilde\rho_A}+e^{-\tilde\rho_B}\over e^{\i\tilde\phi}-1}+{e^{-\rho_A}+e^{-\rho_B}\over e^{\i \phi}-1}+e^{-\rho_A}+e^{-\tilde\rho_B})+{\tilde\Delta\over 4 }({\rho_A+\rho_B+\tilde\rho_A+\tilde\rho_B\over 2}+\log \sin{\phi\over 2}\sin{\tilde\phi\over 2})\\
&-2\i e^{-{\tilde\rho_A\over 2}-{\rho_B\over 2}+3\i {\theta\over 2}-3\i{\phi+\tilde\phi\over 4}}\cosh\xi-2e^{-{\rho_A\over 2}-{\tilde\rho_B\over 2}+3\i{\theta\over 2}+3\i {\phi+\tilde\phi\over 4}}\cosh\tilde\xi\\
&-\i (e^{-\tilde\rho_B}+e^{-\rho_A})e^{\i \theta+\i{\phi+\tilde\phi\over 2}}-\i (e^{-\rho_B}+e^{-\tilde\rho_A})e^{\i \theta-\i{\phi+\tilde\phi\over 2}}.
\end{split}
\ee
Here $\xi$ and $\tilde\xi$ are the same parameters that enter in Bessel K function of the WdW wavefunction (see equation (\ref{eqn:propatator2})), and again we used the permutation symmetry to reduce down to two variables. Here the first two lines of the action do not decay with $T$, and the last two lines are the leading corrections which decay exponentially in time.

Second, we can work out the equations of motion.  For that we just need to vary the first two lines in the action, and we get:
\be
\begin{split}
&{\xi-\i \pi\over {\beta\over 4}+\i T}+e^{-\ell/2}\sinh\xi=0;~~~~~~{\tilde\xi-\i \pi\over {\beta\over 4}-\i T}+e^{-\tilde\ell/2}\sinh\tilde\xi=0;~~~e^{-\ell/2}\cosh\xi+e^{-\tilde\ell/2}\cosh\tilde\xi=0;\\
&-{1\over 2} e^{-\tilde\ell/2}\cosh\tilde \xi+\i e^{-\rho_A} {e^{\i\phi}\over 1-e^{\i \phi}}+{\tilde\Delta\over 8}=0;~~~-{1\over 2} e^{-\tilde\ell/2}\cosh\tilde \xi+\i e^{-\tilde \rho_B} {e^{\i\tilde\phi}\over 1-e^{\i \tilde\phi}}+{\tilde\Delta\over 8}=0;\\
&-{1\over 2} e^{-\ell/2}\cosh \xi+\i e^{-\rho_B} {1\over 1-e^{\i \phi}}+{\tilde\Delta\over 8}=0;~~~-{1\over 2} e^{-\ell/2}\cosh \xi+\i e^{-\tilde \rho_A} {1\over 1-e^{\i \tilde\phi}}+{\tilde\Delta\over 8}=0;\\
&{\i\over 4} e^{-\tilde\ell/2}\cosh\tilde\xi-{\i\over 4}e^{-\ell/2}\cosh\xi+{\tilde\Delta\over 8}{1\over \tan{\phi\over 2}}-{e^{-\rho_A}+e^{-\rho_B}\over 4\sin^2{\phi\over 2}}=0;\\
&{\i\over 4} e^{-\tilde\ell/2}\cosh\tilde\xi-{\i\over 4}e^{-\ell/2}\cosh\xi+{\tilde\Delta\over 8}{1\over \tan{\tilde \phi\over 2}}-{e^{-\tilde \rho_A}+e^{-\tilde \rho_B}\over 4\sin^2{\tilde \phi\over 2}}=0,
\end{split}
\ee
where we introduce the notation for the geodesic lengths:
\be
e^{-\ell/2}=-2\i e^{-{\tilde\rho_A\over 2}-{\rho_B\over 2}+\i {\theta\over 2}-\i{\phi+\tilde\phi\over 4}};~~~e^{-\tilde\ell/2}=2e^{-{\rho_A\over 2}-{\tilde\rho_B\over 2}+\i{\theta\over 2}+\i {\phi+\tilde\phi\over 4}}.
\ee
To solve these equations, we can first introduce effective temperature parameters $s,\tilde s$ so that:
\be
\xi=\i \pi+{\i s\over 2}({\beta\over 4}+\i T);~~~\tilde\xi=\i \pi +{\i \tilde s\over 2}({\beta\over 4}-\i T).
\ee
Then the first line of equations requires $s=\tilde s$ and
\be\label{eqn:energyeqn}
e^{-\ell/2}\sinh\xi=-\i{ s\over 2},~~~e^{-\tilde\ell/2}\sinh\tilde\xi=-\i{ s\over 2}.
\ee
Plugging these into the rest of the equations we can easily get:
\be
\begin{split}
&e^{-\rho_A}={\sin{\phi\over 2}\over 4} e^{-\i \phi/2}(\tilde\Delta+2\i s);~~~e^{-\rho_B}={\sin{\phi\over 2}\over 4} e^{\i \phi/2}(\tilde\Delta-2\i s)\\
&e^{-\tilde\rho_A}={\sin{\tilde\phi\over 2}\over 4} e^{\i \tilde\phi/2}(\tilde\Delta-2\i s);~~~e^{-\tilde\rho_B}={\sin{\tilde\phi\over 2}\over 4} e^{-\i \tilde\phi/2}(\tilde\Delta+2\i s).
\end{split}
\ee
In particular, the last two equations which are the equation of motion for $\phi$ and $\tilde\phi$ are automatically satisfied. 
This means that the two nearly zero modes are $\phi$ and $\tilde\phi$.
One can also see that the geodesic distances $\ell_{V,W}$ are fixed:
\be
e^{-\ell_{V}/2}={e^{-\rho_A/2-\rho_B/2}\over \sin{\phi\over 2}}={\sqrt{4s^2+\tilde\Delta^2}\over 4};~~~e^{-\ell_{W}/2}={e^{-\tilde\rho_A/2-\tilde\rho_B/2}\over \sin{\tilde\phi\over 2}}={\sqrt{4s^2+\tilde\Delta^2}\over 4}.
\ee
After plugging these into (\ref{eqn:energyeqn}), we can determine $s$ and $\theta$:
\be\label{eqn:effectivetemp}
\theta=\i s T+\i \log \sin{\phi\over 2}\sin{\tilde\phi\over 2}-\i \log{4s^2\over \tilde\Delta^2+4 s^2}+{\pi\over 2};~~~2\arctan{2s\over \tilde\Delta}-{\pi\over 2}+{s\beta\over 4}=0.
\ee

Geometrically, these two nearly zero modes $\phi$ and $\tilde\phi$ represent the fluctuations of the geodesic length of the circles crossing $\ell_{V,W}$ which is roughly equal to $\rho_A+\rho_B=\ell_V-2\log \sin{\phi\over 2}$ and $\tilde\rho_A+\tilde\rho_B=\ell_W-2\log \sin{\tilde\phi\over 2}$.
Physically, we can understand their appearance as due to the large bulk disk like region with effective temperature $\beta_{\text{eff}}={2\pi\over s}$ in the late time handle-disk geometry, such region describes the middle transition region in the Brownian SYK saddles (see (\ref{exampleSol})).

Finally, we want to evaluate the dependence of the action on the soft modes. In order to simplify the formulas, we will consider the special case $\beta=0$, where we have the solution $s={\tilde \Delta\over 2}$. 
This gives the action:
\be\label{eqn:handlediskresult}
I={4\Delta}(1+{3\over 2}\log2-\log{\Delta\over N})-2\Delta e^{-{\Delta T\over 2N }}(1-\cot{\phi\over 2})(1-\cot{\tilde\phi\over 2}),
\ee
where we have used $\tilde{\Delta}=\frac{\Delta}{N}$. The second term is the Dray-t'Hooft shockwave action with effective temperature $\beta_{\text{eff}}={2\pi\over s}$, which is the main result of this appendix.

To make the connection between our saddle point analysis with the exact path integral result (\ref{eqn:WVWVHD}) in section \ref{sec:handle-disk}, we can use the large $s$ and $\Delta$ approximation of the $\Gamma$ functions:
\be
{\Gamma(\Delta)^2\over 2^{2\Delta+1}\Gamma(2\Delta)}|\Gamma(\Delta-2\i S)|^2\rightarrow  \exp\left(-4s\tan^{-1}{2S\over \Delta}+\Delta\log(\Delta^2+4 S^2)-2\Delta-2\log 2\Delta\right).
\ee
This leads to the saddle point equation for $S$:
\be
-8\tan^{-1}{2S\over \Delta}+2\pi-{\beta\over N}S=0
\ee
With the replacement of $\Delta=N\tilde\Delta$ and $S=N s$, we recover the equation for the effective temperature (\ref{eqn:effectivetemp}).
The leading piece in (\ref{eqn:handlediskresult}) just matches with the leading approximation of the exact result when $\beta=0$:
\be
\langle WVWV\rangle_{\text{handle-disk}}\approx \rho(S)\langle S|VV|S\rangle\langle S|WW|S\rangle|_{S={\Delta\over 2}}\approx e^{-4\Delta(1+{3\over 2}\log2-\log{\Delta\over N})}.
\ee

\bibliography{references}

\providecommand{\href}[2]{#2}\begingroup\raggedright\begin{thebibliography}{10}

\bibitem{Hayden:2007cs}
P.~Hayden and J.~Preskill, ``{Black holes as mirrors: Quantum information in
  random subsystems},''
  \href{http://dx.doi.org/10.1088/1126-6708/2007/09/120}{{\em JHEP} {\bfseries
  09} (2007) 120},
\href{http://arxiv.org/abs/0708.4025}{{\ttfamily arXiv:0708.4025 [hep-th]}}.

\bibitem{Sekino:2008he}
Y.~Sekino and L.~Susskind, ``{Fast Scramblers},''
  \href{http://dx.doi.org/10.1088/1126-6708/2008/10/065}{{\em JHEP} {\bfseries
  10} (2008) 065}, \href{http://arxiv.org/abs/0808.2096}{{\ttfamily
  arXiv:0808.2096 [hep-th]}}.

\bibitem{Hosur:2015ylk}
P.~Hosur, X.-L. Qi, D.~A. Roberts, and B.~Yoshida, ``{Chaos in quantum
  channels},'' \href{http://dx.doi.org/10.1007/JHEP02(2016)004}{{\em JHEP}
  {\bfseries 02} (2016) 004}, \href{http://arxiv.org/abs/1511.04021}{{\ttfamily
  arXiv:1511.04021 [hep-th]}}.

\bibitem{nahum2017quantum}
A.~Nahum, J.~Ruhman, S.~Vijay, and J.~Haah, ``Quantum entanglement growth under
  random unitary dynamics,'' {\em Physical Review X} {\bfseries 7} no.~3,
  (2017) 031016.

\bibitem{von2018operator}
C.~Von~Keyserlingk, T.~Rakovszky, F.~Pollmann, and S.~L. Sondhi, ``Operator
  hydrodynamics, otocs, and entanglement growth in systems without conservation
  laws,'' {\em Physical Review X} {\bfseries 8} no.~2, (2018) 021013.

\bibitem{nahum2018operator}
A.~Nahum, S.~Vijay, and J.~Haah, ``Operator spreading in random unitary
  circuits,'' {\em Physical Review X} {\bfseries 8} no.~2, (2018) 021014.

\bibitem{collins2003moments}
B.~Collins, ``Moments and cumulants of polynomial random variables on
  unitarygroups, the itzykson-zuber integral, and free probability,'' {\em
  International Mathematics Research Notices} {\bfseries 2003} no.~17, (2003)
  953--982.

\bibitem{collins2006integration}
B.~Collins and P.~{\'S}niady, ``Integration with respect to the haar measure on
  unitary, orthogonal and symplectic group,'' {\em Communications in
  Mathematical Physics} {\bfseries 264} no.~3, (2006) 773--795.

\bibitem{Almheiri:2019qdq}
A.~Almheiri, T.~Hartman, J.~Maldacena, E.~Shaghoulian, and A.~Tajdini,
  ``{Replica Wormholes and the Entropy of Hawking Radiation},''
  \href{http://dx.doi.org/10.1007/JHEP05(2020)013}{{\em JHEP} {\bfseries 05}
  (2020) 013}, \href{http://arxiv.org/abs/1911.12333}{{\ttfamily
  arXiv:1911.12333 [hep-th]}}.

\bibitem{Penington:2019kki}
G.~Penington, S.~H. Shenker, D.~Stanford, and Z.~Yang, ``{Replica wormholes and
  the black hole interior},'' \href{http://arxiv.org/abs/1911.11977}{{\ttfamily
  arXiv:1911.11977 [hep-th]}}.

\bibitem{Page:1993df}
D.~N. Page, ``{Average entropy of a subsystem},''
  \href{http://dx.doi.org/10.1103/PhysRevLett.71.1291}{{\em Phys. Rev. Lett.}
  {\bfseries 71} (1993) 1291--1294},
\href{http://arxiv.org/abs/gr-qc/9305007}{{\ttfamily arXiv:gr-qc/9305007
  [gr-qc]}}.

\bibitem{Penington:2019npb}
G.~Penington, ``{Entanglement Wedge Reconstruction and the Information
  Paradox},''
\href{http://arxiv.org/abs/1905.08255}{{\ttfamily arXiv:1905.08255 [hep-th]}}.

\bibitem{Almheiri:2019psf}
A.~Almheiri, N.~Engelhardt, D.~Marolf, and H.~Maxfield, ``{The entropy of bulk
  quantum fields and the entanglement wedge of an evaporating black hole},''
\href{http://arxiv.org/abs/1905.08762}{{\ttfamily arXiv:1905.08762 [hep-th]}}.

\bibitem{Roberts:2016hpo}
D.~A. Roberts and B.~Yoshida, ``{Chaos and complexity by design},''
  \href{http://dx.doi.org/10.1007/JHEP04(2017)121}{{\em JHEP} {\bfseries 04}
  (2017) 121}, \href{http://arxiv.org/abs/1610.04903}{{\ttfamily
  arXiv:1610.04903 [quant-ph]}}.

\bibitem{Yoshida:2017non}
B.~Yoshida and A.~Kitaev, ``{Efficient decoding for the Hayden-Preskill
  protocol},'' \href{http://arxiv.org/abs/1710.03363}{{\ttfamily
  arXiv:1710.03363 [hep-th]}}.

\bibitem{Teitelboim:1983ux}
C.~Teitelboim, ``{Gravitation and Hamiltonian Structure in Two Space-Time
  Dimensions},''
\href{http://dx.doi.org/10.1016/0370-2693(83)90012-6}{{\em Phys. Lett.}
  {\bfseries B126} (1983) 41--45}.

\bibitem{Jackiw:1984je}
R.~Jackiw, ``{Lower Dimensional Gravity},''
\href{http://dx.doi.org/10.1016/0550-3213(85)90448-1}{{\em Nucl. Phys.}
  {\bfseries B252} (1985) 343--356}.

\bibitem{almheiri2015models}
A.~Almheiri and J.~Polchinski, ``Models of ads 2 backreaction and holography,''
  {\em Journal of High Energy Physics} {\bfseries 2015} no.~11, (2015) 14.

\bibitem{Blommaert:2020seb}
A.~Blommaert, ``{Dissecting the ensemble in JT gravity},''
  \href{http://arxiv.org/abs/2006.13971}{{\ttfamily arXiv:2006.13971
  [hep-th]}}.

\bibitem{Saad:2019pqd}
P.~Saad, ``{Late Time Correlation Functions, Baby Universes, and ETH in JT
  Gravity},''
\href{http://arxiv.org/abs/1910.10311}{{\ttfamily arXiv:1910.10311 [hep-th]}}.

\bibitem{Saad:2018bqo}
P.~Saad, S.~H. Shenker, and D.~Stanford, ``{A semiclassical ramp in SYK and in
  gravity},''
\href{http://arxiv.org/abs/1806.06840}{{\ttfamily arXiv:1806.06840 [hep-th]}}.

\bibitem{Sachdev:1992fk}
S.~Sachdev and J.-w. Ye, ``{Gapless spin fluid ground state in a random,
  quantum Heisenberg magnet},''
  \href{http://dx.doi.org/10.1103/PhysRevLett.70.3339}{{\em Phys. Rev. Lett.}
  {\bfseries 70} (1993) 3339},
\href{http://arxiv.org/abs/cond-mat/9212030}{{\ttfamily arXiv:cond-mat/9212030
  [cond-mat]}}.

\bibitem{KitaevTalks}
A.~Kitaev, ``A simple model of quantum holography
  \href{http://online.kitp.ucsb.edu/online/entangled15/kitaev/}{talk1} and
  \href{http://online.kitp.ucsb.edu/online/entangled15/kitaev2/}{talk2}.''.
  Talks at KITP, April 7, 2015 and May 27, 2015.

\bibitem{Kitaev:2017awl}
A.~Kitaev and S.~J. Suh, ``{The soft mode in the Sachdev-Ye-Kitaev model and
  its gravity dual},'' \href{http://dx.doi.org/10.1007/JHEP05(2018)183}{{\em
  JHEP} {\bfseries 05} (2018) 183},
\href{http://arxiv.org/abs/1711.08467}{{\ttfamily arXiv:1711.08467 [hep-th]}}.

\bibitem{tHooft:1990fkf}
G.~'t~Hooft, ``{The black hole interpretation of string theory},''
  \href{http://dx.doi.org/10.1016/0550-3213(90)90174-C}{{\em Nucl. Phys. B}
  {\bfseries 335} (1990) 138--154}.

\bibitem{Kabat:1992tb}
D.~N. Kabat and M.~Ortiz, ``{Eikonal quantum gravity and Planckian
  scattering},'' \href{http://dx.doi.org/10.1016/0550-3213(92)90627-N}{{\em
  Nucl. Phys. B} {\bfseries 388} (1992) 570--592},
  \href{http://arxiv.org/abs/hep-th/9203082}{{\ttfamily arXiv:hep-th/9203082}}.

\bibitem{Dray:1984ha}
T.~Dray and G.~'t~Hooft, ``{The Gravitational Shock Wave of a Massless
  Particle},'' \href{http://dx.doi.org/10.1016/0550-3213(85)90525-5}{{\em Nucl.
  Phys. B} {\bfseries 253} (1985) 173--188}.

\bibitem{Gu:2021xaj}
Y.~Gu, A.~Kitaev, and P.~Zhang, ``{A two-way approach to out-of-time-order
  correlators},'' \href{http://arxiv.org/abs/2111.12007}{{\ttfamily
  arXiv:2111.12007 [hep-th]}}.

\bibitem{Shenker:2013pqa}
S.~H. Shenker and D.~Stanford, ``{Black holes and the butterfly effect},''
  \href{http://dx.doi.org/10.1007/JHEP03(2014)067}{{\em JHEP} {\bfseries 03}
  (2014) 067}, \href{http://arxiv.org/abs/1306.0622}{{\ttfamily arXiv:1306.0622
  [hep-th]}}.

\bibitem{kitaevfundamental}
A.~Kitaev. \url{https://www.youtube.com/watch?v=OQ9qN8j7EZI}.
\newblock Talk given at the Fundamental Physics Prize Symposium, Nov. 10, 2014.

\bibitem{Shenker:2014cwa}
S.~H. Shenker and D.~Stanford, ``{Stringy effects in scrambling},''
  \href{http://dx.doi.org/10.1007/JHEP05(2015)132}{{\em JHEP} {\bfseries 05}
  (2015) 132}, \href{http://arxiv.org/abs/1412.6087}{{\ttfamily arXiv:1412.6087
  [hep-th]}}.

\bibitem{Maldacena:2016upp}
J.~Maldacena, D.~Stanford, and Z.~Yang, ``{Conformal symmetry and its breaking
  in two dimensional Nearly Anti-de-Sitter space},''
  \href{http://dx.doi.org/10.1093/ptep/ptw124}{{\em PTEP} {\bfseries 2016}
  no.~12, (2016) 12C104}, \href{http://arxiv.org/abs/1606.01857}{{\ttfamily
  arXiv:1606.01857 [hep-th]}}.

\bibitem{Haehl:2021dto}
F.~M. Haehl, A.~Streicher, and Y.~Zhao, ``{Six-point functions and collisions
  in the black hole interior},''
  \href{http://arxiv.org/abs/2105.12755}{{\ttfamily arXiv:2105.12755
  [hep-th]}}.

\bibitem{Yang:2018gdb}
Z.~Yang, ``{The Quantum Gravity Dynamics of Near Extremal Black Holes},''
  \href{http://dx.doi.org/10.1007/JHEP05(2019)205}{{\em JHEP} {\bfseries 05}
  (2019) 205},
\href{http://arxiv.org/abs/1809.08647}{{\ttfamily arXiv:1809.08647 [hep-th]}}.

\bibitem{Saad:2019lba}
P.~Saad, S.~H. Shenker, and D.~Stanford, ``{JT gravity as a matrix integral},''
\href{http://arxiv.org/abs/1903.11115}{{\ttfamily arXiv:1903.11115 [hep-th]}}.

\bibitem{Maldacena:2018lmt}
J.~Maldacena and X.-L. Qi, ``{Eternal traversable wormhole},''
\href{http://arxiv.org/abs/1804.00491}{{\ttfamily arXiv:1804.00491 [hep-th]}}.

\bibitem{Sunderhauf:2019djv}
C.~S\"underhauf, L.~Piroli, X.-L. Qi, N.~Schuch, and J.~I. Cirac, ``{Quantum
  chaos in the Brownian SYK model with large finite $N$: OTOCs and tripartite
  information},'' \href{http://dx.doi.org/10.1007/JHEP11(2019)038}{{\em JHEP}
  {\bfseries 11} (2019) 038}, \href{http://arxiv.org/abs/1908.00775}{{\ttfamily
  arXiv:1908.00775 [quant-ph]}}.

\bibitem{Jian:2020krd}
S.-K. Jian and B.~Swingle, ``{Note on entropy dynamics in the Brownian SYK
  model},'' \href{http://arxiv.org/abs/2011.08158}{{\ttfamily arXiv:2011.08158
  [cond-mat.stat-mech]}}.

\bibitem{Stanford:2019vob}
D.~Stanford and E.~Witten, ``{JT Gravity and the Ensembles of Random Matrix
  Theory},'' \href{http://arxiv.org/abs/1907.03363}{{\ttfamily arXiv:1907.03363
  [hep-th]}}.

\bibitem{Kitaev:2018wpr}
A.~Kitaev and S.~J. Suh, ``{Statistical mechanics of a two-dimensional black
  hole},'' \href{http://dx.doi.org/10.1007/JHEP05(2019)198}{{\em JHEP}
  {\bfseries 05} (2019) 198},
\href{http://arxiv.org/abs/1808.07032}{{\ttfamily arXiv:1808.07032 [hep-th]}}.

\end{thebibliography}\endgroup

\bibliographystyle{utphys}

\end{document}